# A Quantitative Model of Charge Injection by Ruthenium Chromophores Connecting Femtosecond to Continuous Irradiance Conditions

Thomas P. Cheshire[1], Jéa Shetler-Boodry[1,2], Erin A. Kober[3], M. Kyle Brennaman[3], Paul G. Giokas[4], David F. Zigler[5], Andrew M. Moran[3], John M. Papanikolas[3], Gerald J. Meyer[3], Thomas J. Meyer[3], and Frances A. Houle[1]*

[1] Chemical Sciences Division, Lawrence Berkeley National Laboratory Berkeley, CA 94720

[2] Department of Chemical and Biochemical Engineering, University of California, Berkeley, CA 94720

[3] Department of Chemistry, University of North Carolina at Chapel Hill, Chapel Hill, NC 27599

[4] Coherent Inc., Santa Clara, CA 95054

[5] Chemistry & Biochemistry Department, California Polytechnic State University, San Luis Obispo, CA 93407

**ABSTRACT:** A kinetic framework for the ultrafast photophysics of tris(2,2-bipyridine)ruthenium(II) phosphonated and methyl-phosphonated derivatives is used as a basis for modeling charge injection by ruthenium dyes into a semiconductor substrate. By including the effects of light scattering, dye diffusion and adsorption kinetics during sample preparation, and the optical response of oxidized dyes, quantitative agreement with multiple transient absorption datasets is achieved on timescales spanning femtoseconds to nanoseconds. In particular, quantitative agreement with important spectroscopic handles—decay of an excited state absorption signal component associated with charge injection in the UV region of the spectrum, and the dynamical redshift of an approximately 500 nm isosbestic point—validates our kinetic model. Pseudo-first-order rate coefficients for charge injection are estimated in this work, with an order of magnitude ranging $10^{11}$ s$^{-1}$ to $10^{12}$ s$^{-1}$. The model makes the minimalist assumption that all excited states of a particular dye have the same charge injection coefficient, an assumption that would benefit from additional theoretical and experimental exploration. We have adapted this kinetic model to predict charge injection under continuous solar irradiation, and find that as many as 68 electron transfer events per dye per second take place, significantly more than prior estimates in the literature.

## 1. Introduction

Molecule-semiconductor electron transfer plays a principal role in solar energy conversion for dye-sensitized solar cells (DSSC) and dye-sensitized photoelectrosynthesis cells (DSPEC).[1-10] Quantifying charge injection frequency and efficiency have been of particular interest, leveraging macroscopic observations to report such values.[11-17] Numerous theoretical and experimental studies[18-28] have attempted to determine the primary steps that govern charge injection efficiency, employing well-founded theory to interpret lifetimes extracted from transient absorption (TA) signals with femtosecond[29-32] to picosecond[33-36] resolution. In these studies, the observed spectral features were assigned to the relevant chemical species (e.g., oxidized form of the dye, molecular excited state, or the injected electron) and the kinetics of the observed absorption changes with time were typically modelled by using sum-of-exponential (SOE) fits. Attempts were then made to distinguish the magnitudes of often overlapping contributions to the observed TA signal. Yet, there are two key limitations of this SOE data treatment: 1) lifetimes extracted from SOE fits of observed TA signals can only be resolved for events well-separated in time and 2) exponential prefactors, often referred to as amplitudes, can only be quantitatively assigned to a specific kinetic step of a mechanism for the simplest of kinetic schemes.[37] We propose a quantitative model of the molecular photophysics and charge injection of a set of ruthenium dyes commonly studied for DSSCs and DSPECs that is not hindered by the pitfalls of SOE analysis. The development of the model in Section 3 below demonstrates the importance of the many factors that must be considered to interpret spectroscopic signals of systems involving molecule-semiconductor electron transfer. These factors reveal key considerations for efficient dye sensitization in DSSCs and DSPECs, and we use our model to make specific predictions for charge injection rates under solar irradiation conditions. Finally, we consider ways in which simpler model analyses could be implemented in an effective manner.

From the earliest studies of tris(2,2-bipyridine)ruthenium(II) (RuBPY),[38-42] this dye has been recognized as a promising charge-transfer agent due to long-lived states assigned to be triplets. TA and transient grating (TG) techniques have been employed to measure and identify charge injection of RuBPY derivatives on a $TiO_2$ substrate. Picosecond excited state lifetimes attributed to charge injection were extracted from simple single exponential fits of excited state absorption (ESA) decays.[30,

[32] In a more rigorous treatment, a two-state electron injection model was proposed and a closed solution was derived for the purpose of fitting TA data.[33] Asbury et al found RuN3—cis-bis(isothiocyanato)bis(2,2'-bipyridyl-4,4'-dicarboxylato)ruthenium(II)—to have biphasic decay from fits of IR spectra of injected electrons for three excitation wavelengths, a <100 fs decay and slower decay of approximately 20 ps. Their use of the electron absorption cross-section makes an explicit connection between the assumed model and the observed signal. These studies are successful at extracting characteristic lifetimes of charge injection, and the latter case, model rate coefficients for total injected electrons. Uncovering the primary ultrafast photophysics that leads to charge injection and the role of the particular excited states in charge injection is not so readily discoverable using these methods, however. A full treatment of the optical transitions and excited state relaxation pathways, with minimal assumptions, is appropriate for building a fundamental understanding of the dyes' role in DSSC and DSPEC, and that is the subject of this work.

## 2. Methods

This study is primarily computational, but is strongly rooted in experiments including data from two previously published investigations and new spectroelectrochemical and reflectance measurements made as needed for testing assumptions and validating the calculations.

### 2.1. Experimental Data Used in this Work

Phosphonated and methyl-phosphonated tris(2,2-bipyridine)ruthenium(2+) (RuBPY) derivatives—RuP, RuP2, RuP3, RuCP, RuCP2, and RuCP3 (Supporting Information (SI) Section S1 Figure S1)— relevant to this and previous studies are referred to as a set **6-Ru**. Experimental data were compiled from previous spectroscopic investigations conducted on **6-Ru** adsorbed to $ZrO_2$ and $TiO_2$ in methanol[29] and acetonitrile,[36] referred to here as **Ru-G** (electrodes in methanol; by Giokas et al) and **Ru-Z** (electrodes in acetonitrile; by Zigler et al) respectively.

### 2.1.1 Transient Absorption Spectra

Sample preparation, storage, and measurement conditions are described briefly. Further details are presented in Ref. 29 for **Ru-G** and Ref. 36 for **Ru-Z**.

*Materials:* Sample properties for **Ru-G** and **Ru-Z** dyes adsorbed to approximately 15-20 nm diameter $ZrO_2$ and $TiO_2$ nanoparticle-based films on fluorine-doped $SnO_2$ (FTO) glass slides are detailed in Table 1 below. The samples were prepared by fabricating the nanoparticle film, then soaking it in a dye solution followed by rinsing to remove as much unbound dye as possible.

**Table 1. Material properties of samples used in experiments yielding datasets Ru-G and Ru-Z**

|  | **Ru-G**[29] | **Ru-Z**[36] |
|---|---|---|
| Film Thickness (μm) | 7 | 2-4 |
| Dyes on $ZrO_2$ | RuP | **6-Ru** |
| Dyes on $TiO_2$ | **6-Ru** | **6-Ru** |
| Dye concentration in solution[a] (mM) | 0.1 | 1 |
| Time film soaked in solution (h) | 2 | 24 |
| Film rinse | 0.1 M $HClO_4$ aqueous solution | 0.1 M $HClO_4$ aqueous solution[b] |
| Storage | under nitrogen in the dark | in 0.1 M $HClO_4$ aqueous solution in the dark |

[a] 0.1 M $HClO_4$ aqueous solution.

[b] New solution was added at least once.

*Experimental Conditions*: In previous studies,[29, 36] TA signals were measured by broadband probe pulses, following an experimentally controlled delay time after a pump pulse. Slides were held in cuvettes and continuously moved during experiments. Linear absorption (LA) spectra were measured before and after acquiring TA signals to ensure the films did not degrade during the course of the experiments. Differences in the experimental conditions for **Ru-G** and **Ru-Z** are detailed in Table 2.

**Table 2: Experimental conditions for TA measurements used for datasets Ru-G and Ru-Z**

|  | **Ru-G**[29,a] | **Ru-Z**[36] |
|---|---|---|
| Solution conditions | 0.08 mM triethanolamine 0.1 M LiClO$_4$ aqueous solution | argon deoxygenated 0.1 M HClO$_4$ aqueous solution |
| Pump duration (fs) | 45-55 | 200 |
| Pump carrier wavelength (nm) | 400 | various (420 to 535) |
| Pump energy (µJ) | 1.5 | ≤0.1 |
| Probe spot size (µm) | 300 | 150 |
| Timescale (s) | $10^{-14}$-$10^{-12}$ | $10^{-13}$-$10^{-9}$ |

[a] Excitation linewidth was approximately 1000 cm$^{-1}$.

### 2.1.2. Reflectance

Transmittance and transflectance measurements were recorded for the present work using a double-beam absorption spectrophotometer (Cary 5000) equipped with an external diffuse reflectance accessory (eDRA-2500) that features a 150-mm diameter integrating sphere outfitted with two built-in detectors, a photomultiplier tube and a PbS unit, to cover the wavelength range from 250 nm to 2500 nm. A small-spot kit was used to limit the size of the incident beam (~3 mm diameter) to be smaller than the thin film samples.

Sample films were held in solution in a 1-cm quartz cuvette at a 45° angle. The reflectance was calculated using Equation (1) below. The cuvette was placed in the center of the integrating sphere to measure the transflectance, SI Section S2 Figure S2A. The transmittance was measured by placing the cuvette at the entrance of the integrating sphere, SI Section S2 Figure S2B.

$$\text{Reflectance} = \text{Transflectance} - \text{Transmittance} \quad (1)$$

**2.1.3 Spectroelectrochemical Measurements**

The synthesis of the mesoporous indium tin oxide (ITO) thin films used for spectroelectrochemistry has been previously reported. A viscous solution of ITO nanoparticles was doctor-bladed onto a masked FTO substrate followed by heating at 450°C under an $O_2$ atmosphere to combust the organic components of the solution. Profilometry measurements established a film thickness of 3-4 μm. The thin films appeared lightly yellow colored with high transparency in the visible region.

The RuP complexes were linked to the ITO nanoparticles by overnight reaction of the ITO thin film with a 30 mM solution of the complex dissolved in 0.1 M $HClO_4$. Removal of the thin film on its substrate from the solution revealed an intense red color. The samples were stored in 0.1 M $LiClO_4/CH_3CN$ until use.

Spectroelectrochemical analysis was conducted with an Avantes AvaLight DHc light source coupled to an Avantes StarLine AvaSpec-2048 ultraviolet-visible spectrophotometer. All experiments were performed in a standard three-electrode cell with an ITO thin film working electrode placed at 90° angle to the light beam. A Pt mesh counter electrode and a silver in 0.1 M $LiClO_4/CH_3CN$ reference electrode were kept proximate to the working electrode yet outside of the optical path. The reference electrode was calibrated by cyclic voltammetry measurements of the ferrocenium/ferrocene ($Fc^{+/0}$ = 630 mV vs NHE) redox waves in a 0.1M *tert*-butylammonium perchlorate/acetonitrile. This $Fc^{+/0}$ redox couple was also measured after spectroelectrochemical studies to ensure negligible drift in the reference. Solutions were sparged with argon gas for 30 min prior to experiments.

In a typical experiment, the applied potential was stepped from +800 to +1400 mV vs NHE in 20 mV increments. A 30 s dwell time was taken between each step during which time the visible absorption spectrum was recorded. This dwell time was sufficient to provide a time independent steady state spectrum. The appearance of isosbestic points indicated that oxidation of $Ru^{II}P$ to $Ru^{III}P$ occurred quantitatively. The spectra measured at all potentials were modelled as a weighted sum of the absorption spectra of the reduced and oxidized complexes from which the mole fraction of each species was extracted. The equilibrium potential where the two redox states were present in equal mole fractions is reported as the standard $E^o$(Ru(III/II)) reduction potential.

## 2.2. Simulation Methods

Due to the multiscale nature of the models in this study, we use a stochastic simulator for the coupled reaction-diffusion kinetics that readily accommodates stiff problems and reaction schema whose details are not completely characterized. Results of the kinetic simulations yield insights into physical and chemical populations and intermediates, and are used to directly simulate steady-state and time-resolved optical signals.

*Kinetiscope:* The open access software Kinetiscope[43] was used to simulate the diffusion of dyes from bulk solution into nanoscopic cavities, adsorption to semiconductor nanoparticles, and the kinetics underpinning the ultrafast molecular photophysics that leads to charge injection. Originally introduced by Bunker,[44] Gillespie[45] fully developed the formalism used in the core algorithm that generates a rigorous solution to the master equation for the system. Random event selections are made from an ensemble in event space, unique from conventional kinetic Monte Carlo which makes selections from an ensemble in real space. The advantages of stochastic chemical kinetics simulations include generation of an absolute time base for direct comparison to experimental data when accurate mechanisms and rate coefficients are used. The simulator uses particles to represent species involved in the reactions, and these particles are automatically self-conserving. Unlike ordinary differential equation solvers for kinetics systems, mass and energy balances are not tracked. This allows introduction of non-chemical marker species into the kinetics steps to gain additional insights into how physiochemical systems evolve in time (See SI Section S3).[37, 46-48]

*Optical simulations:* The time-dependent pump-probe signal is computed from the results of kinetics simulations of dye photophysical processes with and without an initial excitation step. The negative log of the ratio of light-matter interactions, Equation (2), gives the $\Delta A(\tau; \lambda)$ at experimentally controlled delay times $\tau$, such that $I_0(\lambda)$, $I_{on}(\tau; \lambda)$, and $I_{off}(\lambda)$ are the wavelength $\lambda$-dependent probe intensities, signal intensities with an initial pump excitation, and signal intensities without an initial excitation respectively.

$$\Delta A(\tau; \lambda) = -\log_{10}\left|\frac{I_0(\lambda)-I_{on}(\tau;\lambda)}{I_0(\lambda)-I_{off}(\lambda)}\right| \qquad (2)$$

Signal intensities are proportional to the number of photons *n* at a given wavelength, Equation (3).

$$I \propto n\frac{hc}{\lambda} \qquad (3)$$

Marker species are used in the simulations to count interactions of the probe beam with the populations of the dye states, providing a measure of the total number of photons involved in each type of interaction. Further details about the methods used for simulating LA and TA spectra from kinetic data are given in Ref. 37 and Ref. 46.

## 3. Model development
### 3.1. Dye Photophysical Kinetics

The success of our kinetic framework for **6-Ru** in solution in reproducing both steady-state and dynamical spectroscopic signals has laid the necessary foundation for constructing a comprehensive framework for simulating the photophysics of dyes on substrates. Briefly, in Ref 37 (also seen in Figure 1, Jablonski Diagram), we introduced explicit optical interactions into a kinetic scheme of a four-level system—ground state $|S_0\rangle$, highest energy singlet state $|Y\rangle$, lowest energy singlet state $|X\rangle$, and triplet state $|T\rangle$—with non-radiative relaxation pathways between states (black arrows) and an incoherent emission (rose arrow). The optical transitions included the common TA signal components: a) ground state bleach (GSB, blue arrows), b) excited state emission (ESE, blue arrows), and c) excited state absorption (ESA, orange arrows) from probe interactions as well as absorption from $|S_0\rangle$ to excited states from both pump and probe interactions (blue arrows). The pump pulse interactions were treated as a square wave in the time-domain with a constant rate of excitation, and the probe pulse was assumed to have constant intensity across all wavelengths.

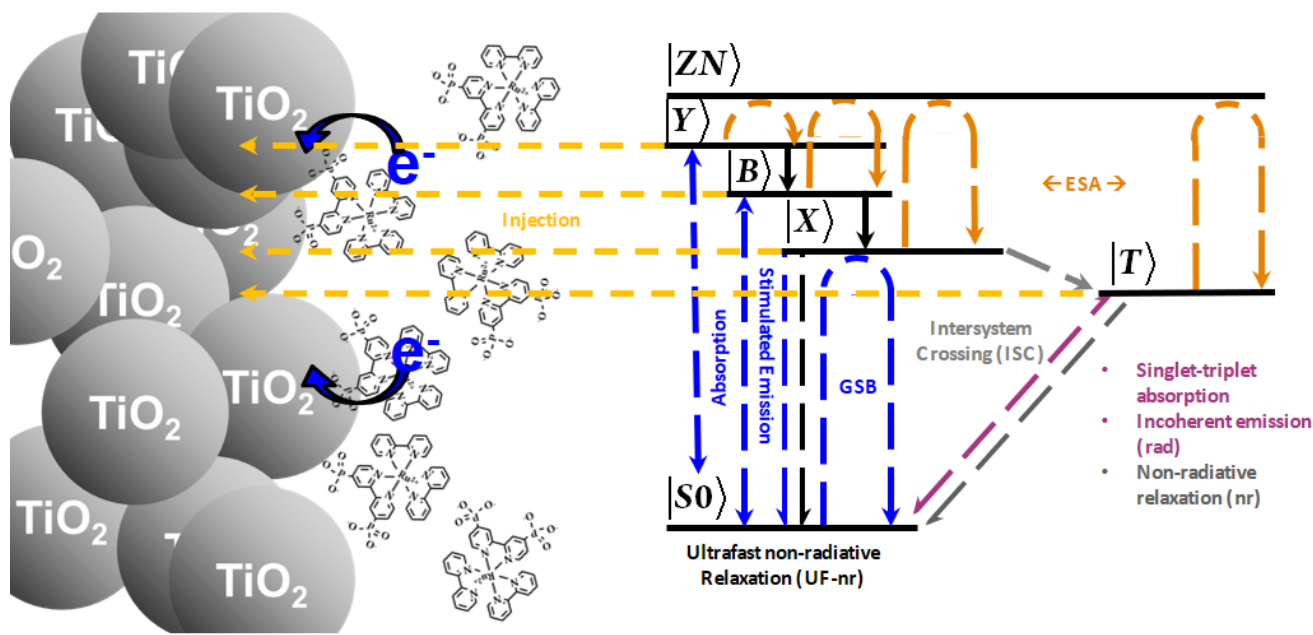

Figure 1. Jablonski diagram illustrating the molecular optical and non-radiative transitions necessary for simulating the photophysics for set of six structurally related dyes, 6-Ru, and the excited states involved in charge injection (gold arrows). Transitions between the electronic ground state and the electronically excited singlet and triplet manifolds are represented in blue and rose, respectively. ESAs are orange, and non-radiative pathways are black or gray. Model 0τ (see text) was constructed to simulate transient measurements without intermediate state |B⟩. This figure was adapted with permission from Cheshire, T. P.; Houle, F. A., Ruthenium Dye Excitations and Relaxations in Natural Sunlight. J. Phys. Chem. A 2021, doi.org/10.1021/acs.jpca.1c02386 (Ref 46), Copyright 2021, American Chemical Society.

Marker species for optical transitions were used to calculate decays of TA signal components for comparison to experimental measurements. By requiring that we have quantitative agreement between experimental and simulated signals, we were able to predict 1) rate coefficients for non-radiative relaxation channels, including an ultrafast relaxation pathway from excited state $|X\rangle$ to $|S_0\rangle$ that was previously unreported, 2) intersystem crossing (ISC) efficiencies for set **6-Ru**, and 3) transition dipole moment magnitudes for ESAs. Lastly, we confirmed that the proposed ultrafast photophysics has a negligible effect on the microsecond dynamics (i.e. non-radiative relaxation and incoherent emission from $|T\rangle$).

In Ref. 46, we extended our model of the **6-Ru** photophysics to simulate the steady-state dynamics under solar irradiation. We adapted the pump interactions in the TA model to use the signal components (e.g. GSBs and ESAs) to produce the optical rate coefficients as seen in Equation (4); in which $k_{i,f}$ are optical rate coefficients, $I_{Pump:\,i,f}(\lambda)$ are laser pump intensities, and $k_{Pump:\,i,f}(\lambda)$ are optical rate coefficients from initial state $i \in \{Y, B, X, T\}$ to final state $f \in \{Y, B, X, T\}$.

$$k_{Pump:\,i,f} = k_{i,f} \cdot \int I_{Pump:\,i,f}(\lambda)\, d\lambda \tag{4}$$

To reproduce the full UV-Vis LA spectra (SI Section S4 Figure S4), a bridge state $|B\rangle$ between singlet states $|Y\rangle$ and $|X\rangle$ was introduced (a five-level system) with the simple assumption that the timescale of the $|Y\rangle$ to $|X\rangle$ transition should remain constant between models (i.e. $\frac{1}{k_{YX}} = \frac{1}{k_{YB}} + \frac{1}{k_{BX}}$ and $k_{YB} = k_{BX}$). The addition of the intermediate state had a negligible effect on the full time-dependent TA spectra. Lastly, solar rate coefficients ($k_{Solar:\,i,f}$) computed using the optical transition coefficients $k_{i,f}$, Equation (5), and the AM 1.5 global tilt spectrum ($I_{Solar:\,i,f}(\lambda)$),[49-51] were used to count the number of optical and non-radiative transitions per dye per second (dye[-1] s[-1]) for dyes in **6-Ru** under 1-sun condition from the simulations:

$$k_{Solar:\,i,f} = k_{i,f} \cdot \int I_{Solar:\,i,f}(\lambda)\, d\lambda \tag{5}$$

These basic solution-phase kinetic schemes were used as starting points for models of photophysics of dyes adsorbed to metal oxide substrates.

### 3.2. Dye on Substrate Model

Mesoporous thin films are inherently an inhomogeneous distribution of semiconductor and cavity concentrations and morphologies. The effects of such distributions complicates how dye concentrations and light-matter interactions are handled, which are critical details for a robust treatment of dye on substrate photophysics. In this section we describe how we accounted for sample inhomogeneities by simulations of the dye sorption process.

### 3.2.1. Dye Concentrations in the Mesoporous Films

Sample composition and concentrations are necessary for the quantitative kinetic interpretation of spectroscopic data. Systems containing chromophores that exhibit similar absorption features and photophysics involve non-trivial overlaps in optical response and their kinetics must be separately tracked. In the specific case of dyes that impregnate mesoporous thin films, dyes that are unbound or bound to the nanoparticles are present, and both homogeneous and heterogeneous dye states contribute to measured LA and TA signals. Typically, the surface coverage of a dye on a semiconductor is estimated using the Langmuir equation, Equation (6).[12]

$$\Gamma = \Gamma_{MAX} \frac{K_{Adsorption}[Dye]}{1+K_{Adsorption}[Dye]} \tag{6}$$

The surface concentration $\Gamma$ is calculated from the concentration of the dye in bulk solution, the adsorption equilibrium constant $K_{Adsorption}$, and the maximum observed surface concentration $\Gamma_{MAX}$, calculated from the Beer-Lambert law. This approach rests on the assumptions that 1) the Langmuir model is appropriate for high surface area mesoporous samples and 2) the presence of a substrate does not have an effect on the molecular photophysics so that absorptions are proportional to coverages alone.

To bypass the first of these assumptions, in the present study we did not use Eq 5. Rather, a 3D model based on the framework described in Refs. 47, 48 was developed to simulate dye diffusion into and adsorption onto a semiconductor mesoporous thin film following the preparation used in Ref. 29. As shown in Figure 2A, 7-μm nanocavities with cross-sections of 100 nm², 400 nm², and 1600 nm² and 1-nm thick solution-semiconductor interfaces represent the geometry of the nanoparticle film. The density of dye adsorption sites in these materials is unknown, so two cases are assumed, 0.1 and 1 sites/nm². The bulk solution phase has a depth of ~1 cm, the width of a cuvette. The structure in Figure 2A is periodic, using wrap-around diffusion paths to convert the 2-pore system into a semi-infinite array. Adsorption and desorption of dyes from solution follow the reaction step in Equation (7), with the associated rate coefficients $k_{Adsorption}$ = 0.86 M$^{-1}$ s$^{-1}$ and $k_{Desorption}$ = 5.0·10$^{-5}$ s$^{-1}$.[12]

$$Unbound + Site \rightleftarrows Bound \tag{7}$$

These values are for RuP; parameters for the full **6-Ru** series are within an order of magnitude of each other.[12]

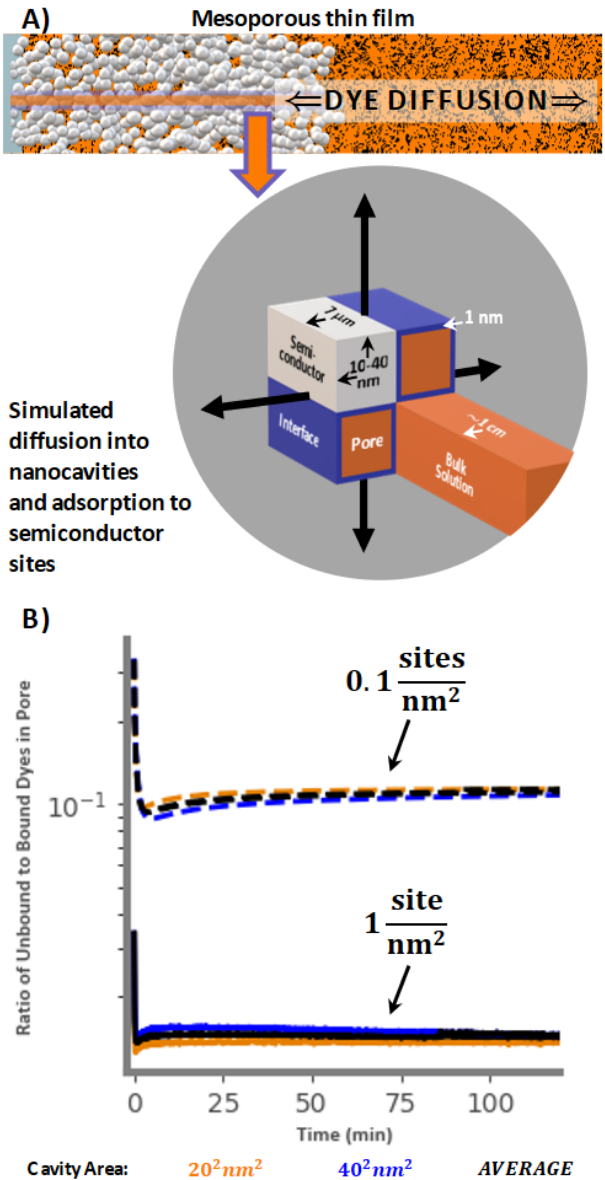

Figure 2. A) Dye diffusion into nanocavities of mesoporous thin films and adsorption to semiconductor sites is modeled with wraparound diffusion paths and 1-nm interfaces, where dye adsorption occurs, between the solutions in the pores and the semiconductor. B) Simulated observable unbound-to-bound dye concentrations following films with 100-nm², 400-nm², and 1600-nm² pores soaking in dye solution for 2 h and soaking in fresh solution for an additional 2 h.

Four cases (two adsorption site densities for two pore sizes) were examined using conditions in Table 1 for **Ru-G**. Simulations of the soaking process for all cases were made for an initial dye concentration in the bulk solution of 0.1 mM. Following a simulated time of 2 h, the concentrations of unbound and bound dyes and free sites were used as starting conditions for follow-on simulations of adsorption-desorption using pure solvent only for an additional 2h. The resulting ratios of unbound-to-bound dyes are shown in Figure 2B.

Though it is clear the pore size has a negligible effect on the ratio of unbound and bound dyes, the effect of the concentration of adsorption sites is significant. Without knowledge of the distribution, or at least the average, adsorption site concentration, we must make the approximation that the fraction of unbound to bound dyes falls between $10^{-2}$ and $10^{-1}$. These simulations are directly relevant only to dataset **Ru-G**. As shown in Table 1, **Ru-Z** films were soaked in solution with a higher concentration of dyes and for a longer time, which will allow more dyes to diffuse deeper into pores. However, rinsing multiple times in clean solutions can be expected to be more effective for removing unbound dyes. Therefore, for dataset **Ru-Z**, we make the assumption that the concentration of unbound dyes in the pores is negligible due to the higher concentration of dyes in the soaking solution and the longer time the films were soaked. We assume that rinsing the films affects dye concentrations less than the initial adsorption cycle. Our simulations show that this latter assumption is justified because re-adsorption of dyes released from nanoparticle surfaces is found to out-compete diffusion through the pores into the bulk solution, whereas diffusion from bulk solution during the initial adsorption cycle into the pore is aided by adsorption since diffusion is gradient driven.

### 3.2.2. Influence of Nanoporous Film Reflectance on Pump and Probe Pulses

$ZrO_2$ and $TiO_2$ films prepared from ~20 nm diameter nanoparticles exhibit significant diffuse light scattering at ultraviolet and blue wavelengths (measurements shown in SI Section S2 Figure S3).[52, 53] The increased reflectance effectively increases the electromagnetic field strength around the 400 nm to 500 nm spectral region, thereby increasing the number of optical interactions. As a result, pump laser pulses produce a higher number of dye excitations in the films than in solution. Probe laser pulses also generate more light-matter interactions in the films than in solution, and such interactions generate signals that propagate both in the direction of the spectrometer and in the direction of the scattered probe light. In both **Ru-G** and **Ru-Z**, the probe spot size is smaller than that of the pump pulse (Table 2), making it a fair approximation that all of the direct and scattered pump light impinges on the dye molecules measured by the probe pulse, illustrated in Figure 3A.

The effect of pump light scatter is included in the rate coefficients used in our model as shown in Equation (8), integrating over the product of 1 plus the fraction of light reflected, $\phi_{\text{Reflectance}}(\lambda)$, and the wavelength-dependent rate coefficient for the pump excitation of each absorptive signal component $\tilde{k}_{Pump,i}(\lambda)$, where $i \in \{Y, B, X, T\}$.

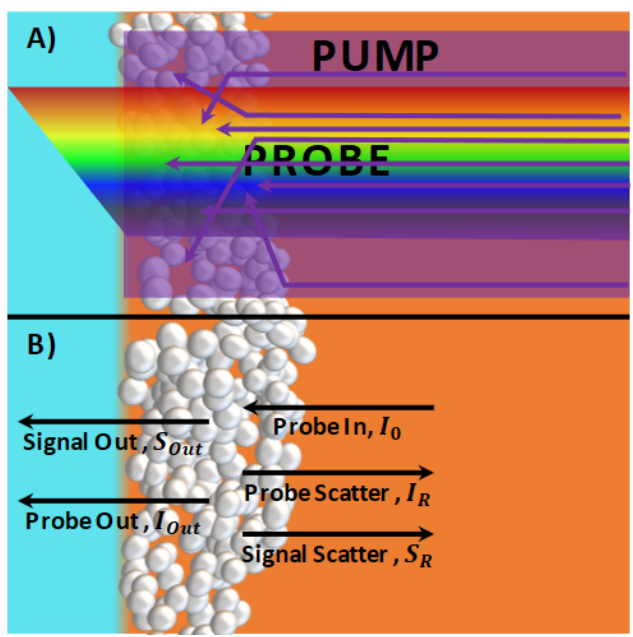

Figure 3. Depicted is a mesoporous thin film of semiconductor nanoparticles (white spheres) on an FTO slide (light blue) in solvent (orange). A) Illustration of pump scatter into region of sample measured by probe laser pulse. B) Probe and signal propagations.

$$k_{Pump,i} = \int (1 + \phi_{\text{Reflectance}}(\lambda)) \cdot \tilde{k}_{Pump,i}(\lambda) \, d\lambda \qquad (8)$$

Caution is taken in considering the effect of probe light scatter, as not all signal is directed toward the spectrometer. Figure 3B depicts the paths of probe and signal light. The incident probe $I_0(\lambda)$ 1) passes through the sample with intensity $I_{Out}(\lambda)$, 2) interacts with a dye molecule, or 3) scatters with reflectance intensity $I_R(\lambda)$. The signal $S$ generated from $I_0(\lambda)$ is either measured, $S_{Out}(\lambda)$, or scattered away from the detector, $S_R(\lambda)$. From Equation (2), $I_0(\lambda)$, $I_{On}(\tau;\lambda)$, and $I_{Off}(\lambda)$—producing the measured $I_{Out}(\lambda)$ and $S_{Out}(\tau;\lambda)$ for both pump on and pump off measurements—will have the same scatter intensity offset $S_R(\lambda)$ originating from $\phi_{\text{Reflectance}}(\lambda)$. Therefore, it is not necessary to modify how $\Delta A(\tau;\lambda)$ is computed from Eq. 2, as the scattered component cancels out. Although the fraction of $S_R(\lambda)$ from $I_R(\lambda)$ is not part of the observed TA signal, dye molecules interact with the reflected light and generate signal that propagates in directions other than the spectrometer. Interactions with scattered light do influence the probabilities of events, however, and are included in the kinetic scheme to quantitatively simulate the dye photophysics.

### 3.2.3. Transient Absorption of Dyes on ZrO$_2$

The kinetic scheme for dye absorptions and relaxations in solution (SI Section 5 Table S1) is the starting point for the scheme for dyes on ZrO$_2$, where the photophysics may be perturbed by the presence of the substrate but charge injection does not occur. Examples of the reaction steps needed to expand the scheme to simulate dyes on ZrO$_2$ are given in Equations (9)-(11). To account for bound and unbound dyes, we include steps for the two initial populations throughout the scheme, delineating the marker species appropriately. In this example, equations (9) and (10) represent probe interactions that generate ESA signal components from coherences between excited state $|X\rangle$ and an implicit higher energy excited state $|Z_i\rangle$ of bound and unbound dyes respectively. Such equations are necessary for both dye adsorption states for all radiative and non-radiative transitions (e.g. GSB, ESE, and ISC). Equation (11) is similar to Equation (9), involving scattered probe light in place of direct probe light. We only include kinetic steps involving scattered probe light for bound dyes, not unbound dyes which have a lower concentration thereby making the equivalent steps kinetically insignificant. For the full kinetic scheme see SI Section S6 Tables S2 and S3.

$$X_{Bound} \xrightarrow{Z_{i,Bound}, I_{Probe}(\tau)} X_{Bound} + ESA_{i,Bound} \qquad (9)$$

$$X_{Unbound} \xrightarrow{Z_{i,Unbound}, I_{Probe}(\tau)} X_{Unbound} + ESA_{i,Unbound} \qquad (10)$$

$$X_{Bound} \xrightarrow{Z_{i,Bound}, I_{Reflect}(\tau)} X_{Bound} + ESA_{i,Bound,Reflect} \qquad (11)$$

### 3.3. Injection
### 3.3.1. The General Injection Step

It is straightforward to extend the model thus far developed to include charge injection kinetics. We are well justified to assume that only excited states of semiconductor bound dyes have non-negligible probability for electron transfer into the TiO$_2$ conduction band. The correct reaction steps for charge injection are second-order steps, but can be treated as pseudo-first-order steps under the assumption that the number of conduction band acceptor states greatly exceeds the number of molecular excited states, and therefore the TiO$_2$ acceptor state concentration would be approximately constant. The pseudo-first-order rate coefficients are then the product of the second-order rate coefficient and the TiO$_2$ acceptor state concentration. So-called trap states are not well-quantified and likely vary greatly between samples, and are thus neglected in our model. As part of the charge transfer step given by Equations (12)-(15), we use marker species to track the number of oxidized dyes (Ru$^{III}$), injected electrons in the conduction band (TiO$_2^*$), and the molecular state from which the electron was injected (*Injection$_i$*). Finally, back electron transfer (BET) is neglected because the electron concentration in acceptor states would be low following electron diffusion into bulk semiconductor states.[54, 55]

$$Y_{Bound} + TiO_2 \rightarrow Ru^{III} + TiO_2^* + Injection_Y \qquad (12)$$

$$B_{Bound} + TiO_2 \rightarrow Ru^{III} + TiO_2^* + Injection_B \qquad (13)$$

$$X_{Bound} + TiO_2 \rightarrow Ru^{III} + TiO_2^* + Injection_X \qquad (14)$$

$$T_{Bound} + TiO_2 \rightarrow Ru^{III} + TiO_2^* + Injection_T \quad (15)$$

The rate coefficients for charge injection have been estimated using a simple decay of exponentials analysis,[29-33] but are not independently known. The goal of the present study is to determine them within the full quantitative photophysical kinetic scheme. The simplest approximation is to set the primary pseudo-first-order rate coefficients for injection from each singlet excited state and the triplet state $k_{Injection,i}$ to be equal to the product of the primary second-order rate coefficient for charge injection and the TiO$_2$ conduction band density of states (DOS), Equation (16).

$$k_{Injection,Y} = k_{Injection,B} = k_{Injection,X} = k_{Injection,T} = k'_{Injection}[TiO_2] \quad (16)$$

The primary pseudo-first-order rate coefficients encode DOS, therefore knowledge of the specific DOS is necessary to extract the primary second-order injection coefficient $k'_{Injection}$.

### 3.3.2. Optical Response of Oxidized Molecular Species

The population of oxidized ruthenium, Ru$^{III}$, increasingly interacts with impinging light as the Ru$^{III}$ population grows following charge injection. Figure S5 of SI section S7 shows the relative absorption of the Ru$^{II}$ (blue) and Ru$^{III}$ (orange) species. Though the LA intensity of the ground state species is generally more intense than the absorption of the oxidized species, there is non-negligible absorption by Ru$^{III}$ in the blue to UV region of the spectra. The ratio of the oxidized dye absorption intensity to that of the ground state dye is given in SI Figure S6. With the exception of RuCP3 (Figure S6F, SI Section S7), the absorption intensity of the oxidized dye is within 10-20% that of the ground state dye absorption near the peak of the visible metal-to-ligand charge transfer (MLCT) band. However, the net absorption intensity of Ru$^{III}$ relative to the ground state is substantially greater in both near-UV and NIR regions, a characteristic that gives rise to a positive contribution to the TA signal in each of these spectral regions. The optical transition in Equation (17) captures the combined Ru$^{III}$ TA signal contribution. The rate coefficient for the oxidized dye response ($k_{Oxidized}$) is given by Equation (18). The product of the absorption spectrum of the oxidized dye normalized by the MLCT peak of the absorption spectrum (SI Section S7 Figure S5, $\tilde{\sigma}_{Oxidized}(\lambda)$), multiplied with the rate coefficient for absorption to the $|X\rangle$ state $(k_{Abs,X})$, and summed to determine $k_{Oxidized}$.

$$Ru^{III} \xrightarrow{Ru^{III*}, I_{Probe}(\tau)} Ru^{III} + Oxidized\ Response \tag{17}$$

$$k_{Oxidized} = \sum k_{Abs,X} \cdot \tilde{\sigma}_{Oxidized}(\lambda) \tag{18}$$

## 4. Results and Discussion

### 4.1. Dyes on ZrO$_2$

The experimental and simulated full TA spectra for dye RuP in dataset **Ru-G** are shown in Figure 4 panels A and B, respectively, and a comparison of experimental (grays) and simulated (blues) TA lineshapes at delay times of 0 fs, 100 fs, and 500 fs are shown in panel C. The two important adaptations to the solution phase model for the dye on ZrO$_2$ model to achieve the remarkable quantitative agreement are to account for: 1) the dye concentration in the film using adsorption simulations and 2) incorporating scattered pump light into the photophysics. The fraction of unbound to bound dyes is not relevant for dyes on ZrO$_2$ as the dyes on ZrO$_2$ are presumed to have a negligible probability for charge injection upon MLCT photoexcitation. A key success is that the isosbestic point located at ~500 nm for the dyes of set **6-Ru** is redshifted in the simulated spectrum as is observed in the experimental signal.

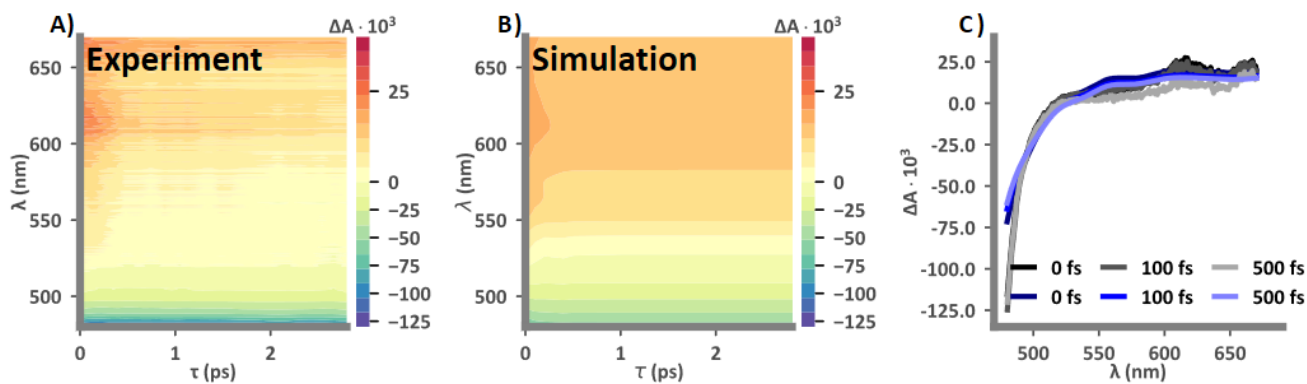

Figure 4. A) Experimental and B) simulated TA spectra of dye RuP on ZrO$_2$ from dataset **Ru-G**. C) Direct comparison of experimental (grays) and simulated (blues) TA lineshapes at delay times of 0 fs, 100 fs, and 500 fs.

Dataset **Ru-Z** contains significantly more measurements for all of the dyes in **6-Ru** on ZrO$_2$ than the **Ru-G** dataset, making it ideal for testing our model against a wide range of conditions (e.g. dye, pump wavelength, and film thickness). Figure 5 shows experimental and simulated results for RuP pumped at 420 nm. The film thickness for each experimental trial is unknown, but can be matched using simulations of films 1 μm (top row), 2 μm (middle row), and 4 μm (bottom row) thick respectively. The probe spectrum for dataset **Ru-Z** spanned approximately 380 nm to 700 nm, illuminating a high energy ESA not observed in dataset **Ru-G**. It is notable that by achieving agreement with the intensities of the high energy ESA and the GSB, the second isosbestic point at ~400 nm is also in good agreement for all experimental delays out to 300 ps, as exhibited in panels C, F, and I. The predicted low energy ESAs are not observed in the experimental measurements, as is evident in panels A, D, and G. We assume the lack of positive signal in the **Ru-Z** measurements at wavelengths greater than 525 nm is due to measurement conditions.

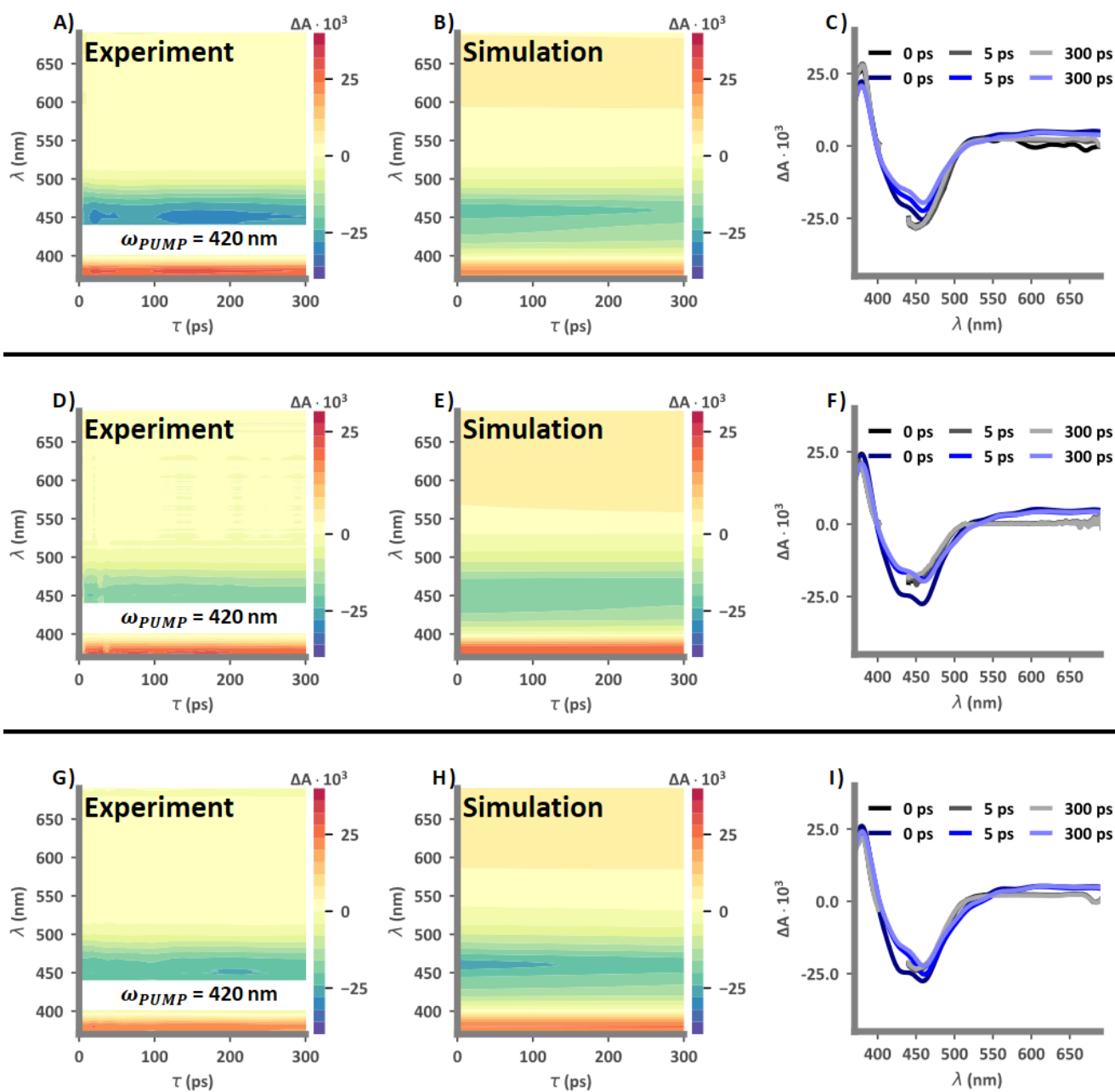

Figure 5. Each row is a different experimental signal obtained using films of different thicknesses, simulated as 1 μm (panels A, B, C), 2 μm (panels D, E, F), and 4 μm (panels G, H, I) thick films. Experimental (panels A, D, and G) and simulated (panels B, E, and H) TA spectra of dye RuP on $ZrO_2$ from dataset **Ru-Z**. C, F, and I: Direct comparison of experimental (grays) and simulated (blues) TA lineshapes at delay times of 0 fs, 5 ps, and 300 ps.

SI Section S8 presents the experimental and simulated spectroscopic data for the full set of dyes **6-Ru** on $ZrO_2$. Figure S5 in Section S8.1 shows RuP pumped at 470 nm (top row) and 535 nm (bottom row). The results are similar to Figure 5 above; there is good agreement at wavelengths < 500 nm and no ESA are exhibited in the experimental data at wavelengths >500 nm. In Sections S6.2-S6.6, Figures S6-S10, the experimental and simulated spectra of the remaining dyes pumped at 420 nm (top rows) and 535 nm (bottom rows) show the same patterns as those of RuP. By using rate coefficients specific to each dye in the set, the pump wavelength used, and the thickness of the films, our model can be used to simulate a vast swath of parameter space.

**4.2. Dyes on $TiO_2$**

Having demonstrated that the photophysics of dyes on a metal oxide substrate where no injection occurs are very similar to those in solution, we extend the model for the dyes on $ZrO_2$ to include charge injection into $TiO_2$ by simply including the reaction steps in Equations (12)-(18) (SI Section S9 Tables S4 and S5). It should be noted that we find the ultrafast decay mechanism, necessary to describe the molecular photophysics of **6-Ru** in solution and on substrates not associated with charge injection, is not required to successfully simulate TA signals for RuP and RuP2. This decay pathway is overcome by charge injection kinetics for dyes RuP and RuP2, effectively making it negligible. For the remaining dyes, the ultrafast relaxation pathway is necessary to quantitatively simulate the TA spectra. Figure 6 and Figure 7 present A) experimental and B) simulated TA signals, and C) spectra comparing the two signals at various delay points for RuP from datasets **Ru-G** and **Ru-Z,** respectively. Panels A) and B) in Figure 6 exhibit a spectral handle unique to the dye RuP on $TiO_2$. The experimental isosbestic point for RuP on $ZrO_2$ is at approximately 520 nm for all delays (Figure 4A) while it starts at approximately 575 nm for RuP on $TiO_2$ and redshifts within 1 ps (Figure 6A). In the simulated spectrum (Figure 6B), the corresponding isosbestic point redshifts ~1000 $cm^{-1}$ within 1 ps, which is not observed for the simulated spectrum of RuP on $ZrO_2$ (Figure 4A). In Figure 7, the simulated signal reaches an asymptotic limit of this redshift much earlier than the experimental signal, though final wavelengths are in agreement. It is additionally observed that there is a decay of the ESA located at ~380 nm for RuP on TiO$_2$ that is absent for RuP on ZrO$_2$ (Figure 5).

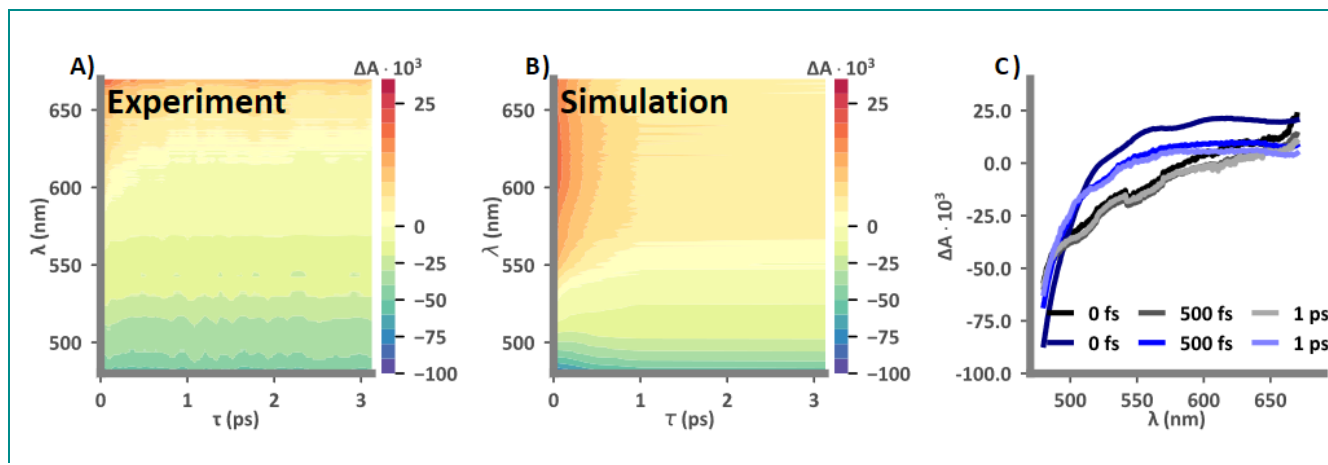

Figure 6. A) Experimental and B) simulated TA spectra of dye RuP on TiO$_2$ from dataset **Ru-G**. C) Direct comparison of experimental (grays) and simulated (blues) TA lineshapes at delay times of 0 fs, 500 fs, and 1 ps.

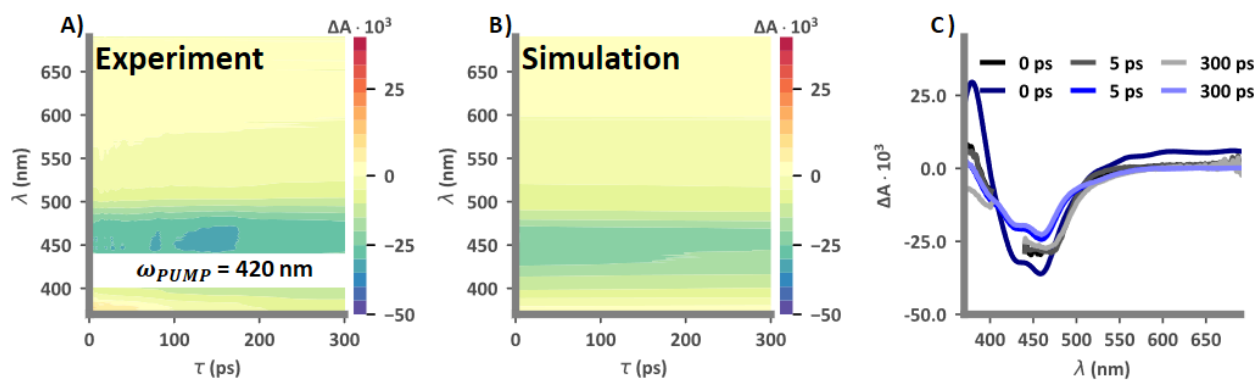

Figure 7. A) Experimental and B) simulated TA spectra of dye RuP on TiO$_2$ from dataset **Ru-Z**. C) Direct comparison of experimental (grays) and simulated (blues) TA lineshapes at delay times of 0 fs, 500 fs, and 1 ps.

Giokas et al used the redshift of the isosbestic point located in the green region of the spectrum to characterize the ultrafast charge injection of the dyes in set **6-Ru**,[29] such a shift was not observed with dyes on ZrO$_2$. Charge injection leads to a decrease in ESA and an increase in GSB and ESE signal components, exhibited as the redshift of the isosbestic point. If charge injection were the only event that drove this shift, then it would be expected there would be a near perfect correlation between the total charge injection rate and the energy change associated with the redshift of the isosbestic point. Whether this is true for RuP is examined in Figure 8. Figure 8A shows the change in the difference between the final isosbestic point frequency $\omega(t)$ and the time-dependent isosbestic point frequency $\omega(-\infty)$ in wavenumbers; $\Delta\omega(\tau) = \omega(t) - \omega(-\infty)$ as a function of time. The experimental and simulated shifts are in remarkable agreement. Panel B) shows how the molecular excited states are predicted to contribute to the total injection rate over time, as determined using markers in the simulations. The inset of Figure 8B shows that within 100 fs the majority of electrons transferred from dye to semiconductor come from $|T\rangle$. A correlation between the total charge injection rate and the energy change associated with the isosbestic point shift is highlighted in Figure 8C by the diagonal line, and compared to data taken from experimental and simulated spectra. There appears to be near perfect correlation confirming that charge injection is the only process contributing to the red shift. This is because there is no ultrafast decay to $|S_0\rangle$ competing with charge injection for dye RuP—relaxation to $|S_0\rangle$ from $|T\rangle$ is negligibly slow.

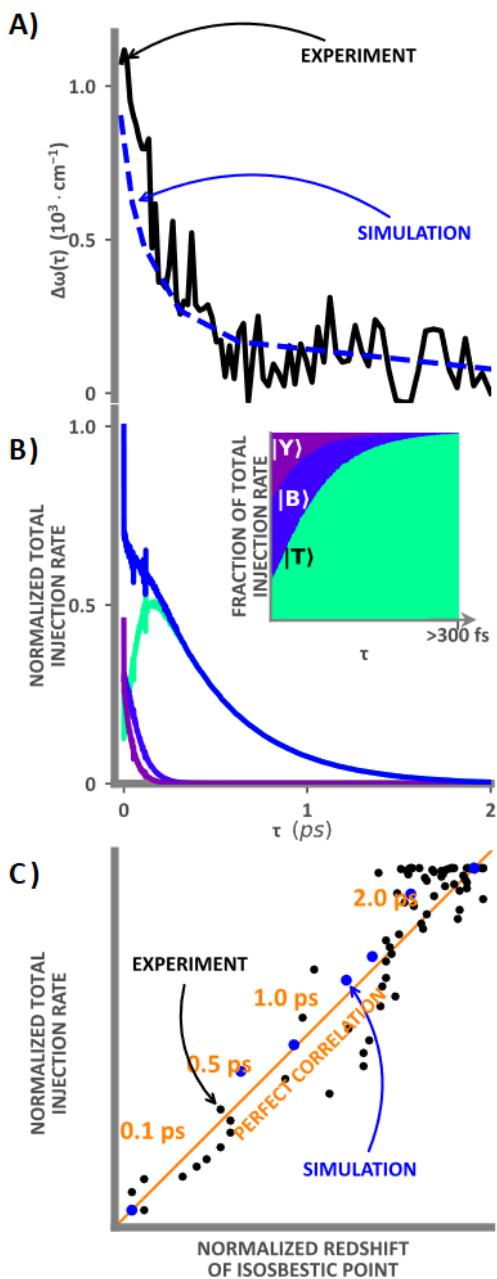

Figure 8. A) The change in experimental (black) and simulated (blue) difference between the final isosbestic point frequency $\omega(t)$ and the time-dependent isosbestic point frequency $\omega(-\infty)$ in wavenumbers. B) Normalized total simulated injection rate (blue) and contributions of the total normalized injection rate for states $|Y\rangle$ (violet), $|B\rangle$ (indigo), and $|T\rangle$ (cyan); charge injection from state $|X\rangle$ was not observed in simulations. The inset shows the fraction of the total normalized injection rate for states $|Y\rangle$ (violet), $|B\rangle$ (indigo), and $|T\rangle$ (cyan). C) Normalized total simulated injection rate plotted against the experimental (black) and simulated (blue) normalized redshift of the isosbestic point. The points are clustered along the diagonal (orange), indicating near perfect correlation.

A full discussion of the remaining dyes of **6-Ru** can be found in SI Section S10, and is summarized briefly here. In general, the spectra of the three phosphonated derivatives of RuBPY are similar, though the initial and asymptotic wavelengths of the isosbestic points for RuP2 and RuP3 (SI Sections S10.1 and S10.2) differ from that of RuP. The simulated results for RuP2 in dataset **Ru-G** agree with experimental observations. However, while simulations for RuP3 show initial intensities in agreement with experiment for dataset **Ru-G**, the GSB dominates the ESAs more than expected at longer probe delays for both **Ru-G** and **Ru-Z**. Simulations of both dyes from dataset **Ru-Z** exhibit faster decay of the ~380 nm ESA compared to experiment, though the initial and final intensities are well matched.

The methyl-phosphonated derivatives of RuBPY (SI Section S10.3-S10.5), like RuP3, are in good agreement with experiments when the ultrafast relaxation pathway from $|X\rangle$ to $|S_0\rangle$ found in the solution phase and on $ZrO_2$ is included in the mechanism. The isosbestic point observed in each of the experimental TA spectra of RuCP, RuCP2, and RuCP3 exhibits a red-shift on the picosecond time scale, a characteristic that is reproduced well in the simulated TA spectra. The ultrafast decay pathway represents an additional pathway for depopulation of the molecular excited state by returning to $|S_0\rangle$. By including this transition, the excited-state population entering $|T\rangle$ is decreased and, as a result, the observed rate of charge injection from $|T\rangle$ is reduced. Zero intensity wavelengths in the simulated spectra are slightly to the red of those observed in experiment, though the redshift timescales are in agreement. Simulations for dataset **Ru-Z** overestimate the intensity of the GSB in the 500 nm to 700 nm region of the spectra within experimental error. Experimental and simulated intensities of the high energy ESAs agree well, although they decay faster in the simulated spectra. Analysis of the decay of the high energy ESA to extract the total rate of charge injection as done in Ref 36 could be a viable alternative to analysis of the lower energy isosbestic point, however resolving the relative contribution of the excited states to this signal component would be necessary to make such an analysis feasible using our simulations. Finally, because of the ultrafast relaxation pathway, using the redshift of the isosbestic point to extract the rate of charge injection becomes a challenge. The influence of the methylene group can be assessed by comparing SI Section S10.3 Figure S18 for RuCP to Figure 8 for RuP. Overall the figures are similar, however the change in the frequency of the isosbestic point is slower and the fraction of triplet contribution to the total rate of injection dominates much earlier in the simulation for RuCP than for RuP. This suggests that the total rate of charge injection and the change in the isosbestic point frequency are much less correlated in RuCP than in RuP.

Notable differences in the models and results for each dye on TiO$_2$ are reported in Table 3 below. As previously stated, the injection rate coefficients were the same for each excited state, and the ultrafast relaxation pathway from $|X\rangle$ to $|S_0\rangle$ was only omitted for dyes RuP and RuP2. We find that for RuP and RuP2, which do not have an ultrafast decay pathway, the order of magnitude of the injection rate coefficient is $10^{12}$ s$^{-1}$. The injection rate coefficients for the remaining four dyes are about a factor of 10 smaller. It is noteworthy that the injection rate coefficient for RuP3 is intermediate in value although almost as large as the other two phosphonated derivatives, suggesting that the methylene spacer indeed plays a significant role in reducing the probability of molecule-semiconductor charge transfer. The branching ratios for charge injection (Equations (19)-(22))), $\gamma_{Injection,i}$, are small relative to relaxations for the excited singlet states of all dyes, though particularly small for the dyes that include the ultrafast decay pathway to $|S_0\rangle$.

$$\gamma_{Injection,Y} = \frac{k_{Injection}}{k_{Injection}+k_{Y \to B}} \tag{19}$$

$$\gamma_{Injection,B} = \frac{k_{Injection}}{k_{Injection}+k_{Rad}+k_{nr}} \tag{20}$$

$$\gamma_{Injection,X} = \frac{k_{Injection}}{k_{Injection}+k_{ISC}+k_{Ultrafast}} \tag{21}$$

$$\gamma_{Injection,T} = \frac{k_{Injection}}{k_{Injection}+k_{B \to X}} \tag{22}$$

Table 3. Model values for charge injection of dyes in set 6-Ru on TiO$_2$: The simplest model assumption is for the injection rate coefficients to be the same from all excited states of the same dye; more information would be needed to resolve the values of these rate coefficients for the different states. After 0.5 ps charge injection is solely from $|T\rangle$.

| Dye | Injection Rate Coefficient (s$^{-1}$) | Ultrafast Decay from Singlet Manifold | $\gamma_{Injection,Y}$ [a] | $\gamma_{Injection,B}$ [b] | $\gamma_{Injection,X}$ [c] | $\gamma_{Injection,T}$ [d] |
|---|---|---|---|---|---|---|
| RuP | 2.4·10$^{12}$ | 0.0 | 0.118 | 0.118 | 0.057 | 1.00 |
| RuP2 | 1.0·10$^{12}$ | 0.0 | 0.040 | 0.040 | 0.024 | 1.00 |
| RuP3 | 8.0·10$^{11}$ | 8.0·10$^{13}$ | 0.005 | 0.005 | 0.007 | 1.00 |
| RuCP | 1.6·10$^{11}$ | 1.6·10$^{13}$ | 0.003 | 0.003 | 0.004 | 1.00 |
| RuCP2 | 1.6·10$^{11}$ | 1.6·10$^{13}$ | 0.003 | 0.003 | 0.004 | 1.00 |
| RuCP3 | 2.0·10$^{11}$ | 2.0·10$^{13}$ | 0.003 | 0.003 | 0.004 | 1.00 |

[a] Equation (19)

[b] Equation (20)

[c] Equation (21)

[d] Equation (22)

Excited state relaxation within the singlet manifold, ISC, and ultrafast decay to $|S_0\rangle$ dominate charge injection with rate coefficients on the order of 10$^{13}$ s$^{-1}$ (SI Section S9 Table S4), one to two orders of magnitude larger than for charge injection. Charge injection from singlet states of RuP3 and the methyl-phosphonated derivatives is negligible, less than 1% of the excited state population undergoes electron transfer. For RuP and RuP2, charge injection from the two, higher-energy singlet-states occurs in 12% and 4% of the populations, respectively. From $|X\rangle$, the percentages drop to of 6% and 2%, respectively. For all of the dyes in set **6-Ru**, charge injection from the triplet state is effectively the only available pathway for excited state populations to take, as charge injection out-competes incoherent emission and slow non-radiative relaxation back to $|S_0\rangle$.

### 4.3. Solar Irradiance of Dyes on TiO$_2$

Using the photophysical kinetics extracted from TA studies of dyes on TiO$_2$, we predict charge injection under 1-sun conditions for **Ru-G** (7 μm film, 4·10$^{-8}$ mol cm$^{-2}$ unbound dye). At steady-state, the fraction of the molecular excited state population in the singlet states is as much as 1/5$^{th}$ (RuP, Figure 9) and as little as 1/100$^{th}$ (RuCP, RuCP2, and RuCP3; SI Section S11 Figure S24) of the molecular excited state population in $|T\rangle$. Such a population imbalance ensures the dominance of the triplet contribution to charge injection, which is already observed in our simulations of pulsed laser experiments. We observe a large number of charge injection events per dye per second for all of the dyes, despite the ultrafast relaxation pathway included in the models for RuP3 and the three methyl-phosphonated dyes, reported in Table 4 below. Relaxation from $|T\rangle$ (i.e. incoherent emission and non-radiative relaxation) occurs on the microsecond timescale; there is a much lower probability for triplet to ground state transitions than for triplet to semiconductor charge transfer. Figure 10 demonstrates this point more starkly, the dyes of set **6-Ru** under 1-sun conditions inject less than 10% of the total number of electrons from the singlet manifold, with the methyl-phosphonated dyes injecting less than 1% of the total number of electrons from the singlet manifold. Though the number of injected electrons per dye per second does not follow a pattern for the number of phosphonated or methyl-phosphonated ligands, it is notable that the pattern does agree with the pattern of the shortest lifetimes extracted by Giokas et al using the isosbestic point analysis.[29] The shortest lifetimes were found to have the largest phenomenological rate coefficients. This trend is in general agreement with that found with the rate coefficients we report for charge injection; the rate coefficients for optical transitions and relaxation within the excited state manifold necessarily define the specific observed patterns.

In all cases, the number of injected electrons per dye per second in sunlight exceeds the often assumed value of 1 electron per dye per second[1] by a factor of 20-60. This has important implications for expectations of energy conversion efficiency using this class of dyes. For example, unassisted water oxidation catalysis, which requires 4 photons to complete a cycle, can have turnover frequencies of up to 15 per second per dye under solar illumination if charge injection is entirely rate determining. In practice, the observed photocurrents are much lower, and the $O_2$ generation rate is likely to be lower still depending on the efficiency of the catalyst. As an example, Swierk et al reported a photocurrent of about 200 µA cm$^{-2}$ for RuP2 dye at a similar concentration on $TiO_2$ as has been investigated here.[56] Charge injection alone could provide up to about 160 mA/cm$^2$ as estimated from the simulation results, indicating that in the experiments the overall photocurrent generation efficiency is around 10$^{-3}$ of a potential maximum. A quantitative investigation of the phenomena responsible for loss of photogenerated electrons in real dye sensitized systems could provide insights that will enable solar energy utilization to be improved. It is clear that the dye photophysics are far from being the limiting factor.

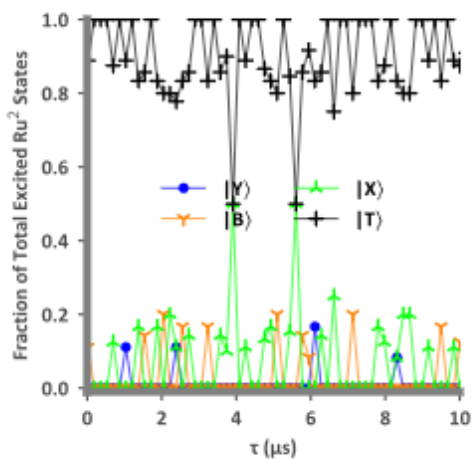

Figure 9. Fraction of excited state populations to total number of excited states across 0.1 µs of solar illumination for dye RuP.

**Table 4. Charge Injection Events under 1-sun Conditions**

| Dye | e⁻ dye⁻¹ s⁻¹ | e⁻ s⁻¹ cm⁻² ᵃ |
|---|---|---|
| RuP | 55 | 1.0·10¹⁸ |
| RuP2 | 58 | 1.1·10¹⁸ |
| RuP3 | 41 | 7.5·10¹⁷ |
| RuCP | 64 | 1.2·10¹⁸ |
| RuCP2 | 22 | 4.0·10¹⁷ |
| RuCP3 | 53 | 9.6·10¹⁷ |

ᵃ 7 μm film, 4·10⁻⁸ mol cm⁻² dye coverage

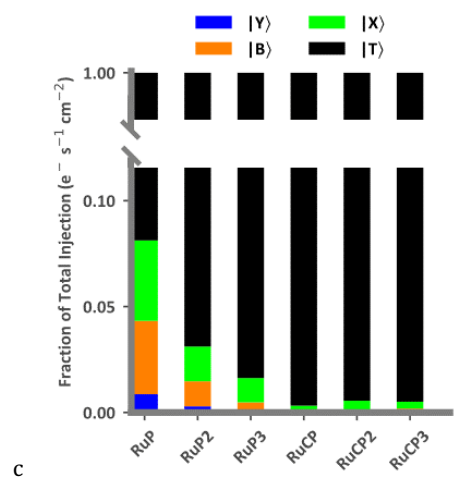

Figure 10. Fraction of injections from excited states to total number of injections from dyes in set **6-Ru** under 1-sun conditions.

Our simulations of dyes under solar irradiance provide the opportunity to examine charge injection efficiencies from the individual excited states as well as the total charge injection efficiency. These efficiencies are the ratio of electron transfer events from specific states to the number of photon absorption events populating that state (Equation 24), or the total number of electron transfer events to the total number of photon absorption events leading to excitation (Equation 25), respectively. It is important to note that the sum of the individual charge injection efficiencies does not equal the total charge injection efficiencies, because the latter is computed using a summation in both the numerator and denominator of the equation. Individual and total charge injection efficiencies are reported in Table 5. The total charge injection efficiencies are effectively unity for RuP and RuP2, but drop by approximately 20% to 40% for the dyes that include an ultrafast decay pathway. RuP and RuP2 also have higher charge injection efficiencies from the individual excited states in the singlet manifold, with $|B\rangle$ being the most efficient. The remaining dyes' singlet states are inefficient at charge injection, though $|X\rangle$ is the most efficient. The inefficient charge injection from the singlet states, and specifically from $|X\rangle$, cannot be attributed to competition from the ultrafast decay pathway alone. RuP3 and the methyl-phosphonated derivatives have rate coefficients for decay within the singlet manifold that are two to four times larger than those for RuP and RuP2, which are also larger than their own rate coefficients for ISC and ultrafast decay (SI Section S9 Table S5).

$$Charge\ Injection\ Efficiency\ from\ state\ j = \frac{e^- \ from\ excited\ state\ j}{h\nu\ absorbed\ by\ excited\ state\ j} \qquad (23)$$

$$Total\ Charge\ Injection\ Efficiency = \frac{\sum e^- \ from\ excited\ state\ j}{\sum h\nu\ absorbed\ by\ excited\ state\ j} \qquad (24)$$

**Table 5. Charge injection efficiencies from excited states of ruthenium dyes**

| Dye | $|Y\rangle$ | $|B\rangle$ | $|X\rangle$ | $|T\rangle$ | Total Charge Injection Efficiency |
|---|---|---|---|---|---|
| RuP | 0.118 | 0.151 | 0.092 | 3.21 | 1.00 |
| RuP2 | 0.040 | 0.052 | 0.042 | 3.10 | 1.00 |
| RuP3 | 0.010 | 0.013 | 0.023 | 1.41 | 0.636 |
| RuCP | 0.003 | 0.004 | 0.008 | 1.45 | 0.803 |
| RuCP2 | 0.003 | 0.004 | 0.008 | 2.14 | 0.703 |
| RuCP3 | 0.003 | 0.005 | 0.007 | 2.38 | 0.758 |

Charge injection efficiencies from $|T\rangle$ exceed unity for all dyes in set **6-Ru**. As mentioned above, $|T\rangle$ is populated by both light absorption and relaxation from the singlet manifold, making it impossible to separate charge injection efficiencies from states prepared from light absorption and

states prepared from excited state relaxation, assuming triplet states are indistinguishable regardless of how they are initially prepared. It is clear however that the efficiency of electron transfer from $|T\rangle$ to the semiconductor is crucial to the overall charge injection efficiency.

### 4.4. The Comprehensive Kinetic Framework as a Starting Point

We have found in building a quantitative model of charge injection for the ruthenium dyes in set **6-Ru** that simple 3-level schemes of photoexcitation, excited state relaxation, and charge injection are inadequate to fully capture the photophysics involved in solar energy conversion. The completeness of our kinetic framework yields predictive results, though the process of developing such a robust model is not quickly and easily implemented. RuBPY and its derivatives are more than an interesting set of compounds, they have been the standard for gauging and a template for designing similar transition metal complexes.[57-60] Accordingly, we propose that this model provides a generic framework that should be applicable to related dye families involving a transition metal complex with aromatic ligands exhibiting MLCT features within the visible spectrum. The model framework is agnostic to the specific nature of kinetically significant states and transitions, it incorporates processes as a series of steps that can be modified as needed for other dye families based on experimental observations and general spectroscopic theory. We leave the semantic details of our model, such as the assignments of the specific singlet excited states and excited state absorptions that must be present to reproduce the experiments, for theoretical studies to interpret. Additionally, though we maintain that SOE analysis methods provide limited information about the fundamental kinetics for any dye that does not have simple photophysics, we recognize that in practice it is preferable to have a simple algorithm to interpret experimental data. An iterative global fit analysis yields more robust results than a simple SOE by examining decays across the bandwidth of the probe pulse and spectra at each experimental delay. The generic aspects of the model framework developed here can be used to design the preliminary model for a global fit, and provide a means of treating the experimental conditions and signal generation explicitly.

The success of using kinetics to interpret spectroscopic data shows the potential of the technique. Even with the ability to use our simulations to pick apart the individual excited states' roles in charge injection, which has provided important new insights, it appears that studies such as the present one are only the beginning of what is needed to fully characterize such systems and propose ideal molecular photoabsorbers for solar energy conversion. Models such as the one we propose for ruthenium complexes can be further developed and improved upon through close work with ultrafast multidimensional spectroscopists, optical physicists, and computational chemists to parse the complex signals generated into a plausible scheme of excited state relaxations, to understand the behavior of light in mesoporous semiconductor films, and to propose the nature of excited states and their transitions respectively.

## 5. Conclusion

Starting with our model of **6-Ru** photophysics in solution,[37, 46] we have constructed a model of the same dyes on both $ZrO_2$ and $TiO_2$ that integrates signal generation from bound and unbound dyes, the effects of light scattering, the optical response of oxidized dyes, and charge injection. We compare calculated optical signals with previous experimental results and find good quantitative agreement,[29, 36] Our dye on substrate model ($ZrO_2$) demonstrates that presence of a substrate has an effect on the fraction of excited dyes generated and their signal intensities, but does not significantly influence the fundamental molecular photophysics. The charge injection model we present employs pseudo-first-order charge injection rate coefficients that reproduce experimental observations for a specific dye when they are set to be equal for all excited states of that dye. To estimate the primary second-order rate coefficients for charge injection, direct knowledge of the DOS of $TiO_2$ nanoparticles in the films is required. As mentioned in Ref. 36, the DOS can be sensitive to the fabrication method of the film; we do not have DOS for all of the samples in our analysis. The DOS presented in Figure 7 of Ref. 36 (for a related ruthenium derivative) shows a decrease of approximately an order of magnitude in the DOS between energies associated with the peak of the MLCT band and the peak of the red-wing of the absorption spectrum. Combining the DOS measurements with our model—which shows pseudo-first-order rate coefficients for charge injection appear to be constant for all states— the second-order rate coefficients for charge injection would be greater for the triplet state than higher energy singlet states. This raises the question: why would charge injection from the triplet be more facile than from the singlet manifold? We also find that the ultrafast relaxation pathway observed for all of the dyes in solution[37] is effectively suppressed in RuP and RuP2, but not for RuP3 and the methyl-phosphonated dyes. Finally, we have found that injection from the triplet to always dominate after ~0.5 ps for dyes under pulsed excitation and for dyes under solar irradiance. The prominent role of the triplet, with only slow non-radiative relaxation and incoherent emission to compete with charge injection, leads to the number of electrons per dye per second to be on the order of 20-60, much larger than previous estimates of about 1. If quantitative knowledge of loss mechanisms for these dyes in various sensitization architectures are available and are added to the simulated mechanism, their influence on the number of injected electrons can be estimated.

**ASSOCIATED CONTENT**

**SUPPORTING INFORMATION.**

Presented: Chromophores in **6-Ru**, schematic of reflectance measurements, solution phase model, dye on substrate model, oxidized dye optical response, TA of dyes on $ZrO_2$, dye on $TiO_2$ model, TA of dyes on $TiO_2$, populations of $Ru^{II}$ excited states under solar illumination, and computed errors of simulation values.

**AUTHOR INFORMATION**

**Corresponding Author**

*Author to whom correspondence should be addressed, fahoule@lbl.gov

**Notes**

FAH and TJM conceived of this study, FAH and TPC designed the modeling approach, and TPC led the kinetic modeling of the optical data. AMM, JMP, GJM and TJM provided data and experimental details and participated in interpretation of the simulation results. AMM was the graduate advisor of PGG. JMP was the postdoctoral advisor of DFZ. TPC wrote the first draft of the manuscript. All authors have given approval to the final version of the manuscript.

**ACKNOWLEDGMENT**

This material is based upon work supported by the U.S. Department of Energy, Office of Science, Office of Basic Energy Sciences, Chemical Sciences, Geosciences, and Biosciences Division, in the Solar Photochemistry Program under Contract No. DE-AC02-05CH11231 (supporting TPC and FAH, who developed the kinetic framework for the ultrafast photophysics and charge injection of RuBPY derivatives on metal oxide mesoporous thin films). MKB (reflectance measurements and compiled TA data for analysis and interpreted simulated TA signals of RuBPY derivatives), a study by PGG (performed fs-TA measurements and provided essential experimental details), and a study by DFZ (performed fs-TA measurements and provided essential experimental details) were supported by the Alliance for Molecular PhotoElectrode Design for Solar Fuels (AMPED), an Energy Frontier Research Center (EFRC) funded by the U.S. Department of Energy, Office of Science, Office of Basic Energy Sciences under Award Number DE-SC0001011. EAK (performed spectroelectrochemical measurements of RuBPY derivatives on metal oxide mesoporous thin films) supported by Division of Chemical Sciences, Office of Basic Energy Sciences, Office of Energy Research, US Department of Energy (DE-SC0013461).

**CONFLICTS OF INTEREST**

The authors have no conflicts to disclose.

**REFERENCES**

[1] B. O'Regan, and M. Grätzel, Nature **353** (1991) 737.
[2] M. Grätzel, Progress in Photovoltaics: Research and Applications **8** (2000) 171.
[3] J. B. Asbury *et al.*, J Phys Chem B **105** (2001) 4545.
[4] M. Grätzel, Nature **414** (2001) 338.
[5] Z. Zou *et al.*, Nature **414** (2001) 625.
[6] B. A. Gregg, The Journal of Physical Chemistry B **107** (2003) 4688.
[7] P. V. Kamat, The Journal of Physical Chemistry C **111** (2007) 2834.
[8] J. J. Concepcion *et al.*, Acc Chem Res **42** (2009) 1954.
[9] A. Hagfeldt *et al.*, Chem Rev **110** (2010) 6595.
[10] T. E. Mallouk, J Phys Chem Lett **1** (2010) 2738.
[11] L. Dloczik *et al.*, The Journal of Physical Chemistry B **101** (1997) 10281.
[12] K. Hanson *et al.*, The Journal of Physical Chemistry C **116** (2012) 14837.
[13] R. Katoh *et al.*, Cr Chim **9** (2006) 639.
[14] R. Katoh *et al.*, Sol Energ Mat Sol C **93** (2009) 698.
[15] A. Listorti, B. O'Regan, and J. R. Durrant, Chem Mater **23** (2011) 3381.
[16] J. van de Lagemaat, N. G. Park, and A. J. Frank, The Journal of Physical Chemistry B **104** (2000) 2044.
[17] D. F. Watson, and G. J. Meyer, Annu Rev Phys Chem **56** (2005) 119.
[18] W. R. Duncan, and O. V. Prezhdo, Annu Rev Phys Chem **58** (2007) 143.
[19] E. Jakubikova *et al.*, J Phys Chem A **113** (2009) 12532.
[20] M. K. Brennaman *et al.*, J Am Chem Soc **138** (2016) 13085.
[21] Z. Guo *et al.*, J Phys Chem A **120** (2016) 5773.
[22] G. Hermann, and J. C. Tremblay, J Chem Phys **145** (2016) 174704.
[23] A. Grupp *et al.*, J Phys Chem Lett **8** (2017) 4858.
[24] C. S. Ponseca, Jr. *et al.*, Chem Rev **117** (2017) 10940.
[25] P. J. Holliman *et al.*, Sci Technol Adv Mater **19** (2018) 599.

[26] M. Chergui, Faraday Discuss **216** (2019) 9.
[27] J. D. Elliott *et al.*, J Phys Chem Lett **12** (2021) 7261.
[28] S. Yamijala, and P. Huo, J Phys Chem A **125** (2021) 628.
[29] P. G. Giokas *et al.*, J Phys Chem C **117** (2013) 812.
[30] D. Kuciauskas *et al.*, J Phys Chem B **106** (2002) 9347.
[31] S. A. Miller *et al.*, J Chem Phys **135** (2011) 081101.
[32] Y. Tachibana *et al.*, J Phys Chem-Us **100** (1996) 20056.
[33] J. B. Asbury *et al.*, J Phys Chem B **107** (2003) 7376.
[34] S. E. Bettis *et al.*, J Phys Chem A **118** (2014) 10301.
[35] M. Juozapavicius *et al.*, J Phys Chem C **117** (2013) 25317.
[36] D. F. Zigler *et al.*, J Am Chem Soc **138** (2016) 4426.
[37] T. P. Cheshire *et al.*, J Phys Chem B **124** (2020) 5971.
[38] J. P. Paris, and W. W. Brandt, J Am Chem Soc **81** (1959) 5001.
[39] G. A. Crosby, W. G. Perkins, and D. M. Klassen, Journal of Chemical Physics **43** (1965) 1498.
[40] D. M. Klassen, and G. A. Crosby, Chem Phys Lett **1** (1967) 127.
[41] J. N. Demas, and G. A. Crosby, Journal of Molecular Spectroscopy **26** (1968) 72.
[42] D. M. Klassen, and G. A. Crosby, Journal of Chemical Physics **48** (1968) 1853.
[43] W. D. Hinsberg, and F. A. Houle, (www.hinsberg.net/Kinetiscope, 2022).
[44] D. L. Bunker *et al.*, Combustion and Flame **23** (1974) 373.
[45] D. T. Gillespie, Journal of Computational Physics **22** (1976) 403.
[46] T. P. Cheshire, and F. A. Houle, J. Phys. Chem. A **doi.org/10.1021/acs.jpca.1c02386** (2021)
[47] F. A. Houle, J Phys Chem C **123** (2019) 14459.
[48] F. A. Houle, Chem Sci **12** (2021) 6117.
[49] C. A. Gueymard, Sol Energy **71** (2001) 325.
[50] C. A. Gueymard, Sol Energy **76** (2004) 423.
[51] C. A. Gueymard, D. Myers, and K. Emery, Sol Energy **73** (2002) 443.
[52] D. Chen *et al.*, Advanced Materials **21** (2009) 2206.
[53] W. Zhang *et al.*, Journal of Materials Science: Materials in Electronics **29** (2018) 7356.
[54] E. C. Brigham, and G. J. Meyer, The Journal of Physical Chemistry C **118** (2014) 7886.
[55] J. R. Swierk, N. S. McCool, and T. E. Mallouk, J Phys Chem C **119** (2015) 13858.
[56] J. R. Swierk *et al.*, J Am Chem Soc **136** (2014) 10974.
[57] D. W. Thompson, A. Ito, and T. J. Meyer, Pure Appl Chem **85** (2013) 1257.
[58] L. Kohler *et al.*, Inorg Chem **56** (2017) 12214.
[59] J. M. Cole *et al.*, Chem Rev **119** (2019) 7279.
[60] J. H. Shon, and T. S. Teets, Acs Energy Lett **4** (2019) 558.

# Supporting Information: A Quantitative Model of Charge Injection by Ruthenium Chromophores Connecting Femtosecond to Continuous Irradiance Conditions

Thomas P. Cheshire[1], Jeb Boodry[1,2], Erin A. Kober[3], M. Kyle Brennaman[c3], Paul G. Giokas[4], David F. Zigler[5], Andrew M. Moran[3], John M. Papanikolas[3], Gerald J. Meyer[3], Thomas J. Meyer[3], and Frances A. Houle[1]*

[1] Chemical Sciences Division, Lawrence Berkeley National Laboratory Berkeley, CA 94720

[2] Department of Chemical and Biochemical Engineering, University of California, Berkeley, CA 94720

[3] Department of Chemistry, University of North Carolina at Chapel Hill, Chapel Hill, NC 27599

[4] Coherent Inc., Santa Clara, CA 95054

[5] Chemistry & Biochemistry Department, California Polytechnic State University, San Luis Obispo, CA 93407

*Author to whom correspondence should be addressed: fahoule@lbl.gov

# Table of Contents

1. **Chromophores 6-Ru**

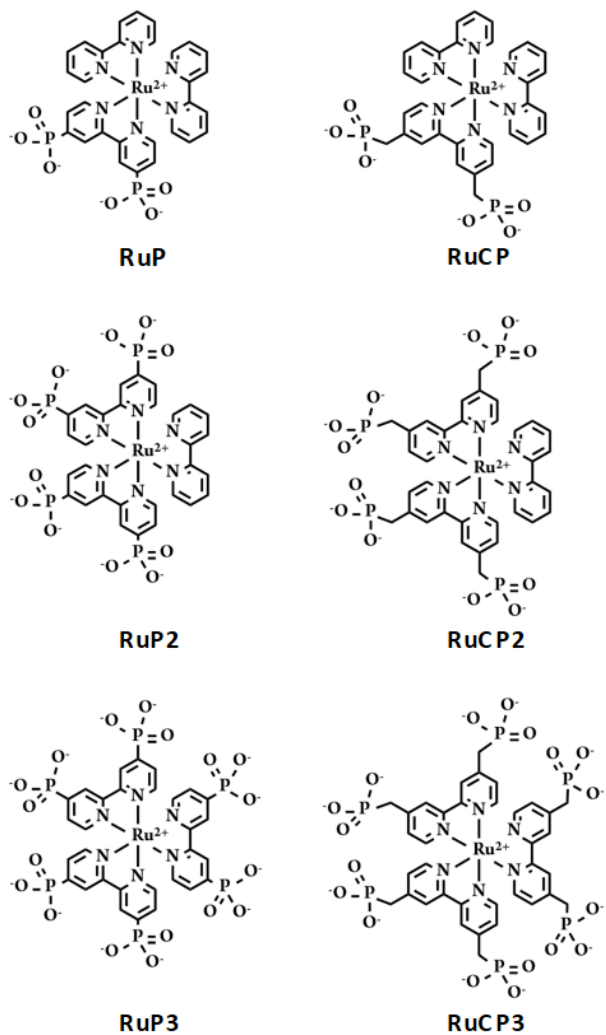

Figure S1. Six ruthenium polypyridyl complexes that make up the set of dyes discussed in this study, **6-Ru**

## 2. Reflectance

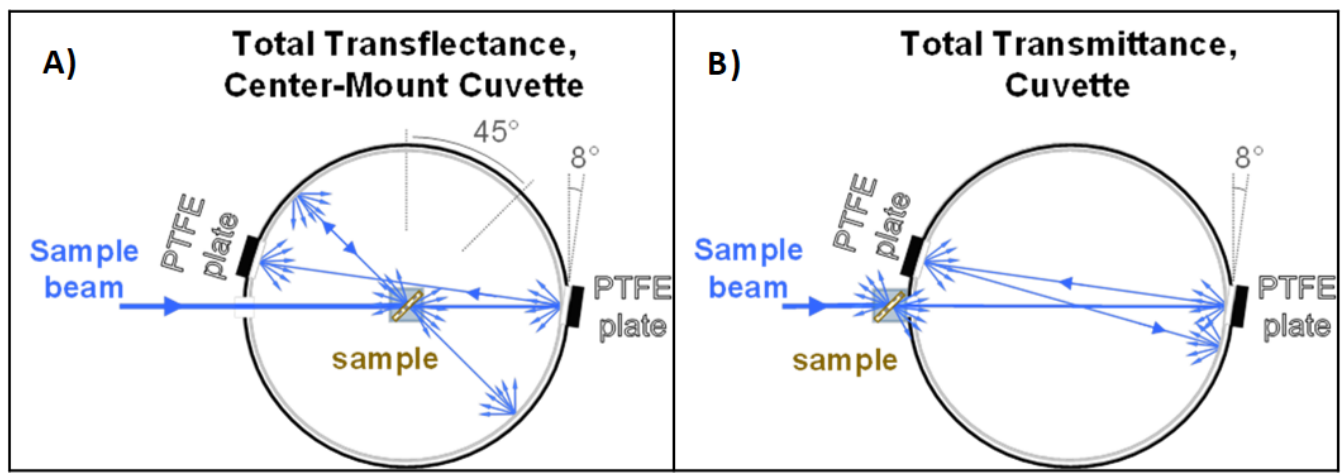

Figure S2. The configuration of the integrating sphere that includes a depiction of the sample, incident light, transmitted light, and scattered light used for A) transflectance and B) transmittance measurements is shown

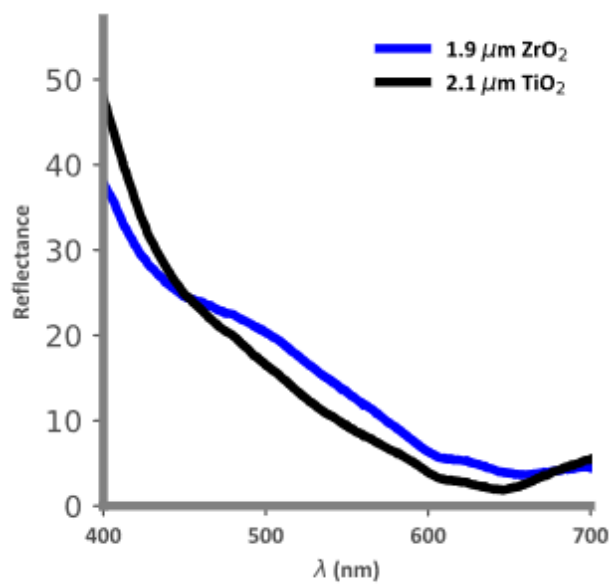

Figure S3. Reflectance measurements of $ZrO_2$ and $TiO_2$ films from ~20 nm in diameter nanoparticles.

## 3. Marker Species

Marker species are pseudo-species used in kinetic simulations to tally the number of a particular event or set of events as needed. The products of a unidirectional reaction step, as in Equation (S25), do not affect the rate equation assuming there are no reaction steps that couple the products to the reactants in a reverse reaction or chain of reactions.

$$aA \rightarrow bB + mM \tag{S25}$$

The rate equations can be written as:

$$\frac{d[A]}{dt} = -k[A]^a = k[B]^b = k[M]^m . \tag{S26}$$

Marker species $M$ can be used to identify the event of particular reaction step, as above, or the total number of events from multiple reaction steps. In practice, m=1 for each marker; multiple markers can be included in each step depending on the information sought. In the case of reversible reactions, as in Equation (S27), the reaction step can be split into forward and reverse steps, seen in Equations (S28) and (S29), to tally the forward and reverse events without affecting the kinetics.

$$aA + m_{forward}M_{forward} \leftrightarrow bB + m_{reverse}M_{reverse} \tag{S27}$$

$$aA \rightarrow bB + m_{forward}M_{forward} \tag{S28}$$

$$bB \rightarrow aA + m_{reverse}M_{reverse} \tag{S29}$$

Lastly, marker species can be used in the reactant side of reaction steps if and only if the concentration is not depleted and the reaction step treats the pseudo-species as a zeroth-order species. In such an event, the marker species is merely a flag to whether the reaction step could occur.

## 4. Linear Absorption Spectrum of dye RuP between 300 nm and 600 nm

The linear absorption spectrum of RuP continues to decrease in intensity from 400 nm to ~360 nm. The increase in intensity in the <350 nm region of the spectrum is significant; however, the solar spectrum intensity becomes vanishingly small before 300 nm. The molecular response below 400 nm is assumed to be negligible.

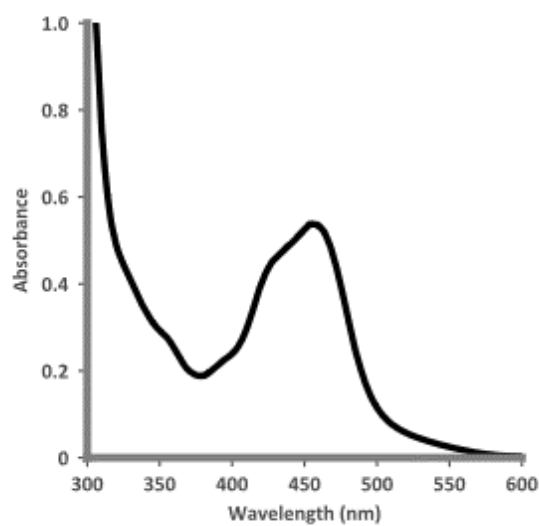

Figure S4. Linear absorption spectrum for dye RuP between 300 nm and 600 nm.

## 5. Solution Phase Model for set 6-Ru

gives the mechanistic steps and rate coefficients for the solution phase kinetic framework for the ultrafast photophysics of set **6-Ru**. Line 1 represents the pump laser pulse interaction with dyes in the ground state. The rate coefficient is determined by the duration of the laser pulse and the fraction of excitation of dyes from a narrow laser pulse centered around 400 nm. For our simulations the pump pulse was 45 fs and excites approximately 10% of the dyes, as was reported for the experiments and previous simulations.[1,2] The time-dependence of the perturbation was removed, and the product of constant irradiance at all wavelengths and the optical rate coefficients from **Model 1τ** give the optical rate coefficients for **Model 1λ**. Rate coefficients for non-radiative transitions and incoherent emission are the same as **Model 1τ**.

**Table S1. Mechanistic steps and rate coefficients for dyes in solution**

| | Rate Coefficients | RuP (s$^{-1}$) | RuP 2 (s$^{-1}$) | RuP3 (s$^{-1}$) | RuCP (s$^{-1}$) | RuCP2 (s$^{-1}$) | RuCP3 (s$^{-1}$) |
|---|---|---|---|---|---|---|---|
| $\lvert S0\rangle \xrightarrow{h\nu_{Pump}} \lvert Y\rangle$ | $k_{Pumps}$ | Variable | Variable | Variable | Variable | Variable | Variable |
| $\lvert S0\rangle \xrightarrow{\lvert Y\rangle, I_{Probe}(\lambda,\tau)} \lvert S0\rangle + GSB_Y$ | $k_{GSB_Y}$ | | | | | | |
| $\lvert S0\rangle \xrightarrow{I_{Probe}(\lambda,\tau)} \lvert Y\rangle + ABS_Y$ | $k_{ABS_Y}$ | 1.8·10$^{13}$ | 2.0·10$^{13}$ | 2.0·10$^{13}$ | 1.9·10$^{13}$ | 1.8·10$^{13}$ | 3.4·10$^{13}$ |
| $\lvert Y\rangle \xrightarrow{I_{Probe}(\lambda,\tau)} \lvert S0\rangle + ESE_Y$ | $k_{ESE_Y}$ | | | | | | |
| $\lvert S0\rangle \xrightarrow{\lvert B\rangle, I_{Probe}(\lambda,\tau)} \lvert S0\rangle + GSB_B$ | $k_{GSB_B}$ | | | | | | |
| $\lvert S0\rangle \xrightarrow{I_{Probe}(\lambda,\tau)} \lvert B\rangle + ABS_B$ | $k_{ABS_B}$ | 3.8·10$^{13}$ | 4.1·10$^{13}$ | 3.3·10$^{13}$ | 3.2·10$^{13}$ | 3.2·10$^{13}$ | 3.7·10$^{13}$ |
| $\lvert B\rangle \xrightarrow{I_{Probe}(\lambda,\tau)} \lvert S0\rangle + ESE_B$ | $k_{ESE_B}$ | | | | | | |
| $\lvert S0\rangle \xrightarrow{\lvert X\rangle, I_{Probe}(\lambda,\tau)} \lvert S0\rangle + GSB_X$ | $k_{GSB_X}$ | | | | | | |
| $\lvert S0\rangle \xrightarrow{I_{Probe}(\lambda,\tau)} \lvert X\rangle + ABS_X$ | $k_{ABS_X}$ | 4.8·10$^{13}$ | 5.0·10$^{13}$ | 4.0·10$^{13}$ | 4.1·10$^{13}$ | 4.1·10$^{13}$ | 4.5·10$^{13}$ |
| $\lvert X\rangle \xrightarrow{I_{Probe}(\lambda,\tau)} \lvert S0\rangle + ESE_X$ | $k_{ESE_X}$ | | | | | | |
| $\lvert S0\rangle \xrightarrow{\lvert T\rangle, I_{Probe}(\lambda,\tau)} \lvert S0\rangle + GSB_T$ | $k_{GSB_T}$ | | | | | | |
| $\lvert S0\rangle \xrightarrow{I_{Probe}(\lambda,\tau)} \lvert T\rangle + ABS_T$ | $k_{ABS_T}$ | 1.8·10$^{13}$ | 1.8·10$^{13}$ | 1.8·10$^{13}$ | 1.8·10$^{13}$ | 1.8·10$^{13}$ | 1.8·10$^{13}$ |
| $\lvert T\rangle \xrightarrow{I_{Probe}(\lambda,\tau)} \lvert S0\rangle + ESE_T$ | $k_{ESE_T}$ | | | | | | |
| $\lvert Y\rangle \to \lvert B\rangle$ | $k_{YB}$ | 1.8·10$^{13}$ | 2.4·10$^{13}$ | 8.0·10$^{13}$ | 6.0·10$^{13}$ | 6.0·10$^{13}$ | 6.0·10$^{13}$ |
| $\lvert B\rangle \to \lvert X\rangle$ | $k_{BX}$ | 1.8·10$^{13}$ | 2.4·10$^{13}$ | 8.0·10$^{13}$ | 6.0·10$^{13}$ | 6.0·10$^{13}$ | 6.0·10$^{13}$ |
| $\lvert X\rangle \to \lvert S0\rangle$ | $k_{UF-nr}$ | 2.4·10$^{13}$ | 1.0·10$^{13}$ | 4.0·10$^{13}$ | 1.6·10$^{13}$ | 1.6·10$^{13}$ | 2.0·10$^{13}$ |
| $\lvert X\rangle \to \lvert T\rangle$ | $k_{ISC}$ | 4.0·10$^{13}$ | 4.0·10$^{13}$ | 2.0·10$^{13}$ | 2.0·10$^{13}$ | 2.0·10$^{13}$ | 3.6·10$^{13}$ |
| $\lvert Y\rangle \xrightarrow{\lvert Z1\rangle, I_{Probe}(\lambda,\tau)} \lvert Y\rangle + ESA_{Y,1}$ | | | | | | | |
| $\lvert B\rangle \xrightarrow{\lvert Z1\rangle, I_{Probe}(\lambda,\tau)} \lvert B\rangle + ESA_{X,1}$ | $k_{ESA_1}$ | 4.0·10$^{13}$ | 3.2·10$^{13}$ | 3.8·10$^{13}$ | 4.8·10$^{13}$ | 4.0·10$^{13}$ | 4.9·10$^{13}$ |
| $\lvert X\rangle \xrightarrow{\lvert Z1\rangle, I_{Probe}(\lambda,\tau)} \lvert X\rangle + ESA_{B,1}$ | | | | | | | |
| $\lvert T\rangle \xrightarrow{\lvert Z1\rangle, I_{Probe}(\lambda,\tau)} \lvert T\rangle + ESA_{T,1}$ | | | | | | | |
| $\lvert Y\rangle \xrightarrow{\lvert Z2\rangle, I_{Probe}(\lambda,\tau)} \lvert Y\rangle + ESA_{Y,2}$ | | | | | | | |
| $\lvert B\rangle \xrightarrow{\lvert Z2\rangle, I_{Probe}(\lambda,\tau)} \lvert B\rangle + ESA_{X,2}$ | $k_{ESA_2}$ | 4.3·10$^{13}$ | 2.8·10$^{13}$ | 4.7·10$^{13}$ | 5.2·10$^{13}$ | 5.7·10$^{13}$ | 5.3·10$^{13}$ |
| $\lvert X\rangle \xrightarrow{\lvert Z2\rangle, I_{Probe}(\lambda,\tau)} \lvert X\rangle + ESA_{B,2}$ | | | | | | | |
| $\lvert T\rangle \xrightarrow{\lvert Z2\rangle, I_{Probe}(\lambda,\tau)} \lvert T\rangle + ESA_{T,2}$ | | | | | | | |
| $\lvert Y\rangle \xrightarrow{\lvert Z3\rangle, I_{Probe}(\lambda,\tau)} \lvert Y\rangle + ESA_{Y,3}$ | | | | | | | |
| $\lvert B\rangle \xrightarrow{\lvert Z3\rangle, I_{Probe}(\lambda,\tau)} \lvert B\rangle + ESA_{X,3}$ | $k_{ESA_3}$ | 4.2·10$^{13}$ | 3.0·10$^{13}$ | 5.3·10$^{13}$ | 5.6·10$^{13}$ | 9.3·10$^{13}$ | 5.7·10$^{13}$ |
| $\lvert X\rangle \xrightarrow{\lvert Z3\rangle, I_{Probe}(\lambda,\tau)} \lvert X\rangle + ESA_{B,3}$ | | | | | | | |
| $\lvert T\rangle \xrightarrow{\lvert Z3\rangle, I_{Probe}(\lambda,\tau)} \lvert T\rangle + ESA_{T,3}$ | | | | | | | |
| $\lvert Y\rangle \xrightarrow{\lvert Z4\rangle, I_{Probe}(\lambda,\tau)} \lvert Y\rangle + ESA_{Y,4}$ | | | | | | | |
| $\lvert B\rangle \xrightarrow{\lvert Z4\rangle, I_{Probe}(\lambda,\tau)} \lvert B\rangle + ESA_{X,4}$ | $k_{ESA_4}$ | 3.9·10$^{13}$ | 3.0·10$^{13}$ | 4.6·10$^{13}$ | 5.7·10$^{13}$ | 11·10$^{13}$ | 5.7·10$^{13}$ |
| $\lvert X\rangle \xrightarrow{\lvert Z4\rangle, I_{Probe}(\lambda,\tau)} \lvert X\rangle + ESA_{B,4}$ | | | | | | | |
| $\lvert T\rangle \xrightarrow{\lvert Z4\rangle, I_{Probe}(\lambda,\tau)} \lvert T\rangle + ESA_{T,4}$ | | | | | | | |
| $\lvert Y\rangle \xrightarrow{\lvert Z5\rangle, I_{Probe}(\lambda,\tau)} \lvert Y\rangle + ESA_{Y,5}$ | | | | | | | |
| $\lvert B\rangle \xrightarrow{\lvert Z5\rangle, I_{Probe}(\lambda,\tau)} \lvert B\rangle + ESA_{X,5}$ | $k_{ESA_5}$ | 5.4·10$^{13}$ | 3.8·10$^{13}$ | 5.0·10$^{13}$ | 7.0·10$^{13}$ | 10.5·10$^{13}$ | 5.0·10$^{13}$ |
| $\lvert X\rangle \xrightarrow{\lvert Z5\rangle, I_{Probe}(\lambda,\tau)} \lvert X\rangle + ESA_{B,5}$ | | | | | | | |
| $\lvert T\rangle \xrightarrow{\lvert Z5\rangle, I_{Probe}(\lambda,\tau)} \lvert T\rangle + ESA_{T,5}$ | | | | | | | |
| $\lvert T\rangle \to \lvert S0\rangle + rad$ | $k_{rad}$ | 10.9·10$^{4}$ | 9.6·10$^{4}$ | 11·10$^{4}$ | 10.6·10$^{4}$ | 10.6·10$^{4}$ | 10.3·10$^{4}$ |
| $\lvert T\rangle \to \lvert S0\rangle$ | $k_{nr}$ | 3.2·10$^{6}$ | 2.6·10$^{6}$ | 2.0·10$^{6}$ | 2.0·10$^{6}$ | 2.2·10$^{6}$ | 2.3·10$^{6}$ |

## 6. Dye on substrate model for set 6-Ru

**Table S2.** Base mechanistic scheme and rate coefficients for dyes in set 6-Ru on 7 μm film of $ZrO_2$ for dataset Ru-G with broadband probe spanning 500 nm to 700 nm. Simulated surface concentration and unbound dye concentration were $4·10^{-8}$ mol cm$^{-2}$ and $4·10^{-9}$ mol cm$^{-2}$ respectively.

| | Rate Coefficients | RuP (s$^{-1}$) | RuP 2 (s$^{-1}$) | RuP3 (s$^{-1}$) | RuCP (s$^{-1}$) | RuCP2 (s$^{-1}$) | RuCP3 (s$^{-1}$) |
|---|---|---|---|---|---|---|---|
| $\|S0_{Bound}\rangle \xrightarrow{h\nu_{Pump}} \|Y_{Bound}\rangle$ | $k_{Pump,Y}$ | Variable | Variable | Variable | Variable | Variable | Variable |
| $\|S0_{Bound}\rangle \xrightarrow{h\nu_{Pump}} \|B_{Bound}\rangle$ | $k_{Pump,B}$ | Variable | Variable | Variable | Variable | Variable | Variable |
| $\|S0_{Bound}\rangle \xrightarrow{h\nu_{Pump}} \|X_{Bound}\rangle$ | $k_{Pump,X}$ | Variable | Variable | Variable | Variable | Variable | Variable |
| $\|S0_{Bound}\rangle \xrightarrow{h\nu_{Pump}} \|T_{Bound}\rangle$ | $k_{Pump,T}$ | Variable | Variable | Variable | Variable | Variable | Variable |
| $\|S0_{Bound}\rangle \xrightarrow{\|Y_{Bound}\rangle,I_{Probe}(\lambda,\tau)} \|S0_{Bound}\rangle$ | $k_{GSB_Y}$ | | | | | | |
| $\|S0_{Bound}\rangle \xrightarrow{I_{Probe}(\lambda,\tau)} \|Y_{Bound}\rangle + ABS_{Y.Bound}$ | $k_{ABS_Y}$ | $1.8·10^{13}$ | $2.0·10^{13}$ | $2.0·10^{13}$ | $1.9·10^{13}$ | $1.8·10^{13}$ | $3.4·10^{13}$ |
| $\|Y_{Bound}\rangle \xrightarrow{I_{Probe}(\lambda,\tau)} \|S0_{Bound}\rangle + ESE_{Y.Bound}$ | $k_{ESE_Y}$ | | | | | | |
| $\|S0_{Bound}\rangle \xrightarrow{\|B_{Bound}\rangle,I_{Probe}(\lambda,\tau)} \|S0_{Bound}\rangle$ | $k_{GSB_B}$ | | | | | | |
| $\|S0_{Bound}\rangle \xrightarrow{I_{Probe}(\lambda,\tau)} \|B_{Bound}\rangle + ABS_{B.Bound}$ | $k_{ABS_B}$ | $3.8·10^{13}$ | $4.1·10^{13}$ | $3.3·10^{13}$ | $3.2·10^{13}$ | $3.2·10^{13}$ | $3.7·10^{13}$ |
| $\|B_{Bound}\rangle \xrightarrow{I_{Probe}(\lambda,\tau)} \|S0_{Bound}\rangle + ESE_{B.Bound}$ | $k_{ESE_B}$ | | | | | | |
| $\|S0_{Bound}\rangle \xrightarrow{\|X_{Bound}\rangle,I_{Probe}(\lambda,\tau)} \|S0_{Bound}\rangle$ | $k_{GSB_X}$ | | | | | | |
| $\|S0_{Bound}\rangle \xrightarrow{I_{Probe}(\lambda,\tau)} \|X_{Bound}\rangle + ABS_{X.Bound}$ | $k_{ABS_X}$ | $4.8·10^{13}$ | $5.0·10^{13}$ | $4.0·10^{13}$ | $4.1·10^{13}$ | $4.1·10^{13}$ | $4.5·10^{13}$ |
| $\|X_{Bound}\rangle \xrightarrow{I_{Probe}(\lambda,\tau)} \|S0_{Bound}\rangle + ESE_{X.Bound}$ | $k_{ESE_X}$ | | | | | | |
| $\|S0_{Bound}\rangle \xrightarrow{\|T_{Bound}\rangle,I_{Probe}(\lambda,\tau)} \|S0_{Bound}\rangle + GSB_{T.Bound}$ | $k_{GSB_T}$ | | | | | | |
| $\|S0_{Bound}\rangle \xrightarrow{I_{Probe}(\lambda,\tau)} \|T_{Bound}\rangle + ABS_{T.Bound}$ | $k_{ABS_T}$ | $1.8·10^{13}$ | $1.8·10^{13}$ | $1.8·10^{13}$ | $1.8·10^{13}$ | $1.8·10^{13}$ | $1.8·10^{13}$ |
| $\|T_{Bound}\rangle \xrightarrow{I_{Probe}(\lambda,\tau)} \|S0_{Bound}\rangle + ESE_{T.Bound}$ | $k_{ESE_T}$ | | | | | | |
| $\|Y_{Bound}\rangle \rightarrow \|B_{Bound}\rangle$ | $k_{YB}$ | $1.8·10^{13}$ | $2.4·10^{13}$ | $8.0·10^{13}$ | $6.0·10^{13}$ | $6.0·10^{13}$ | $6.0·10^{13}$ |
| $\|B_{Bound}\rangle \rightarrow \|X_{Bound}\rangle$ | $k_{BX}$ | $1.8·10^{13}$ | $2.4·10^{13}$ | $8.0·10^{13}$ | $6.0·10^{13}$ | $6.0·10^{13}$ | $6.0·10^{13}$ |
| $\|X_{Bound}\rangle \rightarrow \|S0_{Bound}\rangle$ | $k_{UF-nr}$ | $2.4·10^{13}$ | $1.0·10^{13}$ | $4.0·10^{13}$ | $1.6·10^{13}$ | $1.6·10^{13}$ | $2.0·10^{13}$ |
| $\|X_{Bound}\rangle \rightarrow \|T_{Bound}\rangle$ | $k_{ISC}$ | $4.0·10^{13}$ | $4.0·10^{13}$ | $2.0·10^{13}$ | $2.0·10^{13}$ | $2.0·10^{13}$ | $3.6·10^{13}$ |
| $\|Y_{Bound}\rangle \xrightarrow{\|Z0_{Bound}\rangle,I_{Probe}(\lambda,\tau)} \|Y_{Bound}\rangle + ESA_{Y.0.Bound}$ | | | | | | | |
| $\|B_{Bound}\rangle \xrightarrow{\|Z0_{Bound}\rangle,I_{Probe}(\lambda,\tau)} \|B_{Bound}\rangle + ESA_{B.0.Bound}$ | $k_{ESA_0}$ | $1.6·10^{14}$ | $1.5·10^{14}$ | $1.5·10^{14}$ | $1.8·10^{14}$ | $2.1·10^{14}$ | $1.9·10^{14}$ |
| $\|X_{Bound}\rangle \xrightarrow{\|Z0_{Bound}\rangle,I_{Probe}(\lambda,\tau)} \|X_{Bound}\rangle + ESA_{X.0.Bound}$ | | | | | | | |
| $\|T_{Bound}\rangle \xrightarrow{\|Z0_{Bound}\rangle,I_{Probe}(\lambda,\tau)} \|T_{Bound}\rangle + ESA_{T.0.Bound}$ | | | | | | | |
| $\|Y_{Bound}\rangle \xrightarrow{\|Z1_{Bound}\rangle,I_{Probe}(\lambda,\tau)} \|Y_{Bound}\rangle + ESA_{Y.1.Bound}$ | | | | | | | |
| $\|B_{Bound}\rangle \xrightarrow{\|Z1_{Bound}\rangle,I_{Probe}(\lambda,\tau)} \|B_{Bound}\rangle + ESA_{B.1.Bound}$ | $k_{ESA_1}$ | $4.0·10^{13}$ | $3.2·10^{13}$ | $3.8·10^{13}$ | $4.8·10^{13}$ | $4.0·10^{13}$ | $4.9·10^{13}$ |
| $\|X_{Bound}\rangle \xrightarrow{\|Z1_{Bound}\rangle,I_{Probe}(\lambda,\tau)} \|X_{Bound}\rangle + ESA_{X.1.Bound}$ | | | | | | | |
| $\|T_{Bound}\rangle \xrightarrow{\|Z1_{Bound}\rangle,I_{Probe}(\lambda,\tau)} \|T_{Bound}\rangle + ESA_{T.1.Bound}$ | | | | | | | |
| $\|Y_{Bound}\rangle \xrightarrow{\|Z2_{Bound}\rangle,I_{Probe}(\lambda,\tau)} \|Y_{Bound}\rangle + ESA_{Y.2.Bound}$ | | | | | | | |
| $\|B_{Bound}\rangle \xrightarrow{\|Z2_{Bound}\rangle,I_{Probe}(\lambda,\tau)} \|B_{Bound}\rangle + ESA_{B.2.Bound}$ | $k_{ESA_2}$ | $4.3·10^{13}$ | $2.8·10^{13}$ | $4.7·10^{13}$ | $5.2·10^{13}$ | $5.7·10^{13}$ | $5.3·10^{13}$ |
| $\|X_{Bound}\rangle \xrightarrow{\|Z2_{Bound}\rangle,I_{Probe}(\lambda,\tau)} \|X_{Bound}\rangle + ESA_{X.2.Bound}$ | | | | | | | |
| $\|T_{Bound}\rangle \xrightarrow{\|Z2_{Bound}\rangle,I_{Probe}(\lambda,\tau)} \|T_{Bound}\rangle + ESA_{T.2.Bound}$ | | | | | | | |
| $\|Y_{Bound}\rangle \xrightarrow{\|Z3_{Bound}\rangle,I_{Probe}(\lambda,\tau)} \|Y_{Bound}\rangle + ESA_{Y.3.Bound}$ | | | | | | | |
| $\|B_{Bound}\rangle \xrightarrow{\|Z3_{Bound}\rangle,I_{Probe}(\lambda,\tau)} \|B_{Bound}\rangle + ESA_{B.3.Bound}$ | $k_{ESA_3}$ | $4.2·10^{13}$ | $3.0·10^{13}$ | $5.3·10^{13}$ | $5.6·10^{13}$ | $9.3·10^{13}$ | $5.7·10^{13}$ |
| $\|X_{Bound}\rangle \xrightarrow{\|Z3_{Bound}\rangle,I_{Probe}(\lambda,\tau)} \|X_{Bound}\rangle + ESA_{X.3.Bound}$ | | | | | | | |
| $\|T_{Bound}\rangle \xrightarrow{\|Z3_{Bound}\rangle,I_{Probe}(\lambda,\tau)} \|T_{Bound}\rangle + ESA_{T.3.Bound}$ | | | | | | | |

| Reaction | Rate Coefficient | RuP (s⁻¹) | RuP 2 (s⁻¹) | RuP3 (s⁻¹) | RuCP (s⁻¹) | RuCP2 (s⁻¹) | RuCP3 (s⁻¹) |
|---|---|---|---|---|---|---|---|
| $\|Y_{Bound}\rangle \xrightarrow{\|Z4_{Bound}\rangle, I_{Probe}(\lambda,\tau)} \|Y_{Bound}\rangle + ESA_{Y.4.Bound}$ | $k_{ESA_4}$ | $3.9 \cdot 10^{13}$ | $3.0 \cdot 10^{13}$ | $4.6 \cdot 10^{13}$ | $5.7 \cdot 10^{13}$ | $11 \cdot 10^{13}$ | $5.7 \cdot 10^{13}$ |
| $\|B_{Bound}\rangle \xrightarrow{\|Z4_{Bound}\rangle, I_{Probe}(\lambda,\tau)} \|B_{Bound}\rangle + ESA_{B.4.Bound}$ | | | | | | | |
| $\|X_{Bound}\rangle \xrightarrow{\|Z4_{Bound}\rangle, I_{Probe}(\lambda,\tau)} \|X_{Bound}\rangle + ESA_{X.4.Bound}$ | | | | | | | |
| $\|T_{Bound}\rangle \xrightarrow{\|Z4_{Bound}\rangle, I_{Probe}(\lambda,\tau)} \|T_{Bound}\rangle + ESA_{T.4.Bound}$ | | | | | | | |
| $\|Y_{Bound}\rangle \xrightarrow{\|Z5_{Bound}\rangle, I_{Probe}(\lambda,\tau)} \|Y_{Bound}\rangle + ESA_{Y.5.Bound}$ | $k_{ESA_5}$ | $5.4 \cdot 10^{13}$ | $3.8 \cdot 10^{13}$ | $5.0 \cdot 10^{13}$ | $7.0 \cdot 10^{13}$ | $10.5 \cdot 10^{13}$ | $5.0 \cdot 10^{13}$ |
| $\|B_{Bound}\rangle \xrightarrow{\|Z5_{Bound}\rangle, I_{Probe}(\lambda,\tau)} \|B_{Bound}\rangle + ESA_{B.5.Bound}$ | | | | | | | |
| $\|X_{Bound}\rangle \xrightarrow{\|Z5_{Bound}\rangle, I_{Probe}(\lambda,\tau)} \|X_{Bound}\rangle + ESA_{X.5.Bound}$ | | | | | | | |
| $\|T_{Bound}\rangle \xrightarrow{\|Z5_{Bound}\rangle, I_{Probe}(\lambda,\tau)} \|T_{Bound}\rangle + ESA_{T.5.Bound}$ | | | | | | | |
| $\|T_{Bound}\rangle \rightarrow \|S0_{Bound}\rangle + rad_{Bound}$ | $k_{rad}$ | $10.9 \cdot 10^{4}$ | $9.6 \cdot 10^{4}$ | $11 \cdot 10^{4}$ | $10.6 \cdot 10^{4}$ | $10.6 \cdot 10^{4}$ | $10.3 \cdot 10^{4}$ |
| $\|T_{Bound}\rangle \rightarrow \|S0_{Bound}\rangle$ | $k_{nr}$ | $3.2 \cdot 10^{6}$ | $2.6 \cdot 10^{6}$ | $2.0 \cdot 10^{6}$ | $2.0 \cdot 10^{6}$ | $2.2 \cdot 10^{6}$ | $2.3 \cdot 10^{6}$ |
| $\|S0_{Unbound}\rangle \xrightarrow{h\nu_{Pump}} \|Y_{Unbound}\rangle$ | $k_{Pump,Y}$ | Variable | Variable | Variable | Variable | Variable | Variable |
| $\|S0_{Unbound}\rangle \xrightarrow{h\nu_{Pump}} \|B_{Unbound}\rangle$ | $k_{Pump,B}$ | Variable | Variable | Variable | Variable | Variable | Variable |
| $\|S0_{Unbound}\rangle \xrightarrow{h\nu_{Pump}} \|X_{Unbound}\rangle$ | $k_{Pump,X}$ | Variable | Variable | Variable | Variable | Variable | Variable |
| $\|S0_{Unbound}\rangle \xrightarrow{h\nu_{Pump}} \|T_{Unbound}\rangle$ | $k_{Pump,T}$ | Variable | Variable | Variable | Variable | Variable | Variable |
| $\|S0_{Unbound}\rangle \xrightarrow{\|Y_{Unbound}\rangle, I_{Probe}(\lambda,\tau)} \|S0_{Unbound}\rangle$ | $k_{GSB_Y}$ | $1.8 \cdot 10^{13}$ | $2.0 \cdot 10^{13}$ | $2.0 \cdot 10^{13}$ | $1.9 \cdot 10^{13}$ | $1.8 \cdot 10^{13}$ | $3.4 \cdot 10^{13}$ |
| $\|S0_{Unbound}\rangle \xrightarrow{I_{Probe}(\lambda,\tau)} \|Y_{Unbound}\rangle + ABS_{Y.Unbound}$ | $k_{ABS_Y}$ | | | | | | |
| $\|Y_{Unbound}\rangle \xrightarrow{I_{Probe}(\lambda,\tau)} \|S0_{Unbound}\rangle + ESE_{Y.Unbound}$ | $k_{ESE_Y}$ | | | | | | |
| $\|S0_{Unbound}\rangle \xrightarrow{\|B_{Unbound}\rangle, I_{Probe}(\lambda,\tau)} \|S0_{Unbound}\rangle$ | $k_{GSB_B}$ | $3.8 \cdot 10^{13}$ | $4.1 \cdot 10^{13}$ | $3.3 \cdot 10^{13}$ | $3.2 \cdot 10^{13}$ | $3.2 \cdot 10^{13}$ | $3.7 \cdot 10^{13}$ |
| $\|S0_{Unbound}\rangle \xrightarrow{I_{Probe}(\lambda,\tau)} \|B_{Unbound}\rangle + ABS_{B.Unbound}$ | $k_{ABS_B}$ | | | | | | |
| $\|B_{Unbound}\rangle \xrightarrow{I_{Probe}(\lambda,\tau)} \|S0_{Unbound}\rangle + ESE_{B.Unbound}$ | $k_{ESE_B}$ | | | | | | |
| $\|S0_{Unbound}\rangle \xrightarrow{\|X_{Unbound}\rangle, I_{Probe}(\lambda,\tau)} \|S0_{Unbound}\rangle$ | $k_{GSB_X}$ | $4.8 \cdot 10^{13}$ | $5.0 \cdot 10^{13}$ | $4.0 \cdot 10^{13}$ | $4.1 \cdot 10^{13}$ | $4.1 \cdot 10^{13}$ | $4.5 \cdot 10^{13}$ |
| $\|S0_{Unbound}\rangle \xrightarrow{I_{Probe}(\lambda,\tau)} \|X_{Unbound}\rangle + ABS_{X.Unbound}$ | $k_{ABS_X}$ | | | | | | |
| $\|X_{Unbound}\rangle \xrightarrow{I_{Probe}(\lambda,\tau)} \|S0_{Unbound}\rangle + ESE_{X.Unbound}$ | $k_{ESE_X}$ | | | | | | |
| $\|S0_{Unbound}\rangle \xrightarrow{\|T_{Unbound}\rangle, I_{Probe}(\lambda,\tau)} \|S0_{Unbound}\rangle$ | $k_{GSB_T}$ | $1.8 \cdot 10^{13}$ | $1.8 \cdot 10^{13}$ | $1.8 \cdot 10^{13}$ | $1.8 \cdot 10^{13}$ | $1.8 \cdot 10^{13}$ | $1.8 \cdot 10^{13}$ |
| $\|S0_{Unbound}\rangle \xrightarrow{I_{Probe}(\lambda,\tau)} \|T_{Unbound}\rangle + ABS_{T.Unbound}$ | $k_{ABS_T}$ | | | | | | |
| $\|T_{Unbound}\rangle \xrightarrow{I_{Probe}(\lambda,\tau)} \|S0_{Unbound}\rangle + ESE_{T.Unbound}$ | $k_{ESE_T}$ | | | | | | |
| $\|Y_{Unbound}\rangle \rightarrow \|B_{Unbound}\rangle$ | $k_{YB}$ | $1.8 \cdot 10^{13}$ | $2.4 \cdot 10^{13}$ | $8.0 \cdot 10^{13}$ | $6.0 \cdot 10^{13}$ | $6.0 \cdot 10^{13}$ | $6.0 \cdot 10^{13}$ |
| $\|B_{Unbound}\rangle \rightarrow \|X_{Unbound}\rangle$ | $k_{BX}$ | $1.8 \cdot 10^{13}$ | $2.4 \cdot 10^{13}$ | $8.0 \cdot 10^{13}$ | $6.0 \cdot 10^{13}$ | $6.0 \cdot 10^{13}$ | $6.0 \cdot 10^{13}$ |
| $\|X_{Unbound}\rangle \rightarrow \|S0_{Unbound}\rangle$ | $k_{UF-nr}$ | $2.4 \cdot 10^{13}$ | $1.0 \cdot 10^{13}$ | $4.0 \cdot 10^{13}$ | $1.6 \cdot 10^{13}$ | $1.6 \cdot 10^{13}$ | $2.0 \cdot 10^{13}$ |
| $\|X_{Unbound}\rangle \rightarrow \|T_{Unbound}\rangle$ | $k_{ISC}$ | $4.0 \cdot 10^{13}$ | $4.0 \cdot 10^{13}$ | $2.0 \cdot 10^{13}$ | $2.0 \cdot 10^{13}$ | $2.0 \cdot 10^{13}$ | $3.6 \cdot 10^{13}$ |
| $\|Y_{Unbound}\rangle \xrightarrow{\|Z0_{Unbound}\rangle, I_{Probe}(\lambda,\tau)} \|Y_{Unbound}\rangle$ | $k_{ESA_0}$ | $1.6 \cdot 10^{14}$ | $1.5 \cdot 10^{14}$ | $1.5 \cdot 10^{14}$ | $1.8 \cdot 10^{14}$ | $2.1 \cdot 10^{14}$ | $1.9 \cdot 10^{14}$ |
| $\|B_{Unbound}\rangle \xrightarrow{\|Z0_{Unbound}\rangle, I_{Probe}(\lambda,\tau)} \|B_{Unbound}\rangle$ | | | | | | | |
| $\|X_{Unbound}\rangle \xrightarrow{\|Z0_{Unbound}\rangle, I_{Probe}(\lambda,\tau)} \|X_{Unbound}\rangle$ | | | | | | | |
| $\|T_{Unbound}\rangle \xrightarrow{\|Z0_{Unbound}\rangle, I_{Probe}(\lambda,\tau)} \|T_{Unbound}\rangle$ | | | | | | | |
| $\|Y_{Unbound}\rangle \xrightarrow{\|Z1_{Unbound}\rangle, I_{Probe}(\lambda,\tau)} \|Y_{Unbound}\rangle$ | $k_{ESA_1}$ | $4.0 \cdot 10^{13}$ | $3.2 \cdot 10^{13}$ | $3.8 \cdot 10^{13}$ | $4.8 \cdot 10^{13}$ | $4.0 \cdot 10^{13}$ | $4.9 \cdot 10^{13}$ |
| $\|B_{Unbound}\rangle \xrightarrow{\|Z1_{Unbound}\rangle, I_{Probe}(\lambda,\tau)} \|B_{Unbound}\rangle$ | | | | | | | |
| $\|X_{Unbound}\rangle \xrightarrow{\|Z1_{Unbound}\rangle, I_{Probe}(\lambda,\tau)} \|X_{Unbound}\rangle$ | | | | | | | |
| $\|T_{Unbound}\rangle \xrightarrow{\|Z1_{Unbound}\rangle, I_{Probe}(\lambda,\tau)} \|T_{Unbound}\rangle$ | | | | | | | |

| | Rate Coefficient | RuP (s⁻¹) | RuP 2 (s⁻¹) | RuP3 (s⁻¹) | RuCP (s⁻¹) | RuCP2 (s⁻¹) | RuCP3 (s⁻¹) |
|---|---|---|---|---|---|---|---|
| $\|Y_{Unbound}\rangle \xrightarrow{\|Z2_{Unbound}\rangle, I_{Probe}(\lambda,\tau)} \|Y_{Unbound}\rangle$ <br> $\|B_{Unbound}\rangle \xrightarrow{\|Z2_{Unbound}\rangle, I_{Probe}(\lambda,\tau)} \|B_{Unbound}\rangle$ <br> $\|X_{Unbound}\rangle \xrightarrow{\|Z2_{Unbound}\rangle, I_{Probe}(\lambda,\tau)} \|X_{Unbound}\rangle$ <br> $\|T_{Unbound}\rangle \xrightarrow{\|Z2_{Unbound}\rangle, I_{Probe}(\lambda,\tau)} \|T_{Unbound}\rangle$ | $k_{ESA_2}$ | $4.3 \cdot 10^{13}$ | $2.8 \cdot 10^{13}$ | $4.7 \cdot 10^{13}$ | $5.2 \cdot 10^{13}$ | $5.7 \cdot 10^{13}$ | $5.3 \cdot 10^{13}$ |
| $\|Y_{Unbound}\rangle \xrightarrow{\|Z3_{Unbound}\rangle, I_{Probe}(\lambda,\tau)} \|Y_{Unbound}\rangle$ <br> $\|B_{Unbound}\rangle \xrightarrow{\|Z3_{Unbound}\rangle, I_{Probe}(\lambda,\tau)} \|B_{Unbound}\rangle$ <br> $\|X_{Unbound}\rangle \xrightarrow{\|Z3_{Unbound}\rangle, I_{Probe}(\lambda,\tau)} \|X_{Unbound}\rangle$ <br> $\|T_{Unbound}\rangle \xrightarrow{\|Z3_{Unbound}\rangle, I_{Probe}(\lambda,\tau)} \|T_{Unbound}\rangle$ | $k_{ESA_3}$ | $4.2 \cdot 10^{13}$ | $3.0 \cdot 10^{13}$ | $5.3 \cdot 10^{13}$ | $5.6 \cdot 10^{13}$ | $9.3 \cdot 10^{13}$ | $5.7 \cdot 10^{13}$ |
| $\|Y_{Unbound}\rangle \xrightarrow{\|Z4_{Unbound}\rangle, I_{Probe}(\lambda,\tau)} \|Y_{Unbound}\rangle$ <br> $\|B_{Unbound}\rangle \xrightarrow{\|Z4_{Unbound}\rangle, I_{Probe}(\lambda,\tau)} \|B_{Unbound}\rangle$ <br> $\|X_{Unbound}\rangle \xrightarrow{\|Z4_{Unbound}\rangle, I_{Probe}(\lambda,\tau)} \|X_{Unbound}\rangle$ <br> $\|T_{Unbound}\rangle \xrightarrow{\|Z4_{Unbound}\rangle, I_{Probe}(\lambda,\tau)} \|T_{Unbound}\rangle$ | $k_{ESA_4}$ | $3.9 \cdot 10^{13}$ | $3.0 \cdot 10^{13}$ | $4.6 \cdot 10^{13}$ | $5.7 \cdot 10^{13}$ | $11 \cdot 10^{13}$ | $5.7 \cdot 10^{13}$ |
| $\|Y_{Unbound}\rangle \xrightarrow{\|Z5_{Unbound}\rangle, I_{Probe}(\lambda,\tau)} \|Y_{Unbound}\rangle$ <br> $\|B_{Unbound}\rangle \xrightarrow{\|Z5_{Unbound}\rangle, I_{Probe}(\lambda,\tau)} \|B_{Unbound}\rangle$ <br> $\|X_{Unbound}\rangle \xrightarrow{\|Z5_{Unbound}\rangle, I_{Probe}(\lambda,\tau)} \|X_{Unbound}\rangle$ <br> $\|T_{Unbound}\rangle \xrightarrow{\|Z5_{Unbound}\rangle, I_{Probe}(\lambda,\tau)} \|T_{Unbound}\rangle$ | $k_{ESA_5}$ | $5.4 \cdot 10^{13}$ | $3.8 \cdot 10^{13}$ | $5.0 \cdot 10^{13}$ | $7.0 \cdot 10^{13}$ | $10.5 \cdot 10^{13}$ | $5.0 \cdot 10^{13}$ |
| $\|T_{Unbound}\rangle \to \|S0_{Unbound}\rangle + rad_{Unbound}$ | $k_{rad}$ | $10.9 \cdot 10^{4}$ | $9.6 \cdot 10^{4}$ | $11 \cdot 10^{4}$ | $10.6 \cdot 10^{4}$ | $10.6 \cdot 10^{4}$ | $10.3 \cdot 10^{4}$ |
| $\|T_{Unbound}\rangle \to \|S0_{Unbound}\rangle$ | $k_{nr}$ | $3.2 \cdot 10^{6}$ | $2.6 \cdot 10^{6}$ | $2.0 \cdot 10^{6}$ | $2.0 \cdot 10^{6}$ | $2.2 \cdot 10^{6}$ | $2.3 \cdot 10^{6}$ |
| $\|S0_{Bound}\rangle \xrightarrow{\|Y_{Bound}\rangle, I_{Probe}(\lambda,\tau)} \|S0_{Bound}\rangle$ | $k_{GSB_Y,Reflect}$ | | | | | | |
| $\|S0_{Bound}\rangle \xrightarrow{I_{Probe}(\lambda,\tau)} \|Y_{Bound}\rangle + ABS_{Y.Bound.Reflect}$ | $k_{ABS_Y,Reflect}$ | $1.3 \cdot 10^{10}$ | $5.2 \cdot 10^{9}$ | $5.5 \cdot 10^{9}$ | $3.6 \cdot 10^{9}$ | $1.1 \cdot 10^{6}$ | $2.4 \cdot 10^{10}$ |
| $\|Y_{Bound}\rangle \xrightarrow{I_{Probe}(\lambda,\tau)} \|S0_{Bound}\rangle + ESE_{Y.Bound.Reflect}$ | $k_{ESE_Y,Reflect}$ | | | | | | |
| $\|S0_{Bound}\rangle \xrightarrow{\|B_{Bound}\rangle, I_{Probe}(\lambda,\tau)} \|S0_{Bound}\rangle$ | $k_{GSB_B,Reflect}$ | | | | | | |
| $\|S0_{Bound}\rangle \xrightarrow{I_{Probe}(\lambda,\tau)} \|B_{Bound}\rangle + ABS_{B.Bound.Reflect}$ | $k_{ABS_B,Reflect}$ | $2.4 \cdot 10^{12}$ | $2.5 \cdot 10^{12}$ | $2.5 \cdot 10^{12}$ | $1.6 \cdot 10^{12}$ | $8.7 \cdot 10^{11}$ | $3.3 \cdot 10^{12}$ |
| $\|B_{Bound}\rangle \xrightarrow{I_{Probe}(\lambda,\tau)} \|S0_{Bound}\rangle + ESE_{B.Bound.Reflect}$ | $k_{ESE_B,Reflect}$ | | | | | | |
| $\|S0_{Bound}\rangle \xrightarrow{\|X_{Bound}\rangle, I_{Probe}(\lambda,\tau)} \|S0_{Bound}\rangle$ | $k_{GSB_X,Reflect}$ | | | | | | |
| $\|S0_{Bound}\rangle \xrightarrow{I_{Probe}(\lambda,\tau)} \|X_{Bound}\rangle + ABS_{X.Bound.Reflect}$ | $k_{ABS_X,Reflect}$ | $1.0 \cdot 10^{13}$ | $1.1 \cdot 10^{13}$ | $8.6 \cdot 10^{12}$ | $8.4 \cdot 10^{12}$ | $9.6 \cdot 10^{12}$ | $9.7 \cdot 10^{12}$ |
| $\|X_{Bound}\rangle \xrightarrow{I_{Probe}(\lambda,\tau)} \|S0_{Bound}\rangle + ESE_{X.Bound.Reflect}$ | $k_{ESE_X,Reflect}$ | | | | | | |
| $\|S0_{Bound}\rangle \xrightarrow{\|T_{Bound}\rangle, I_{Probe}(\lambda,\tau)} \|S0_{Bound}\rangle$ | $k_{GSB_T,Reflect}$ | | | | | | |
| $\|S0_{Bound}\rangle \xrightarrow{I_{Probe}(\lambda,\tau)} \|T_{Bound}\rangle + ABS_{T.Bound.Reflect}$ | $k_{ABS_T,Reflect}$ | $3.3 \cdot 10^{12}$ | $3.3 \cdot 10^{12}$ | $2.7 \cdot 10^{12}$ | $2.0 \cdot 10^{12}$ | $3.9 \cdot 10^{12}$ | $3.1 \cdot 10^{12}$ |
| $\|T_{Bound}\rangle \xrightarrow{I_{Probe}(\lambda,\tau)} \|S0_{Bound}\rangle + ESE_{T.Bound.Reflect}$ | $k_{ESE_T,Reflect}$ | | | | | | |
| $\|Y_{Bound}\rangle \xrightarrow{\|Z0_{Bound}\rangle, I_{Reflect}(\lambda,\tau)} \|Y_{Bound}\rangle$ <br> $\|B_{Bound}\rangle \xrightarrow{\|Z0_{Bound}\rangle, I_{Reflect}(\lambda,\tau)} \|B_{Bound}\rangle$ <br> $\|X_{Bound}\rangle \xrightarrow{\|Z0_{Bound}\rangle, I_{Reflect}(\lambda,\tau)} \|X_{Bound}\rangle$ <br> $\|T_{Bound}\rangle \xrightarrow{\|Z0_{Bound}\rangle, I_{Reflect}(\lambda,\tau)} \|T_{Bound}\rangle$ | $k_{ESA_0,Reflect}$ | $6.3 \cdot 10^{8}$ | $1.7 \cdot 10^{7}$ | $7.0 \cdot 10^{9}$ | $6.4 \cdot 10^{8}$ | $2.0 \cdot 10^{8}$ | $2.1 \cdot 10^{8}$ |
| $\|Y_{Bound}\rangle \xrightarrow{\|Z1_{Bound}\rangle, I_{Reflect}(\lambda,\tau)} \|Y_{Bound}\rangle$ <br> $\|B_{Bound}\rangle \xrightarrow{\|Z1_{Bound}\rangle, I_{Reflect}(\lambda,\tau)} \|B_{Bound}\rangle$ <br> $\|X_{Bound}\rangle \xrightarrow{\|Z1_{Bound}\rangle, I_{Reflect}(\lambda,\tau)} \|X_{Bound}\rangle$ <br> $\|T_{Bound}\rangle \xrightarrow{\|Z1_{Bound}\rangle, I_{Reflect}(\lambda,\tau)} \|T_{Bound}\rangle$ | $k_{ESA_1,Reflect}$ | $6.9 \cdot 10^{12}$ | $5.3 \cdot 10^{12}$ | $6.4 \cdot 10^{12}$ | $8.2 \cdot 10^{12}$ | $6.8 \cdot 10^{12}$ | $7.8 \cdot 10^{12}$ |

| | Rate Coefficient | RuP (s$^{-1}$) | RuP 2 (s$^{-1}$) | RuP3 (s$^{-1}$) | RuCP (s$^{-1}$) | RuCP2 (s$^{-1}$) | RuCP3 (s$^{-1}$) |
|---|---|---|---|---|---|---|---|
| $\|Y_{Bound}\rangle \xrightarrow{\|Z2_{Bound}\rangle, I_{Reflect}(\lambda,\tau)} \|Y_{Bound}\rangle$ <br> $\|B_{Bound}\rangle \xrightarrow{\|Z2_{Bound}\rangle, I_{Reflect}(\lambda,\tau)} \|B_{Bound}\rangle$ <br> $\|X_{Bound}\rangle \xrightarrow{\|Z2_{Bound}\rangle, I_{Reflect}(\lambda,\tau)} \|X_{Bound}\rangle$ <br> $\|T_{Bound}\rangle \xrightarrow{\|Z2_{Bound}\rangle, I_{Reflect}(\lambda,\tau)} \|T_{Bound}\rangle$ | $k_{ESA_2,Reflect}$ | 5.2·10$^{12}$ | 3.3·10$^{12}$ | 5.5·10$^{12}$ | 6.0·10$^{12}$ | 5.4·10$^{12}$ | 5.8·10$^{12}$ |
| $\|Y_{Bound}\rangle \xrightarrow{\|Z3_{Bound}\rangle, I_{Reflect}(\lambda,\tau)} \|Y_{Bound}\rangle$ <br> $\|B_{Bound}\rangle \xrightarrow{\|Z3_{Bound}\rangle, I_{Reflect}(\lambda,\tau)} \|B_{Bound}\rangle$ <br> $\|X_{Bound}\rangle \xrightarrow{\|Z3_{Bound}\rangle, I_{Reflect}(\lambda,\tau)} \|X_{Bound}\rangle$ <br> $\|T_{Bound}\rangle \xrightarrow{\|Z3_{Bound}\rangle, I_{Reflect}(\lambda,\tau)} \|T_{Bound}\rangle$ | $k_{ESA_3,Reflect}$ | 2.8·10$^{12}$ | 2.0·10$^{12}$ | 3.3·10$^{12}$ | 3.5·10$^{12}$ | 5.3·10$^{12}$ | 3.6·10$^{12}$ |
| $\|Y_{Bound}\rangle \xrightarrow{\|Z4_{Bound}\rangle, I_{Reflect}(\lambda,\tau)} \|Y_{Bound}\rangle$ <br> $\|B_{Bound}\rangle \xrightarrow{\|Z4_{Bound}\rangle, I_{Reflect}(\lambda,\tau)} \|B_{Bound}\rangle$ <br> $\|X_{Bound}\rangle \xrightarrow{\|Z4_{Bound}\rangle, I_{Reflect}(\lambda,\tau)} \|X_{Bound}\rangle$ <br> $\|T_{Bound}\rangle \xrightarrow{\|Z4_{Bound}\rangle, I_{Reflect}(\lambda,\tau)} \|T_{Bound}\rangle$ | $k_{ESA_4,Reflect}$ | 1.8·10$^{12}$ | 1.4·10$^{12}$ | 1.9·10$^{12}$ | 2.5·10$^{12}$ | 4.7·10$^{12}$ | 2.5·10$^{12}$ |
| $\|Y_{Bound}\rangle \xrightarrow{\|Z5_{Bound}\rangle, I_{Reflect}(\lambda,\tau)} \|Y_{Bound}\rangle$ <br> $\|B_{Bound}\rangle \xrightarrow{\|Z5_{Bound}\rangle, I_{Reflect}(\lambda,\tau)} \|B_{Bound}\rangle$ <br> $\|X_{Bound}\rangle \xrightarrow{\|Z5_{Bound}\rangle, I_{Reflect}(\lambda,\tau)} \|X_{Bound}\rangle$ <br> $\|T_{Bound}\rangle \xrightarrow{\|Z5_{Bound}\rangle, I_{Reflect}(\lambda,\tau)} \|T_{Bound}\rangle$ | $k_{ESA_5,Reflect}$ | 1.5·10$^{12}$ | 1.1·10$^{12}$ | 1.5·10$^{12}$ | 1.9·10$^{12}$ | 3.4·10$^{12}$ | 1.4·10$^{12}$ |

**Table S3.** Base mechanistic scheme and rate coefficients for dyes in set 6-Ru on 4 μm film of ZrO$_2$ for dataset Ru-Z with broadband probe spanning 350 nm to 700 nm. Simulated surface concentration was $2.5 \cdot 10^{-8}$ mol cm$^{-2}$.

| | Rate Coefficients | RuP (s$^{-1}$) | RuP 2 (s$^{-1}$) | RuP3 (s$^{-1}$) | RuCP (s$^{-1}$) | RuCP2 (s$^{-1}$) | RuCP3 (s$^{-1}$) |
|---|---|---|---|---|---|---|---|
| $\|S0_{Bound}\rangle \xrightarrow{h\nu_{Pump}} \|Y_{Bound}\rangle$ | $k_{Pump,Y}$ | Variable | Variable | Variable | Variable | Variable | Variable |
| $\|S0_{Bound}\rangle \xrightarrow{h\nu_{Pump}} \|B_{Bound}\rangle$ | $k_{Pump,B}$ | Variable | Variable | Variable | Variable | Variable | Variable |
| $\|S0_{Bound}\rangle \xrightarrow{h\nu_{Pump}} \|X_{Bound}\rangle$ | $k_{Pump,X}$ | Variable | Variable | Variable | Variable | Variable | Variable |
| $\|S0_{Bound}\rangle \xrightarrow{h\nu_{Pump}} \|T_{Bound}\rangle$ | $k_{Pump,T}$ | Variable | Variable | Variable | Variable | Variable | Variable |
| $\|S0_{Bound}\rangle \xrightarrow{\|Y_{Bound}\rangle, I_{Probe}(\lambda,\tau)} \|S0_{Bound}\rangle$ | $k_{GSB_Y}$ | | | | | | |
| $\|S0_{Bound}\rangle \xrightarrow{I_{Probe}(\lambda,\tau)} \|Y_{Bound}\rangle + ABS_{Y.Bound}$ | $k_{ABS_Y}$ | $1.8 \cdot 10^{13}$ | $2.0 \cdot 10^{13}$ | $2.0 \cdot 10^{13}$ | $1.9 \cdot 10^{13}$ | $1.8 \cdot 10^{13}$ | $3.4 \cdot 10^{13}$ |
| $\|Y_{Bound}\rangle \xrightarrow{I_{Probe}(\lambda,\tau)} \|S0_{Bound}\rangle + ESE_{Y.Bound}$ | $k_{ESE_Y}$ | | | | | | |
| $\|S0_{Bound}\rangle \xrightarrow{\|B_{Bound}\rangle, I_{Probe}(\lambda,\tau)} \|S0_{Bound}\rangle$ | $k_{GSB_B}$ | | | | | | |
| $\|S0_{Bound}\rangle \xrightarrow{I_{Probe}(\lambda,\tau)} \|B_{Bound}\rangle + ABS_{B.Bound}$ | $k_{ABS_B}$ | $3.8 \cdot 10^{13}$ | $4.1 \cdot 10^{13}$ | $3.3 \cdot 10^{13}$ | $3.2 \cdot 10^{13}$ | $3.2 \cdot 10^{13}$ | $3.7 \cdot 10^{13}$ |
| $\|B_{Bound}\rangle \xrightarrow{I_{Probe}(\lambda,\tau)} \|S0_{Bound}\rangle + ESE_{B.Bound}$ | $k_{ESE_B}$ | | | | | | |
| $\|S0_{Bound}\rangle \xrightarrow{\|X_{Bound}\rangle, I_{Probe}(\lambda,\tau)} \|S0_{Bound}\rangle$ | $k_{GSB_X}$ | | | | | | |
| $\|S0_{Bound}\rangle \xrightarrow{I_{Probe}(\lambda,\tau)} \|X_{Bound}\rangle + ABS_{X.Bound}$ | $k_{ABS_X}$ | $4.8 \cdot 10^{13}$ | $5.0 \cdot 10^{13}$ | $4.0 \cdot 10^{13}$ | $4.1 \cdot 10^{13}$ | $4.1 \cdot 10^{13}$ | $4.5 \cdot 10^{13}$ |
| $\|X_{Bound}\rangle \xrightarrow{I_{Probe}(\lambda,\tau)} \|S0_{Bound}\rangle + ESE_{X.Bound}$ | $k_{ESE_X}$ | | | | | | |
| $\|S0_{Bound}\rangle \xrightarrow{\|T_{Bound}\rangle, I_{Probe}(\lambda,\tau)} \|S0_{Bound}\rangle + GSB_{T.Bound}$ | $k_{GSB_T}$ | | | | | | |
| $\|S0_{Bound}\rangle \xrightarrow{I_{Probe}(\lambda,\tau)} \|T_{Bound}\rangle + ABS_{T.Bound}$ | $k_{ABS_T}$ | $1.8 \cdot 10^{13}$ | $1.8 \cdot 10^{13}$ | $1.8 \cdot 10^{13}$ | $1.8 \cdot 10^{13}$ | $1.8 \cdot 10^{13}$ | $1.8 \cdot 10^{13}$ |
| $\|T_{Bound}\rangle \xrightarrow{I_{Probe}(\lambda,\tau)} \|S0_{Bound}\rangle + ESE_{T.Bound}$ | $k_{ESE_T}$ | | | | | | |
| $\|Y_{Bound}\rangle \rightarrow \|B_{Bound}\rangle$ | $k_{YB}$ | $1.8 \cdot 10^{13}$ | $2.4 \cdot 10^{13}$ | $8.0 \cdot 10^{13}$ | $6.0 \cdot 10^{13}$ | $6.0 \cdot 10^{13}$ | $6.0 \cdot 10^{13}$ |
| $\|B_{Bound}\rangle \rightarrow \|X_{Bound}\rangle$ | $k_{BX}$ | $1.8 \cdot 10^{13}$ | $2.4 \cdot 10^{13}$ | $8.0 \cdot 10^{13}$ | $6.0 \cdot 10^{13}$ | $6.0 \cdot 10^{13}$ | $6.0 \cdot 10^{13}$ |
| $\|X_{Bound}\rangle \rightarrow \|S0_{Bound}\rangle$ | $k_{UF-nr}$ | $2.4 \cdot 10^{13}$ | $1.0 \cdot 10^{13}$ | $4.0 \cdot 10^{13}$ | $1.6 \cdot 10^{13}$ | $1.6 \cdot 10^{13}$ | $2.0 \cdot 10^{13}$ |
| $\|X_{Bound}\rangle \rightarrow \|T_{Bound}\rangle$ | $k_{ISC}$ | $4.0 \cdot 10^{13}$ | $4.0 \cdot 10^{13}$ | $2.0 \cdot 10^{13}$ | $2.0 \cdot 10^{13}$ | $2.0 \cdot 10^{13}$ | $3.6 \cdot 10^{13}$ |
| $\|Y_{Bound}\rangle \xrightarrow{\|Z0_{Bound}\rangle, I_{Probe}(\lambda,\tau)} \|Y_{Bound}\rangle + ESA_{Y.0.Bound}$ | | | | | | | |
| $\|B_{Bound}\rangle \xrightarrow{\|Z0_{Bound}\rangle, I_{Probe}(\lambda,\tau)} \|B_{Bound}\rangle + ESA_{B.0.Bound}$ | $k_{ESA_0}$ | $1.6 \cdot 10^{14}$ | $1.5 \cdot 10^{14}$ | $1.5 \cdot 10^{14}$ | $1.8 \cdot 10^{14}$ | $2.1 \cdot 10^{14}$ | $1.9 \cdot 10^{14}$ |
| $\|X_{Bound}\rangle \xrightarrow{\|Z0_{Bound}\rangle, I_{Probe}(\lambda,\tau)} \|X_{Bound}\rangle + ESA_{X.0.Bound}$ | | | | | | | |
| $\|T_{Bound}\rangle \xrightarrow{\|Z0_{Bound}\rangle, I_{Probe}(\lambda,\tau)} \|T_{Bound}\rangle + ESA_{T.0.Bound}$ | | | | | | | |
| $\|Y_{Bound}\rangle \xrightarrow{\|Z1_{Bound}\rangle, I_{Probe}(\lambda,\tau)} \|Y_{Bound}\rangle + ESA_{Y.1.Bound}$ | | | | | | | |
| $\|B_{Bound}\rangle \xrightarrow{\|Z1_{Bound}\rangle, I_{Probe}(\lambda,\tau)} \|B_{Bound}\rangle + ESA_{B.1.Bound}$ | $k_{ESA_1}$ | $4.0 \cdot 10^{13}$ | $3.2 \cdot 10^{13}$ | $3.8 \cdot 10^{13}$ | $4.8 \cdot 10^{13}$ | $4.0 \cdot 10^{13}$ | $4.9 \cdot 10^{13}$ |
| $\|X_{Bound}\rangle \xrightarrow{\|Z1_{Bound}\rangle, I_{Probe}(\lambda,\tau)} \|X_{Bound}\rangle + ESA_{X.1.Bound}$ | | | | | | | |
| $\|T_{Bound}\rangle \xrightarrow{\|Z1_{Bound}\rangle, I_{Probe}(\lambda,\tau)} \|T_{Bound}\rangle + ESA_{T.1.Bound}$ | | | | | | | |
| $\|Y_{Bound}\rangle \xrightarrow{\|Z2_{Bound}\rangle, I_{Probe}(\lambda,\tau)} \|Y_{Bound}\rangle + ESA_{Y.2.Bound}$ | | | | | | | |
| $\|B_{Bound}\rangle \xrightarrow{\|Z2_{Bound}\rangle, I_{Probe}(\lambda,\tau)} \|B_{Bound}\rangle + ESA_{B.2.Bound}$ | $k_{ESA_2}$ | $4.3 \cdot 10^{13}$ | $2.8 \cdot 10^{13}$ | $4.7 \cdot 10^{13}$ | $5.2 \cdot 10^{13}$ | $5.7 \cdot 10^{13}$ | $5.3 \cdot 10^{13}$ |
| $\|X_{Bound}\rangle \xrightarrow{\|Z2_{Bound}\rangle, I_{Probe}(\lambda,\tau)} \|X_{Bound}\rangle + ESA_{X.2.Bound}$ | | | | | | | |
| $\|T_{Bound}\rangle \xrightarrow{\|Z2_{Bound}\rangle, I_{Probe}(\lambda,\tau)} \|T_{Bound}\rangle + ESA_{T.2.Bound}$ | | | | | | | |
| $\|Y_{Bound}\rangle \xrightarrow{\|Z3_{Bound}\rangle, I_{Probe}(\lambda,\tau)} \|Y_{Bound}\rangle + ESA_{Y.3.Bound}$ | | | | | | | |
| $\|B_{Bound}\rangle \xrightarrow{\|Z3_{Bound}\rangle, I_{Probe}(\lambda,\tau)} \|B_{Bound}\rangle + ESA_{B.3.Bound}$ | $k_{ESA_3}$ | $4.2 \cdot 10^{13}$ | $3.0 \cdot 10^{13}$ | $5.3 \cdot 10^{13}$ | $5.6 \cdot 10^{13}$ | $9.3 \cdot 10^{13}$ | $5.7 \cdot 10^{13}$ |
| $\|X_{Bound}\rangle \xrightarrow{\|Z3_{Bound}\rangle, I_{Probe}(\lambda,\tau)} \|X_{Bound}\rangle + ESA_{X.3.Bound}$ | | | | | | | |
| $\|T_{Bound}\rangle \xrightarrow{\|Z3_{Bound}\rangle, I_{Probe}(\lambda,\tau)} \|T_{Bound}\rangle + ESA_{T.3.Bound}$ | | | | | | | |

| Reaction | Rate Coefficient | RuP (s⁻¹) | RuP 2 (s⁻¹) | RuP3 (s⁻¹) | RuCP (s⁻¹) | RuCP2 (s⁻¹) | RuCP3 (s⁻¹) |
|---|---|---|---|---|---|---|---|
| $\|Y_{Bound}\rangle \xrightarrow{\|Z4_{Bound}\rangle, I_{Probe}(\lambda,\tau)} \|Y_{Bound}\rangle + ESA_{Y.4.Bound}$ | $k_{ESA_4}$ | $3.9 \cdot 10^{13}$ | $3.0 \cdot 10^{13}$ | $4.6 \cdot 10^{13}$ | $5.7 \cdot 10^{13}$ | $11 \cdot 10^{13}$ | $5.7 \cdot 10^{13}$ |
| $\|B_{Bound}\rangle \xrightarrow{\|Z4_{Bound}\rangle, I_{Probe}(\lambda,\tau)} \|B_{Bound}\rangle + ESA_{B.4.Bound}$ | | | | | | | |
| $\|X_{Bound}\rangle \xrightarrow{\|Z4_{Bound}\rangle, I_{Probe}(\lambda,\tau)} \|X_{Bound}\rangle + ESA_{X.4.Bound}$ | | | | | | | |
| $\|T_{Bound}\rangle \xrightarrow{\|Z4_{Bound}\rangle, I_{Probe}(\lambda,\tau)} \|T_{Bound}\rangle + ESA_{T.4.Bound}$ | | | | | | | |
| $\|Y_{Bound}\rangle \xrightarrow{\|Z5_{Bound}\rangle, I_{Probe}(\lambda,\tau)} \|Y_{Bound}\rangle + ESA_{Y.5.Bound}$ | $k_{ESA_5}$ | $5.4 \cdot 10^{13}$ | $3.8 \cdot 10^{13}$ | $5.0 \cdot 10^{13}$ | $7.0 \cdot 10^{13}$ | $10.5 \cdot 10^{13}$ | $5.0 \cdot 10^{13}$ |
| $\|B_{Bound}\rangle \xrightarrow{\|Z5_{Bound}\rangle, I_{Probe}(\lambda,\tau)} \|B_{Bound}\rangle + ESA_{B.5.Bound}$ | | | | | | | |
| $\|X_{Bound}\rangle \xrightarrow{\|Z5_{Bound}\rangle, I_{Probe}(\lambda,\tau)} \|X_{Bound}\rangle + ESA_{X.5.Bound}$ | | | | | | | |
| $\|T_{Bound}\rangle \xrightarrow{\|Z5_{Bound}\rangle, I_{Probe}(\lambda,\tau)} \|T_{Bound}\rangle + ESA_{T.5.Bound}$ | | | | | | | |
| $\|T_{Bound}\rangle \to \|S0_{Bound}\rangle + rad_{Bound}$ | $k_{rad}$ | $10.9 \cdot 10^4$ | $9.6 \cdot 10^4$ | $11 \cdot 10^4$ | $10.6 \cdot 10^4$ | $10.6 \cdot 10^4$ | $10.3 \cdot 10^4$ |
| $\|T_{Bound}\rangle \to \|S0_{Bound}\rangle$ | $k_{nr}$ | $3.2 \cdot 10^6$ | $2.6 \cdot 10^6$ | $2.0 \cdot 10^6$ | $2.0 \cdot 10^6$ | $2.2 \cdot 10^6$ | $2.3 \cdot 10^6$ |
| $\|S0_{Unbound}\rangle \xrightarrow{h\nu_{Pump}} \|Y_{Unbound}\rangle$ | $k_{Pump,Y}$ | Variable | Variable | Variable | Variable | Variable | Variable |
| $\|S0_{Unbound}\rangle \xrightarrow{h\nu_{Pump}} \|B_{Unbound}\rangle$ | $k_{Pump,B}$ | Variable | Variable | Variable | Variable | Variable | Variable |
| $\|S0_{Unbound}\rangle \xrightarrow{h\nu_{Pump}} \|X_{Unbound}\rangle$ | $k_{Pump,X}$ | Variable | Variable | Variable | Variable | Variable | Variable |
| $\|S0_{Unbound}\rangle \xrightarrow{h\nu_{Pump}} \|T_{Unbound}\rangle$ | $k_{Pump,T}$ | Variable | Variable | Variable | Variable | Variable | Variable |
| $\|S0_{Unbound}\rangle \xrightarrow{\|Y_{Unbound}\rangle, I_{Probe}(\lambda,\tau)} \|S0_{Unbound}\rangle$ | $k_{GSB_Y}$ | $1.8 \cdot 10^{13}$ | $2.0 \cdot 10^{13}$ | $2.0 \cdot 10^{13}$ | $1.9 \cdot 10^{13}$ | $1.8 \cdot 10^{13}$ | $3.4 \cdot 10^{13}$ |
| $\|S0_{Unbound}\rangle \xrightarrow{I_{Probe}(\lambda,\tau)} \|Y_{Unbound}\rangle + ABS_{Y.Unbound}$ | $k_{ABS_Y}$ | | | | | | |
| $\|Y_{Unbound}\rangle \xrightarrow{I_{Probe}(\lambda,\tau)} \|S0_{Unbound}\rangle + ESE_{Y.Unbound}$ | $k_{ESE_Y}$ | | | | | | |
| $\|S0_{Unbound}\rangle \xrightarrow{\|B_{Unbound}\rangle, I_{Probe}(\lambda,\tau)} \|S0_{Unbound}\rangle$ | $k_{GSB_B}$ | $3.8 \cdot 10^{13}$ | $4.1 \cdot 10^{13}$ | $3.3 \cdot 10^{13}$ | $3.2 \cdot 10^{13}$ | $3.2 \cdot 10^{13}$ | $3.7 \cdot 10^{13}$ |
| $\|S0_{Unbound}\rangle \xrightarrow{I_{Probe}(\lambda,\tau)} \|B_{Unbound}\rangle + ABS_{B.Unbound}$ | $k_{ABS_B}$ | | | | | | |
| $\|B_{Unbound}\rangle \xrightarrow{I_{Probe}(\lambda,\tau)} \|S0_{Unbound}\rangle + ESE_{B.Unbound}$ | $k_{ESE_B}$ | | | | | | |
| $\|S0_{Unbound}\rangle \xrightarrow{\|X_{Unbound}\rangle, I_{Probe}(\lambda,\tau)} \|S0_{Unbound}\rangle$ | $k_{GSB_X}$ | $4.8 \cdot 10^{13}$ | $5.0 \cdot 10^{13}$ | $4.0 \cdot 10^{13}$ | $4.1 \cdot 10^{13}$ | $4.1 \cdot 10^{13}$ | $4.5 \cdot 10^{13}$ |
| $\|S0_{Unbound}\rangle \xrightarrow{I_{Probe}(\lambda,\tau)} \|X_{Unbound}\rangle + ABS_{X.Unbound}$ | $k_{ABS_X}$ | | | | | | |
| $\|X_{Unbound}\rangle \xrightarrow{I_{Probe}(\lambda,\tau)} \|S0_{Unbound}\rangle + ESE_{X.Unbound}$ | $k_{ESE_X}$ | | | | | | |
| $\|S0_{Unbound}\rangle \xrightarrow{\|T_{Unbound}\rangle, I_{Probe}(\lambda,\tau)} \|S0_{Unbound}\rangle$ | $k_{GSB_T}$ | $1.8 \cdot 10^{13}$ | $1.8 \cdot 10^{13}$ | $1.8 \cdot 10^{13}$ | $1.8 \cdot 10^{13}$ | $1.8 \cdot 10^{13}$ | $1.8 \cdot 10^{13}$ |
| $\|S0_{Unbound}\rangle \xrightarrow{I_{Probe}(\lambda,\tau)} \|T_{Unbound}\rangle + ABS_{T.Unbound}$ | $k_{ABS_T}$ | | | | | | |
| $\|T_{Unbound}\rangle \xrightarrow{I_{Probe}(\lambda,\tau)} \|S0_{Unbound}\rangle + ESE_{T.Unbound}$ | $k_{ESE_T}$ | | | | | | |
| $\|Y_{Unbound}\rangle \to \|B_{Unbound}\rangle$ | $k_{YB}$ | $1.8 \cdot 10^{13}$ | $2.4 \cdot 10^{13}$ | $8.0 \cdot 10^{13}$ | $6.0 \cdot 10^{13}$ | $6.0 \cdot 10^{13}$ | $6.0 \cdot 10^{13}$ |
| $\|B_{Unbound}\rangle \to \|X_{Unbound}\rangle$ | $k_{BX}$ | $1.8 \cdot 10^{13}$ | $2.4 \cdot 10^{13}$ | $8.0 \cdot 10^{13}$ | $6.0 \cdot 10^{13}$ | $6.0 \cdot 10^{13}$ | $6.0 \cdot 10^{13}$ |
| $\|X_{Unbound}\rangle \to \|S0_{Unbound}\rangle$ | $k_{UF-nr}$ | $2.4 \cdot 10^{13}$ | $1.0 \cdot 10^{13}$ | $4.0 \cdot 10^{13}$ | $1.6 \cdot 10^{13}$ | $1.6 \cdot 10^{13}$ | $2.0 \cdot 10^{13}$ |
| $\|X_{Unbound}\rangle \to \|T_{Unbound}\rangle$ | $k_{ISC}$ | $4.0 \cdot 10^{13}$ | $4.0 \cdot 10^{13}$ | $2.0 \cdot 10^{13}$ | $2.0 \cdot 10^{13}$ | $2.0 \cdot 10^{13}$ | $3.6 \cdot 10^{13}$ |
| $\|Y_{Unbound}\rangle \xrightarrow{\|Z0_{Unbound}\rangle, I_{Probe}(\lambda,\tau)} \|Y_{Unbound}\rangle$ | $k_{ESA_0}$ | $1.6 \cdot 10^{14}$ | $1.5 \cdot 10^{14}$ | $1.5 \cdot 10^{14}$ | $1.8 \cdot 10^{14}$ | $2.1 \cdot 10^{14}$ | $1.9 \cdot 10^{14}$ |
| $\|B_{Unbound}\rangle \xrightarrow{\|Z0_{Unbound}\rangle, I_{Probe}(\lambda,\tau)} \|B_{Unbound}\rangle$ | | | | | | | |
| $\|X_{Unbound}\rangle \xrightarrow{\|Z0_{Unbound}\rangle, I_{Probe}(\lambda,\tau)} \|X_{Unbound}\rangle$ | | | | | | | |
| $\|T_{Unbound}\rangle \xrightarrow{\|Z0_{Unbound}\rangle, I_{Probe}(\lambda,\tau)} \|T_{Unbound}\rangle$ | | | | | | | |
| $\|Y_{Unbound}\rangle \xrightarrow{\|Z1_{Unbound}\rangle, I_{Probe}(\lambda,\tau)} \|Y_{Unbound}\rangle$ | $k_{ESA_1}$ | $4.0 \cdot 10^{13}$ | $3.2 \cdot 10^{13}$ | $3.8 \cdot 10^{13}$ | $4.8 \cdot 10^{13}$ | $4.0 \cdot 10^{13}$ | $4.9 \cdot 10^{13}$ |
| $\|B_{Unbound}\rangle \xrightarrow{\|Z1_{Unbound}\rangle, I_{Probe}(\lambda,\tau)} \|B_{Unbound}\rangle$ | | | | | | | |
| $\|X_{Unbound}\rangle \xrightarrow{\|Z1_{Unbound}\rangle, I_{Probe}(\lambda,\tau)} \|X_{Unbound}\rangle$ | | | | | | | |
| $\|T_{Unbound}\rangle \xrightarrow{\|Z1_{Unbound}\rangle, I_{Probe}(\lambda,\tau)} \|T_{Unbound}\rangle$ | | | | | | | |

| Reaction | Rate Coefficient | RuP (s⁻¹) | RuP 2 (s⁻¹) | RuP3 (s⁻¹) | RuCP (s⁻¹) | RuCP2 (s⁻¹) | RuCP3 (s⁻¹) |
|---|---|---|---|---|---|---|---|
| $\|Y_{Unbound}\rangle \xrightarrow{\|Z2_{Unbound}\rangle, I_{Probe}(\lambda,\tau)} \|Y_{Unbound}\rangle$ <br> $\|B_{Unbound}\rangle \xrightarrow{\|Z2_{Unbound}\rangle, I_{Probe}(\lambda,\tau)} \|B_{Unbound}\rangle$ <br> $\|X_{Unbound}\rangle \xrightarrow{\|Z2_{Unbound}\rangle, I_{Probe}(\lambda,\tau)} \|X_{Unbound}\rangle$ <br> $\|T_{Unbound}\rangle \xrightarrow{\|Z2_{Unbound}\rangle, I_{Probe}(\lambda,\tau)} \|T_{Unbound}\rangle$ | $k_{ESA_2}$ | $4.3 \cdot 10^{13}$ | $2.8 \cdot 10^{13}$ | $4.7 \cdot 10^{13}$ | $5.2 \cdot 10^{13}$ | $5.7 \cdot 10^{13}$ | $5.3 \cdot 10^{13}$ |
| $\|Y_{Unbound}\rangle \xrightarrow{\|Z3_{Unbound}\rangle, I_{Probe}(\lambda,\tau)} \|Y_{Unbound}\rangle$ <br> $\|B_{Unbound}\rangle \xrightarrow{\|Z3_{Unbound}\rangle, I_{Probe}(\lambda,\tau)} \|B_{Unbound}\rangle$ <br> $\|X_{Unbound}\rangle \xrightarrow{\|Z3_{Unbound}\rangle, I_{Probe}(\lambda,\tau)} \|X_{Unbound}\rangle$ <br> $\|T_{Unbound}\rangle \xrightarrow{\|Z3_{Unbound}\rangle, I_{Probe}(\lambda,\tau)} \|T_{Unbound}\rangle$ | $k_{ESA_3}$ | $4.2 \cdot 10^{13}$ | $3.0 \cdot 10^{13}$ | $5.3 \cdot 10^{13}$ | $5.6 \cdot 10^{13}$ | $9.3 \cdot 10^{13}$ | $5.7 \cdot 10^{13}$ |
| $\|Y_{Unbound}\rangle \xrightarrow{\|Z4_{Unbound}\rangle, I_{Probe}(\lambda,\tau)} \|Y_{Unbound}\rangle$ <br> $\|B_{Unbound}\rangle \xrightarrow{\|Z4_{Unbound}\rangle, I_{Probe}(\lambda,\tau)} \|B_{Unbound}\rangle$ <br> $\|X_{Unbound}\rangle \xrightarrow{\|Z4_{Unbound}\rangle, I_{Probe}(\lambda,\tau)} \|X_{Unbound}\rangle$ <br> $\|T_{Unbound}\rangle \xrightarrow{\|Z4_{Unbound}\rangle, I_{Probe}(\lambda,\tau)} \|T_{Unbound}\rangle$ | $k_{ESA_4}$ | $3.9 \cdot 10^{13}$ | $3.0 \cdot 10^{13}$ | $4.6 \cdot 10^{13}$ | $5.7 \cdot 10^{13}$ | $11 \cdot 10^{13}$ | $5.7 \cdot 10^{13}$ |
| $\|Y_{Unbound}\rangle \xrightarrow{\|Z5_{Unbound}\rangle, I_{Probe}(\lambda,\tau)} \|Y_{Unbound}\rangle$ <br> $\|B_{Unbound}\rangle \xrightarrow{\|Z5_{Unbound}\rangle, I_{Probe}(\lambda,\tau)} \|B_{Unbound}\rangle$ <br> $\|X_{Unbound}\rangle \xrightarrow{\|Z5_{Unbound}\rangle, I_{Probe}(\lambda,\tau)} \|X_{Unbound}\rangle$ <br> $\|T_{Unbound}\rangle \xrightarrow{\|Z5_{Unbound}\rangle, I_{Probe}(\lambda,\tau)} \|T_{Unbound}\rangle$ | $k_{ESA_5}$ | $5.4 \cdot 10^{13}$ | $3.8 \cdot 10^{13}$ | $5.0 \cdot 10^{13}$ | $7.0 \cdot 10^{13}$ | $10.5 \cdot 10^{13}$ | $5.0 \cdot 10^{13}$ |
| $\|T_{Unbound}\rangle \rightarrow \|S0_{Unbound}\rangle + rad_{Unbound}$ | $k_{rad}$ | $10.9 \cdot 10^{4}$ | $9.6 \cdot 10^{4}$ | $11 \cdot 10^{4}$ | $10.6 \cdot 10^{4}$ | $10.6 \cdot 10^{4}$ | $10.3 \cdot 10^{4}$ |
| $\|T_{Unbound}\rangle \rightarrow \|S0_{Unbound}\rangle$ | $k_{nr}$ | $3.2 \cdot 10^{6}$ | $2.6 \cdot 10^{6}$ | $2.0 \cdot 10^{6}$ | $2.0 \cdot 10^{6}$ | $2.2 \cdot 10^{6}$ | $2.3 \cdot 10^{6}$ |
| $\|S0_{Bound}\rangle \xrightarrow{\|Y_{Bound}\rangle, I_{Probe}(\lambda,\tau)} \|S0_{Bound}\rangle$ | $k_{GSB_Y,Reflect}$ | | | | | | |
| $\|S0_{Bound}\rangle \xrightarrow{I_{Probe}(\lambda,\tau)} \|Y_{Bound}\rangle + ABS_{Y.Bound.Reflect}$ | $k_{ABS_Y,Reflect}$ | $3.8 \cdot 10^{12}$ | $4.6 \cdot 10^{12}$ | $4.4 \cdot 10^{12}$ | $4.3 \cdot 10^{12}$ | $4.0 \cdot 10^{12}$ | $7.4 \cdot 10^{12}$ |
| $\|Y_{Bound}\rangle \xrightarrow{I_{Probe}(\lambda,\tau)} \|S0_{Bound}\rangle + ESE_{Y.Bound.Reflect}$ | $k_{ESE_Y,Reflect}$ | | | | | | |
| $\|S0_{Bound}\rangle \xrightarrow{\|B_{Bound}\rangle, I_{Probe}(\lambda,\tau)} \|S0_{Bound}\rangle$ | $k_{GSB_B,Reflect}$ | | | | | | |
| $\|S0_{Bound}\rangle \xrightarrow{I_{Probe}(\lambda,\tau)} \|B_{Bound}\rangle + ABS_{B.Bound.Reflect}$ | $k_{ABS_B,Reflect}$ | $6.2 \cdot 10^{12}$ | $6.7 \cdot 10^{12}$ | $5.3 \cdot 10^{12}$ | $5.4 \cdot 10^{12}$ | $5.3 \cdot 10^{12}$ | $5.9 \cdot 10^{12}$ |
| $\|B_{Bound}\rangle \xrightarrow{I_{Probe}(\lambda,\tau)} \|S0_{Bound}\rangle + ESE_{B.Bound.Reflect}$ | $k_{ESE_B,Reflect}$ | | | | | | |
| $\|S0_{Bound}\rangle \xrightarrow{\|X_{Bound}\rangle, I_{Probe}(\lambda,\tau)} \|S0_{Bound}\rangle$ | $k_{GSB_X,Reflect}$ | | | | | | |
| $\|S0_{Bound}\rangle \xrightarrow{I_{Probe}(\lambda,\tau)} \|X_{Bound}\rangle + ABS_{X.Bound.Reflect}$ | $k_{ABS_X,Reflect}$ | $6.5 \cdot 10^{12}$ | $6.8 \cdot 10^{12}$ | $5.4 \cdot 10^{12}$ | $5.6 \cdot 10^{12}$ | $5.6 \cdot 10^{12}$ | $5.9 \cdot 10^{12}$ |
| $\|X_{Bound}\rangle \xrightarrow{I_{Probe}(\lambda,\tau)} \|S0_{Bound}\rangle + ESE_{X.Bound.Reflect}$ | $k_{ESE_X,Reflect}$ | | | | | | |
| $\|S0_{Bound}\rangle \xrightarrow{\|T_{Bound}\rangle, I_{Probe}(\lambda,\tau)} \|S0_{Bound}\rangle$ | $k_{GSB_T,Reflect}$ | | | | | | |
| $\|S0_{Bound}\rangle \xrightarrow{I_{Probe}(\lambda,\tau)} \|T_{Bound}\rangle + ABS_{T.Bound.Reflect}$ | $k_{ABS_T,Reflect}$ | $1.9 \cdot 10^{12}$ | $1.9 \cdot 10^{12}$ | $1.9 \cdot 10^{12}$ | $1.9 \cdot 10^{12}$ | $2.4 \cdot 10^{12}$ | $2.0 \cdot 10^{12}$ |
| $\|T_{Bound}\rangle \xrightarrow{I_{Probe}(\lambda,\tau)} \|S0_{Bound}\rangle + ESE_{T.Bound.Reflect}$ | $k_{ESE_T,Reflect}$ | | | | | | |
| $\|Y_{Bound}\rangle \xrightarrow{\|Z0_{Bound}\rangle, I_{Reflect}(\lambda,\tau)} \|Y_{Bound}\rangle$ <br> $\|B_{Bound}\rangle \xrightarrow{\|Z0_{Bound}\rangle, I_{Reflect}(\lambda,\tau)} \|B_{Bound}\rangle$ <br> $\|X_{Bound}\rangle \xrightarrow{\|Z0_{Bound}\rangle, I_{Reflect}(\lambda,\tau)} \|X_{Bound}\rangle$ <br> $\|T_{Bound}\rangle \xrightarrow{\|Z0_{Bound}\rangle, I_{Reflect}(\lambda,\tau)} \|T_{Bound}\rangle$ | $k_{ESA_0,Reflect}$ | $3.8 \cdot 10^{13}$ | $3.6 \cdot 10^{13}$ | $3.6 \cdot 10^{13}$ | $4.3 \cdot 10^{13}$ | $5.1 \cdot 10^{13}$ | $4.6 \cdot 10^{13}$ |
| $\|Y_{Bound}\rangle \xrightarrow{\|Z1_{Bound}\rangle, I_{Reflect}(\lambda,\tau)} \|Y_{Bound}\rangle$ <br> $\|B_{Bound}\rangle \xrightarrow{\|Z1_{Bound}\rangle, I_{Reflect}(\lambda,\tau)} \|B_{Bound}\rangle$ <br> $\|X_{Bound}\rangle \xrightarrow{\|Z1_{Bound}\rangle, I_{Reflect}(\lambda,\tau)} \|X_{Bound}\rangle$ <br> $\|T_{Bound}\rangle \xrightarrow{\|Z1_{Bound}\rangle, I_{Reflect}(\lambda,\tau)} \|T_{Bound}\rangle$ | $k_{ESA_1,Reflect}$ | $3.9 \cdot 10^{12}$ | $3.1 \cdot 10^{12}$ | $3.7 \cdot 10^{12}$ | $4.7 \cdot 10^{12}$ | $3.9 \cdot 10^{12}$ | $4.5 \cdot 10^{12}$ |

| | Rate Coefficient | RuP (s⁻¹) | RuP2 (s⁻¹) | RuP3 (s⁻¹) | RuCP (s⁻¹) | RuCP2 (s⁻¹) | RuCP3 (s⁻¹) |
|---|---|---|---|---|---|---|---|
| $\lvert Y_{Bound}\rangle \xrightarrow{\lvert Z2_{Bound}\rangle, I_{Reflect}(\lambda,\tau)} \lvert Y_{Bound}\rangle$ <br> $\lvert B_{Bound}\rangle \xrightarrow{\lvert Z2_{Bound}\rangle, I_{Reflect}(\lambda,\tau)} \lvert B_{Bound}\rangle$ <br> $\lvert X_{Bound}\rangle \xrightarrow{\lvert Z2_{Bound}\rangle, I_{Reflect}(\lambda,\tau)} \lvert X_{Bound}\rangle$ <br> $\lvert T_{Bound}\rangle \xrightarrow{\lvert Z2_{Bound}\rangle, I_{Reflect}(\lambda,\tau)} \lvert T_{Bound}\rangle$ | $k_{ESA_2,Reflect}$ | $3.0 \cdot 10^{12}$ | $1.9 \cdot 10^{12}$ | $3.9 \cdot 10^{12}$ | $3.4 \cdot 10^{12}$ | $3.7 \cdot 10^{12}$ | $3.3 \cdot 10^{12}$ |
| $\lvert Y_{Bound}\rangle \xrightarrow{\lvert Z3_{Bound}\rangle, I_{Reflect}(\lambda,\tau)} \lvert Y_{Bound}\rangle$ <br> $\lvert B_{Bound}\rangle \xrightarrow{\lvert Z3_{Bound}\rangle, I_{Reflect}(\lambda,\tau)} \lvert B_{Bound}\rangle$ <br> $\lvert X_{Bound}\rangle \xrightarrow{\lvert Z3_{Bound}\rangle, I_{Reflect}(\lambda,\tau)} \lvert X_{Bound}\rangle$ <br> $\lvert T_{Bound}\rangle \xrightarrow{\lvert Z3_{Bound}\rangle, I_{Reflect}(\lambda,\tau)} \lvert T_{Bound}\rangle$ | $k_{ESA_3,Reflect}$ | $1.6 \cdot 10^{12}$ | $1.2 \cdot 10^{12}$ | $1.9 \cdot 10^{12}$ | $2.0 \cdot 10^{12}$ | $3.0 \cdot 10^{12}$ | $2.0 \cdot 10^{12}$ |
| $\lvert Y_{Bound}\rangle \xrightarrow{\lvert Z4_{Bound}\rangle, I_{Reflect}(\lambda,\tau)} \lvert Y_{Bound}\rangle$ <br> $\lvert B_{Bound}\rangle \xrightarrow{\lvert Z4_{Bound}\rangle, I_{Reflect}(\lambda,\tau)} \lvert B_{Bound}\rangle$ <br> $\lvert X_{Bound}\rangle \xrightarrow{\lvert Z4_{Bound}\rangle, I_{Reflect}(\lambda,\tau)} \lvert X_{Bound}\rangle$ <br> $\lvert T_{Bound}\rangle \xrightarrow{\lvert Z4_{Bound}\rangle, I_{Reflect}(\lambda,\tau)} \lvert T_{Bound}\rangle$ | $k_{ESA_4,Reflect}$ | $1.0 \cdot 10^{12}$ | $7.8 \cdot 10^{11}$ | $1.1 \cdot 10^{12}$ | $1.4 \cdot 10^{12}$ | $2.7 \cdot 10^{12}$ | $1.4 \cdot 10^{12}$ |
| $\lvert Y_{Bound}\rangle \xrightarrow{\lvert Z5_{Bound}\rangle, I_{Reflect}(\lambda,\tau)} \lvert Y_{Bound}\rangle$ <br> $\lvert B_{Bound}\rangle \xrightarrow{\lvert Z5_{Bound}\rangle, I_{Reflect}(\lambda,\tau)} \lvert B_{Bound}\rangle$ <br> $\lvert X_{Bound}\rangle \xrightarrow{\lvert Z5_{Bound}\rangle, I_{Reflect}(\lambda,\tau)} \lvert X_{Bound}\rangle$ <br> $\lvert T_{Bound}\rangle \xrightarrow{\lvert Z5_{Bound}\rangle, I_{Reflect}(\lambda,\tau)} \lvert T_{Bound}\rangle$ | $k_{ESA_5,Reflect}$ | $8.5 \cdot 10^{11}$ | $6.2 \cdot 10^{11}$ | $1.6 \cdot 10^{11}$ | $1.1 \cdot 10^{12}$ | $1.9 \cdot 10^{12}$ | $7.8 \cdot 10^{11}$ |

## 7. Oxidized Ruthenium Dye Optical Response

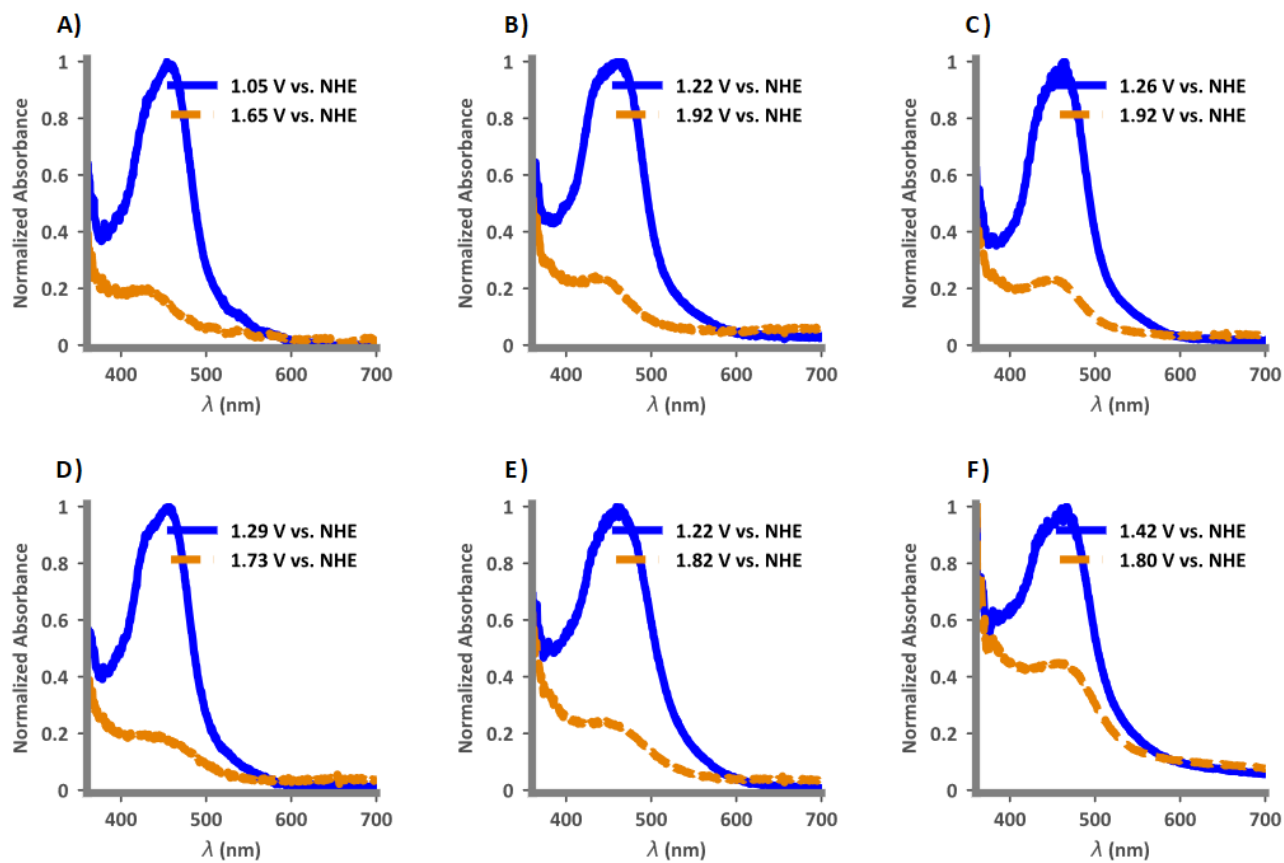

Figure S5. Spectroelectrochemical measurements corresponding to $Ru^{II}$ (blue) and $Ru^{III}$ (orange) oxidation states of the dyes A) RuP, B) RuP2, C) RuP3, D) RuCP, E) RuCP2, and F) RuCP3, normalized to the peak of the $Ru^{II}$ MLCT absorption.

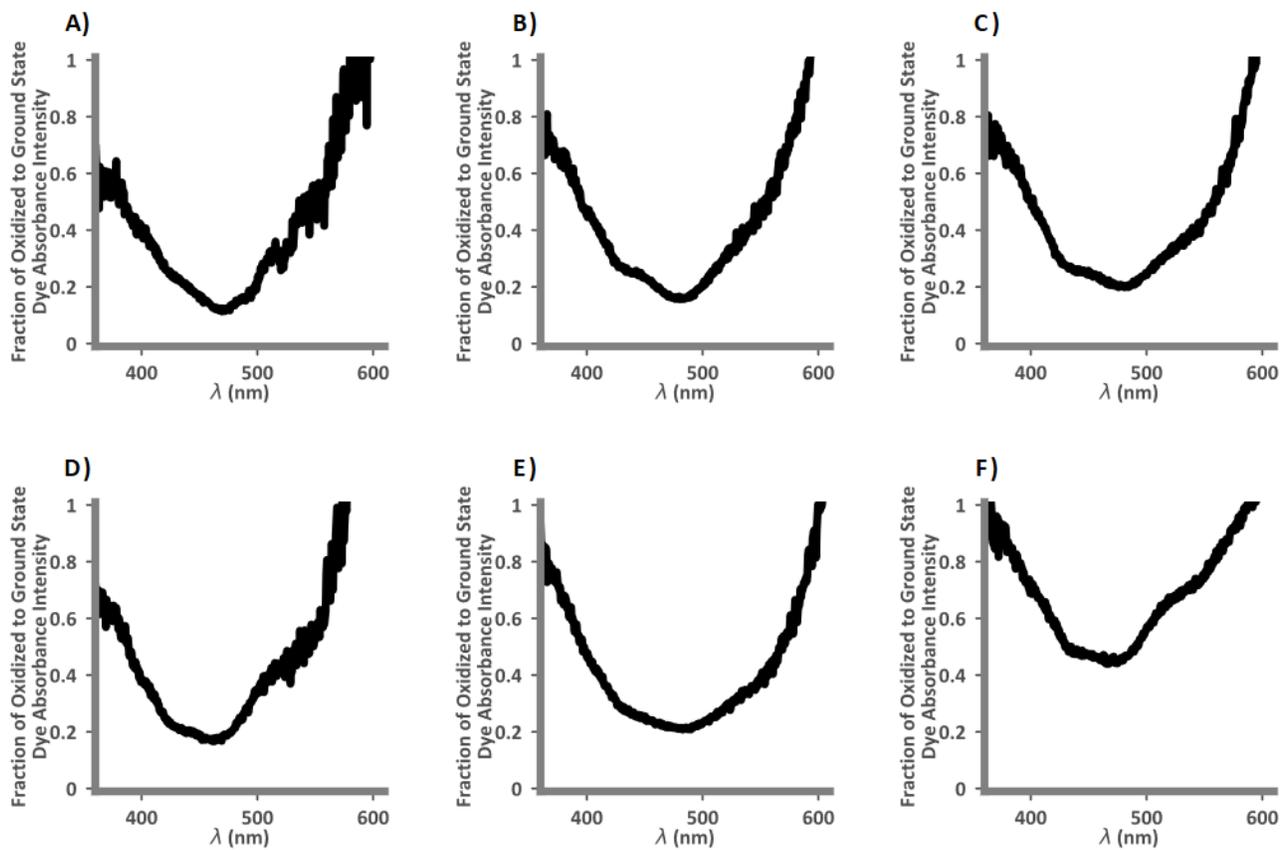

Figure S6. Fraction of Ru$^{III}$ (oxidized) and Ru$^{II}$ (ground state) absorption intensity of the dyes A) RuP, B) RuP2, C) RuP3, D) RuCP, E) RuCP2, and F) RuCP3.

## 8. Transient Absorption of Dyes on ZrO$_2$, Set Ru-Z
### 8.1. RuP, 470 nm and 535 nm Centered Pump Pulse

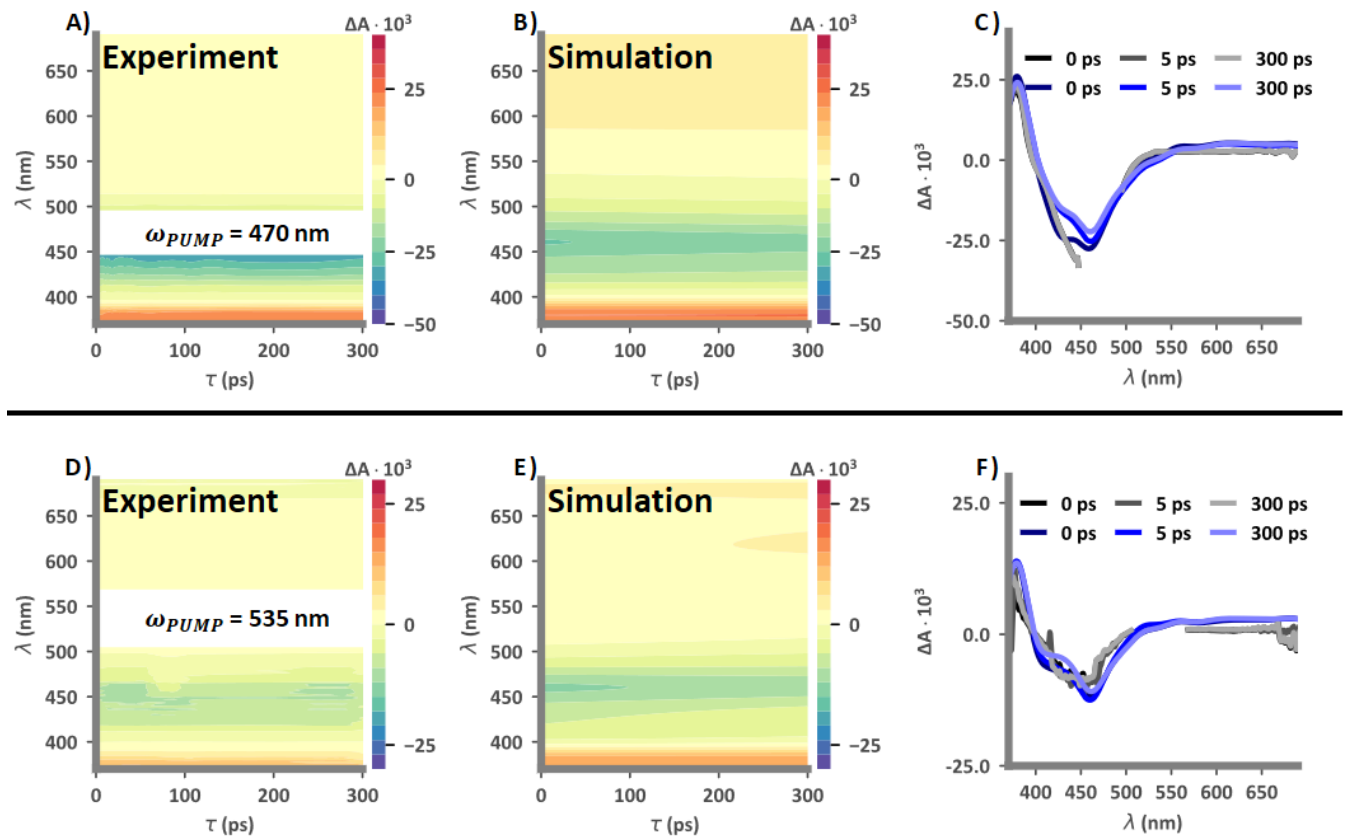

Figure S7. Each row is a different experimental signal, simulated as 4 μm thick films pumped at 470 nm and 535 nm respectively. Experimental (panels A and D) and simulated (panels B and E) TA spectra of dye RuP on ZrO$_2$ from data set **Ru-Z**. C and F: Direct comparison of experimental (grays) and simulated (blues) TA lineshapes at delay times of 0 fs, 5 ps, and 300 ps.

## 8.2. RuP2, 420 nm and 535 nm Centered Pump Pulse

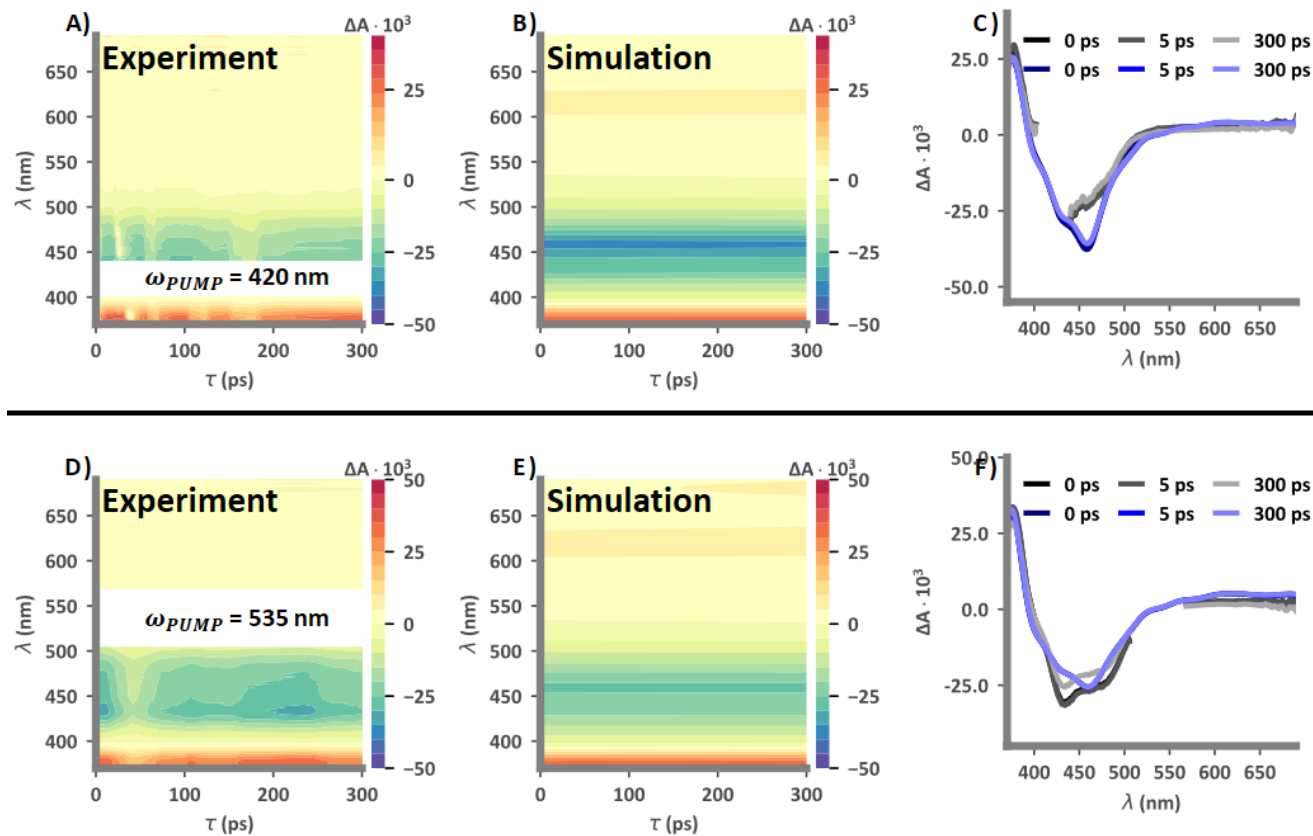

Figure S8. Each row is a different experimental signal, simulated as 4 μm thick films pumped at 420 nm and 535 nm respectively. Experimental (panels A and D) and simulated (panels B and E) TA spectra of dye RuP2 on ZrO$_2$ from data set **Ru-Z**. C and F: Direct comparison of experimental (grays) and simulated (blues) TA lineshapes at delay times of 0 fs, 5 ps, and 300 ps.

## 8.3. RuP3, 420 nm and 535 nm Centered Pump Pulse

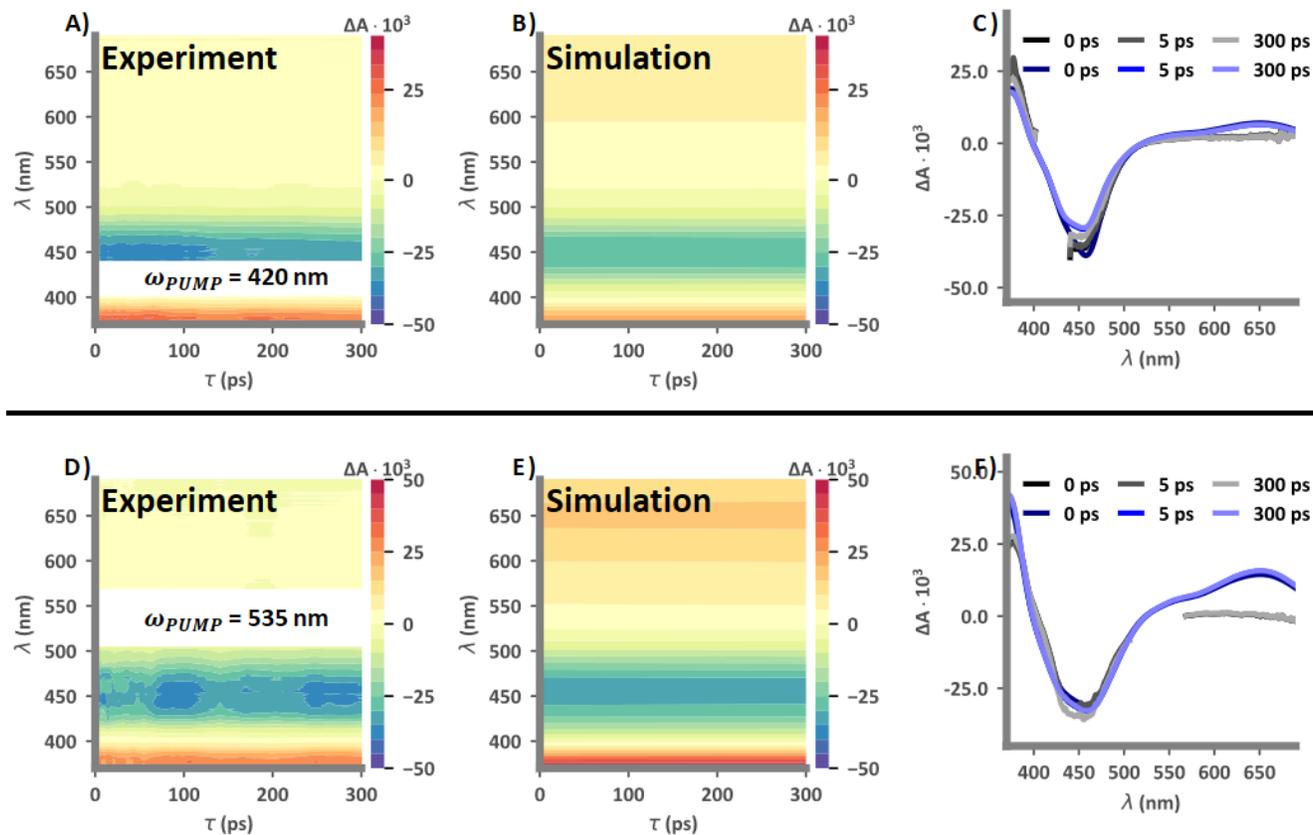

Figure S9. Each row is a different experimental signal, simulated as 4 μm thick films pumped at 420 nm and 535 nm respectively. Experimental (panels A and D) and simulated (panels B and E) TA spectra of dye RuP3 on ZrO$_2$ from data set **Ru-Z**. C and F: Direct comparison of experimental (grays) and simulated (blues) TA lineshapes at delay times of 0 fs, 5 ps, and 300 ps.

## 8.4. RuCP, 420 nm and 535 nm Centered Pump Pulse

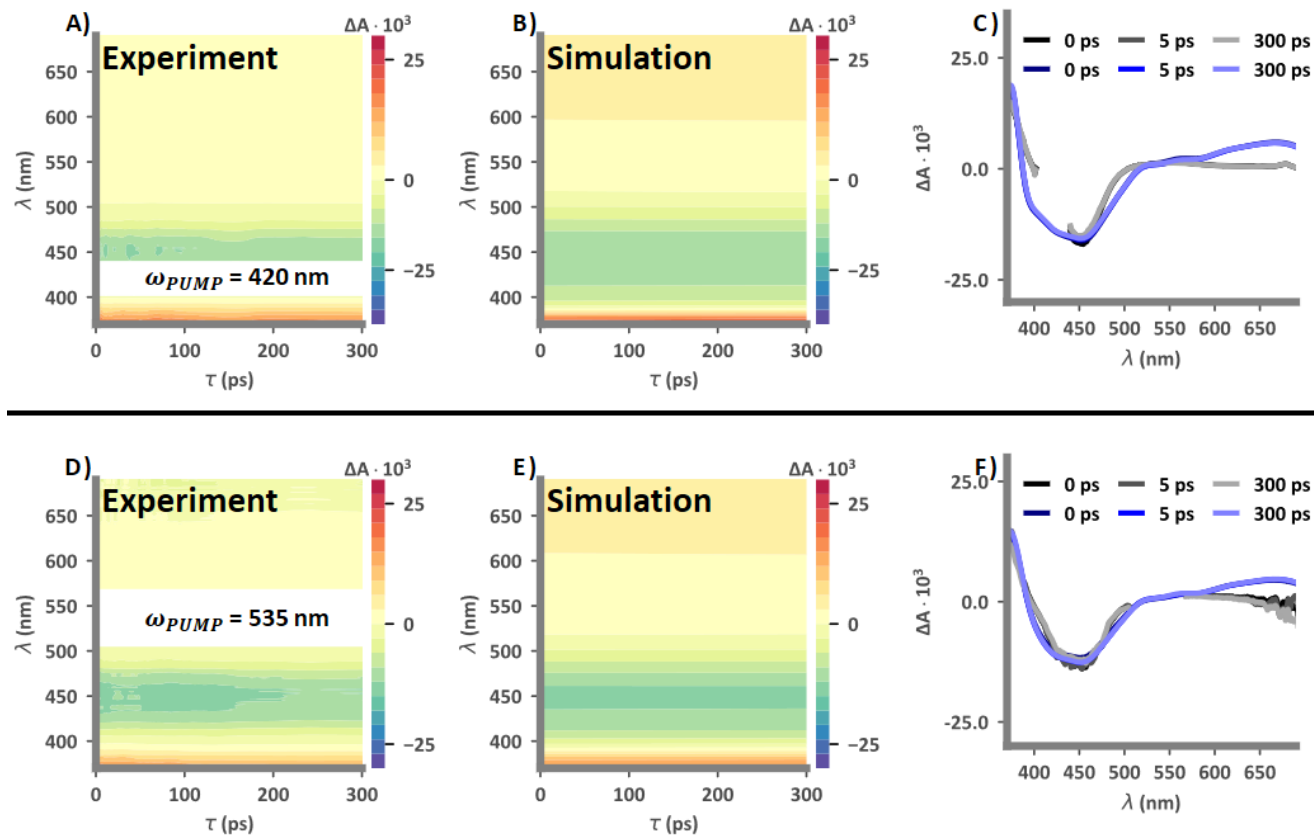

Figure S10. Each row is a different experimental signal, simulated as 4 μm thick films pumped at 420 nm and 535 nm respectively. Experimental (panels A and D) and simulated (panels B and E) TA spectra of dye RuCP on ZrO$_2$ from data set **Ru-Z**. C and F: Direct comparison of experimental (grays) and simulated (blues) TA lineshapes at delay times of 0 fs, 5 ps, and 300 ps.

## 8.5. RuCP2, 420 nm and 535 nm Centered Pump Pulse

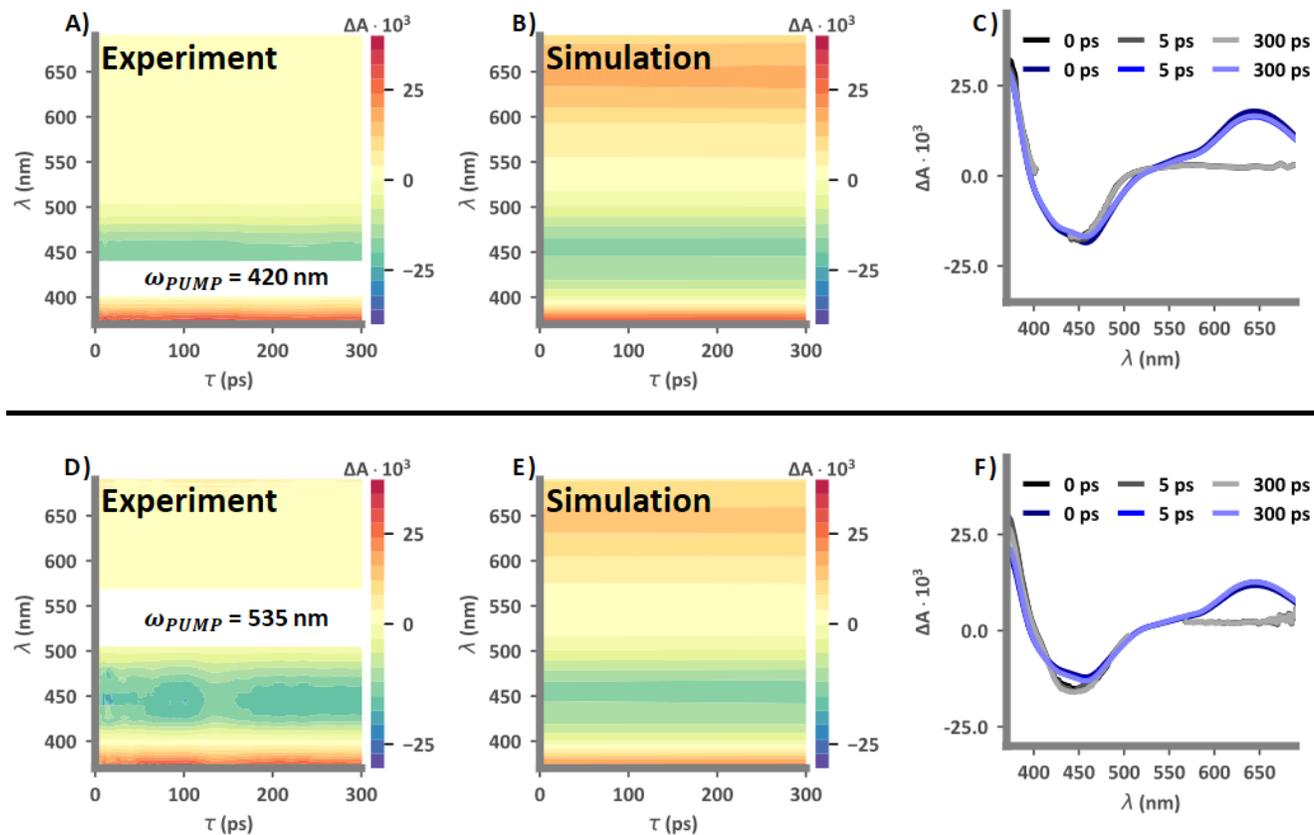

Figure S11. Each row is a different experimental signal, simulated as 4 μm thick films pumped at 420 nm and 535 nm respectively. Experimental (panels A and D) and simulated (panels B and E) TA spectra of dye RuCP2 on ZrO$_2$ from data set **Ru-Z**. C and F: Direct comparison of experimental (grays) and simulated (blues) TA lineshapes at delay times of 0 fs, 5 ps, and 300 ps.

## 8.6. RuCP3, 420 nm and 535 nm Centered Pump Pulse

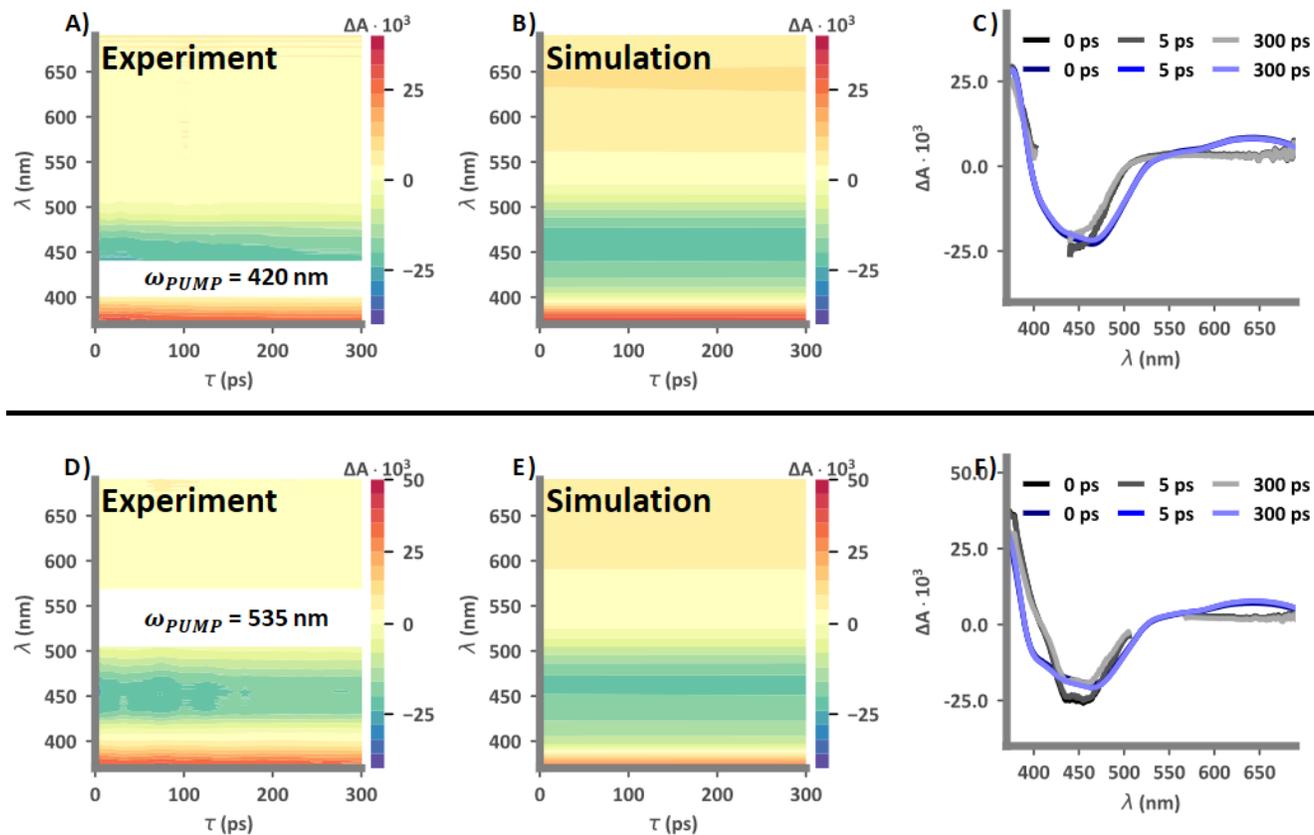

Figure S12. Each row is a different experimental signal, simulated as 4 μm thick films pumped at 420 nm and 535 nm respectively. Experimental (panels A and D) and simulated (panels B and E) TA spectra of dye RuCP3 on $ZrO_2$ from data set **Ru-Z**. C and F: Direct comparison of experimental (grays) and simulated (blues) TA lineshapes at delay times of 0 fs, 5 ps, and 300 ps.

## 9. Dye on TiO$_2$ model for set 6-Ru

**Table S4.** Base mechanistic scheme and rate coefficients for dyes in set 6-Ru on 7 μm film of TiO$_2$ for dataset Ru-G with broadband probe spanning 500 nm to 700 nm. Simulated surface concentration and unbound dye concentration were 4·10$^{-8}$ mol cm$^{-2}$ and 4·10$^{-9}$ mol cm$^{-2}$ respectively.

| | Rate Coefficients | RuP (s$^{-1}$) | RuP 2 (s$^{-1}$) | RuP3 (s$^{-1}$) | RuCP (s$^{-1}$) | RuCP2 (s$^{-1}$) | RuCP3 (s$^{-1}$) |
|---|---|---|---|---|---|---|---|
| $\lvert S0_{Bound}\rangle \xrightarrow{h\nu_{Pump}} \lvert Y_{Bound}\rangle$ | $k_{Pump,Y}$ | Variable | Variable | Variable | Variable | Variable | Variable |
| $\lvert S0_{Bound}\rangle \xrightarrow{h\nu_{Pump}} \lvert B_{Bound}\rangle$ | $k_{Pump,B}$ | Variable | Variable | Variable | Variable | Variable | Variable |
| $\lvert S0_{Bound}\rangle \xrightarrow{h\nu_{Pump}} \lvert X_{Bound}\rangle$ | $k_{Pump,X}$ | Variable | Variable | Variable | Variable | Variable | Variable |
| $\lvert S0_{Bound}\rangle \xrightarrow{h\nu_{Pump}} \lvert T_{Bound}\rangle$ | $k_{Pump,T}$ | Variable | Variable | Variable | Variable | Variable | Variable |
| $\lvert S0_{Bound}\rangle \xrightarrow{\lvert Y_{Bound}\rangle, I_{Probe}(\lambda,\tau)} \lvert S0_{Bound}\rangle$ | $k_{GSB_Y}$ | 1.8·10$^{13}$ | 2.0·10$^{13}$ | 2.0·10$^{13}$ | 1.9·10$^{13}$ | 1.8·10$^{13}$ | 3.4·10$^{13}$ |
| $\lvert S0_{Bound}\rangle \xrightarrow{I_{Probe}(\lambda,\tau)} \lvert Y_{Bound}\rangle + ABS_{Y.Bound}$ | $k_{ABS_Y}$ | | | | | | |
| $\lvert Y_{Bound}\rangle \xrightarrow{I_{Probe}(\lambda,\tau)} \lvert S0_{Bound}\rangle + ESE_{Y.Bound}$ | $k_{ESE_Y}$ | | | | | | |
| $\lvert S0_{Bound}\rangle \xrightarrow{\lvert B_{Bound}\rangle, I_{Probe}(\lambda,\tau)} \lvert S0_{Bound}\rangle$ | $k_{GSB_B}$ | 3.8·10$^{13}$ | 4.1·10$^{13}$ | 3.3·10$^{13}$ | 3.2·10$^{13}$ | 3.2·10$^{13}$ | 3.7·10$^{13}$ |
| $\lvert S0_{Bound}\rangle \xrightarrow{I_{Probe}(\lambda,\tau)} \lvert B_{Bound}\rangle + ABS_{B.Bound}$ | $k_{ABS_B}$ | | | | | | |
| $\lvert B_{Bound}\rangle \xrightarrow{I_{Probe}(\lambda,\tau)} \lvert S0_{Bound}\rangle + ESE_{B.Bound}$ | $k_{ESE_B}$ | | | | | | |
| $\lvert S0_{Bound}\rangle \xrightarrow{\lvert X_{Bound}\rangle, I_{Probe}(\lambda,\tau)} \lvert S0_{Bound}\rangle$ | $k_{GSB_X}$ | 4.8·10$^{13}$ | 5.0·10$^{13}$ | 4.0·10$^{13}$ | 4.1·10$^{13}$ | 4.1·10$^{13}$ | 4.5·10$^{13}$ |
| $\lvert S0_{Bound}\rangle \xrightarrow{I_{Probe}(\lambda,\tau)} \lvert X_{Bound}\rangle + ABS_{X.Bound}$ | $k_{ABS_X}$ | | | | | | |
| $\lvert X_{Bound}\rangle \xrightarrow{I_{Probe}(\lambda,\tau)} \lvert S0_{Bound}\rangle + ESE_{X.Bound}$ | $k_{ESE_X}$ | | | | | | |
| $\lvert S0_{Bound}\rangle \xrightarrow{\lvert T_{Bound}\rangle, I_{Probe}(\lambda,\tau)} \lvert S0_{Bound}\rangle$ | $k_{GSB_T}$ | 1.8·10$^{13}$ | 1.8·10$^{13}$ | 1.8·10$^{13}$ | 1.8·10$^{13}$ | 1.8·10$^{13}$ | 1.8·10$^{13}$ |
| $\lvert S0_{Bound}\rangle \xrightarrow{I_{Probe}(\lambda,\tau)} \lvert T_{Bound}\rangle + ABS_{T.Bound}$ | $k_{ABS_T}$ | | | | | | |
| $\lvert T_{Bound}\rangle \xrightarrow{I_{Probe}(\lambda,\tau)} \lvert S0_{Bound}\rangle + ESE_{T.Bound}$ | $k_{ESE_T}$ | | | | | | |
| $\lvert Y_{Bound}\rangle \to \lvert B_{Bound}\rangle$ | $k_{YB}$ | 1.8·10$^{13}$ | 2.4·10$^{13}$ | 8.0·10$^{13}$ | 6.0·10$^{13}$ | 6.0·10$^{13}$ | 6.0·10$^{13}$ |
| $\lvert B_{Bound}\rangle \to \lvert X_{Bound}\rangle$ | $k_{BX}$ | 1.8·10$^{13}$ | 2.4·10$^{13}$ | 8.0·10$^{13}$ | 6.0·10$^{13}$ | 6.0·10$^{13}$ | 6.0·10$^{13}$ |
| $\lvert X_{Bound}\rangle \to \lvert S0_{Bound}\rangle$ | $k_{UF-nr}$ | N.A | NA | 4.0·10$^{13}$ | 1.6·10$^{13}$ | 1.6·10$^{13}$ | 2.0·10$^{13}$ |
| $\lvert X_{Bound}\rangle \to \lvert T_{Bound}\rangle$ | $k_{ISC}$ | 4.0·10$^{13}$ | 4.0·10$^{13}$ | 2.0·10$^{13}$ | 2.0·10$^{13}$ | 2.0·10$^{13}$ | 3.6·10$^{13}$ |
| $\lvert Y_{Bound}\rangle \xrightarrow{\lvert Z0_{Bound}\rangle, I_{Probe}(\lambda,\tau)} \lvert Y_{Bound}\rangle$ | $k_{ESA_0}$ | 1.6·10$^{14}$ | 1.5·10$^{14}$ | 1.5·10$^{14}$ | 1.8·10$^{14}$ | 2.1·10$^{14}$ | 1.9·10$^{14}$ |
| $\lvert B_{Bound}\rangle \xrightarrow{\lvert Z0_{Bound}\rangle, I_{Probe}(\lambda,\tau)} \lvert B_{Bound}\rangle$ | | | | | | | |
| $\lvert X_{Bound}\rangle \xrightarrow{\lvert Z0_{Bound}\rangle, I_{Probe}(\lambda,\tau)} \lvert X_{Bound}\rangle$ | | | | | | | |
| $\lvert T_{Bound}\rangle \xrightarrow{\lvert Z0_{Bound}\rangle, I_{Probe}(\lambda,\tau)} \lvert T_{Bound}\rangle$ | | | | | | | |
| $\lvert Y_{Bound}\rangle \xrightarrow{\lvert Z1_{Bound}\rangle, I_{Probe}(\lambda,\tau)} \lvert Y_{Bound}\rangle$ | $k_{ESA_1}$ | 4.0·10$^{13}$ | 3.2·10$^{13}$ | 3.8·10$^{13}$ | 4.8·10$^{13}$ | 4.0·10$^{13}$ | 4.9·10$^{13}$ |
| $\lvert B_{Bound}\rangle \xrightarrow{\lvert Z1_{Bound}\rangle, I_{Probe}(\lambda,\tau)} \lvert B_{Bound}\rangle$ | | | | | | | |
| $\lvert X_{Bound}\rangle \xrightarrow{\lvert Z1_{Bound}\rangle, I_{Probe}(\lambda,\tau)} \lvert X_{Bound}\rangle$ | | | | | | | |
| $\lvert T_{Bound}\rangle \xrightarrow{\lvert Z1_{Bound}\rangle, I_{Probe}(\lambda,\tau)} \lvert T_{Bound}\rangle$ | | | | | | | |
| $\lvert Y_{Bound}\rangle \xrightarrow{\lvert Z2_{Bound}\rangle, I_{Probe}(\lambda,\tau)} \lvert Y_{Bound}\rangle$ | $k_{ESA_2}$ | 4.3·10$^{13}$ | 2.8·10$^{13}$ | 4.7·10$^{13}$ | 5.2·10$^{13}$ | 5.7·10$^{13}$ | 5.3·10$^{13}$ |
| $\lvert B_{Bound}\rangle \xrightarrow{\lvert Z2_{Bound}\rangle, I_{Probe}(\lambda,\tau)} \lvert B_{Bound}\rangle$ | | | | | | | |
| $\lvert X_{Bound}\rangle \xrightarrow{\lvert Z2_{Bound}\rangle, I_{Probe}(\lambda,\tau)} \lvert X_{Bound}\rangle$ | | | | | | | |
| $\lvert T_{Bound}\rangle \xrightarrow{\lvert Z2_{Bound}\rangle, I_{Probe}(\lambda,\tau)} \lvert T_{Bound}\rangle$ | | | | | | | |
| $\lvert Y_{Bound}\rangle \xrightarrow{\lvert Z3_{Bound}\rangle, I_{Probe}(\lambda,\tau)} \lvert Y_{Bound}\rangle$ | $k_{ESA_3}$ | 4.2·10$^{13}$ | 3.0·10$^{13}$ | 5.3·10$^{13}$ | 5.6·10$^{13}$ | 9.3·10$^{13}$ | 5.7·10$^{13}$ |
| $\lvert B_{Bound}\rangle \xrightarrow{\lvert Z3_{Bound}\rangle, I_{Probe}(\lambda,\tau)} \lvert B_{Bound}\rangle$ | | | | | | | |
| $\lvert X_{Bound}\rangle \xrightarrow{\lvert Z3_{Bound}\rangle, I_{Probe}(\lambda,\tau)} \lvert X_{Bound}\rangle$ | | | | | | | |
| $\lvert T_{Bound}\rangle \xrightarrow{\lvert Z3_{Bound}\rangle, I_{Probe}(\lambda,\tau)} \lvert T_{Bound}\rangle$ | | | | | | | |

| | Rate Coefficients | RuP (s⁻¹) | RuP 2 (s⁻¹) | RuP3 (s⁻¹) | RuCP (s⁻¹) | RuCP2 (s⁻¹) | RuCP3 (s⁻¹) |
|---|---|---|---|---|---|---|---|
| $\|Y_{Bound}\rangle \xrightarrow{\|Z4_{Bound}\rangle, I_{Probe}(\lambda,\tau)} \|Y_{Bound}\rangle$ | $k_{ESA_4}$ | $3.9 \cdot 10^{13}$ | $3.0 \cdot 10^{13}$ | $4.6 \cdot 10^{13}$ | $5.7 \cdot 10^{13}$ | $11 \cdot 10^{13}$ | $5.7 \cdot 10^{13}$ |
| $\|B_{Bound}\rangle \xrightarrow{\|Z4_{Bound}\rangle, I_{Probe}(\lambda,\tau)} \|B_{Bound}\rangle$ | | | | | | | |
| $\|X_{Bound}\rangle \xrightarrow{\|Z4_{Bound}\rangle, I_{Probe}(\lambda,\tau)} \|X_{Bound}\rangle$ | | | | | | | |
| $\|T_{Bound}\rangle \xrightarrow{\|Z4_{Bound}\rangle, I_{Probe}(\lambda,\tau)} \|T_{Bound}\rangle$ | | | | | | | |
| $\|Y_{Bound}\rangle \xrightarrow{\|Z5_{Bound}\rangle, I_{Probe}(\lambda,\tau)} \|Y_{Bound}\rangle$ | $k_{ESA_5}$ | $5.4 \cdot 10^{13}$ | $3.8 \cdot 10^{13}$ | $5.0 \cdot 10^{13}$ | $7.0 \cdot 10^{13}$ | $10.5 \cdot 10^{13}$ | $5.0 \cdot 10^{13}$ |
| $\|B_{Bound}\rangle \xrightarrow{\|Z5_{Bound}\rangle, I_{Probe}(\lambda,\tau)} \|B_{Bound}\rangle$ | | | | | | | |
| $\|X_{Bound}\rangle \xrightarrow{\|Z5_{Bound}\rangle, I_{Probe}(\lambda,\tau)} \|X_{Bound}\rangle$ | | | | | | | |
| $\|T_{Bound}\rangle \xrightarrow{\|Z5_{Bound}\rangle, I_{Probe}(\lambda,\tau)} \|T_{Bound}\rangle$ | | | | | | | |
| $\|T_{Bound}\rangle \rightarrow \|S0_{Bound}\rangle + rad_{Bound}$ | $k_{rad}$ | $10.9 \cdot 10^4$ | $9.6 \cdot 10^4$ | $11 \cdot 10^4$ | $10.6 \cdot 10^4$ | $10.6 \cdot 10^4$ | $10.3 \cdot 10^4$ |
| $\|T_{Bound}\rangle \rightarrow \|S0_{Bound}\rangle$ | $k_{nr}$ | $3.2 \cdot 10^6$ | $2.6 \cdot 10^6$ | $2.0 \cdot 10^6$ | $2.0 \cdot 10^6$ | $2.2 \cdot 10^6$ | $2.3 \cdot 10^6$ |
| $\|S0_{Unbound}\rangle \xrightarrow{h\nu_{Pump}} \|Y_{Unbound}\rangle$ | $k_{Pump,Y}$ | Variable | Variable | Variable | Variable | Variable | Variable |
| $\|S0_{Unbound}\rangle \xrightarrow{h\nu_{Pump}} \|B_{Unbound}\rangle$ | $k_{Pump,B}$ | Variable | Variable | Variable | Variable | Variable | Variable |
| $\|S0_{Unbound}\rangle \xrightarrow{h\nu_{Pump}} \|X_{Unbound}\rangle$ | $k_{Pump,X}$ | Variable | Variable | Variable | Variable | Variable | Variable |
| $\|S0_{Unbound}\rangle \xrightarrow{h\nu_{Pump}} \|T_{Unbound}\rangle$ | $k_{Pump,T}$ | Variable | Variable | Variable | Variable | Variable | Variable |
| $\|S0_{Unbound}\rangle \xrightarrow{\|Y_{Unbound}\rangle, I_{Probe}(\lambda,\tau)} \|S0_{Unbound}\rangle$ | $k_{GSB_Y}$ | $1.8 \cdot 10^{13}$ | $2.0 \cdot 10^{13}$ | $2.0 \cdot 10^{13}$ | $1.9 \cdot 10^{13}$ | $1.8 \cdot 10^{13}$ | $3.4 \cdot 10^{13}$ |
| $\|S0_{Unbound}\rangle \xrightarrow{I_{Probe}(\lambda,\tau)} \|Y_{Unbound}\rangle$ | $k_{ABS_Y}$ | | | | | | |
| $\|Y_{Unbound}\rangle \xrightarrow{I_{Probe}(\lambda,\tau)} \|S0_{Unbound}\rangle$ | $k_{ESE_Y}$ | | | | | | |
| $\|S0_{Unbound}\rangle \xrightarrow{\|B_{Unbound}\rangle, I_{Probe}(\lambda,\tau)} \|S0_{Unbound}\rangle$ | $k_{GSB_B}$ | $3.8 \cdot 10^{13}$ | $4.1 \cdot 10^{13}$ | $3.3 \cdot 10^{13}$ | $3.2 \cdot 10^{13}$ | $3.2 \cdot 10^{13}$ | $3.7 \cdot 10^{13}$ |
| $\|S0_{Unbound}\rangle \xrightarrow{I_{Probe}(\lambda,\tau)} \|B_{Unbound}\rangle$ | $k_{ABS_B}$ | | | | | | |
| $\|B_{Unbound}\rangle \xrightarrow{I_{Probe}(\lambda,\tau)} \|S0_{Unbound}\rangle$ | $k_{ESE_B}$ | | | | | | |
| $\|S0_{Unbound}\rangle \xrightarrow{\|X_{Unbound}\rangle, I_{Probe}(\lambda,\tau)} \|S0_{Unbound}\rangle$ | $k_{GSB_X}$ | $4.8 \cdot 10^{13}$ | $5.0 \cdot 10^{13}$ | $4.0 \cdot 10^{13}$ | $4.1 \cdot 10^{13}$ | $4.1 \cdot 10^{13}$ | $4.5 \cdot 10^{13}$ |
| $\|S0_{Unbound}\rangle \xrightarrow{I_{Probe}(\lambda,\tau)} \|X_{Unbound}\rangle$ | $k_{ABS_X}$ | | | | | | |
| $\|X_{Unbound}\rangle \xrightarrow{I_{Probe}(\lambda,\tau)} \|S0_{Unbound}\rangle$ | $k_{ESE_X}$ | | | | | | |
| $\|S0_{Unbound}\rangle \xrightarrow{\|T_{Unbound}\rangle, I_{Probe}(\lambda,\tau)} \|S0_{Unbound}\rangle$ | $k_{GSB_T}$ | $1.8 \cdot 10^{13}$ | $1.8 \cdot 10^{13}$ | $1.8 \cdot 10^{13}$ | $1.8 \cdot 10^{13}$ | $1.8 \cdot 10^{13}$ | $1.8 \cdot 10^{13}$ |
| $\|S0_{Unbound}\rangle \xrightarrow{I_{Probe}(\lambda,\tau)} \|T_{Unbound}\rangle$ | $k_{ABS_T}$ | | | | | | |
| $\|T_{Unbound}\rangle \xrightarrow{I_{Probe}(\lambda,\tau)} \|S0_{Unbound}\rangle$ | $k_{ESE_T}$ | | | | | | |
| $\|Y_{Unbound}\rangle \rightarrow \|B_{Unbound}\rangle$ | $k_{YB}$ | $1.8 \cdot 10^{13}$ | $2.4 \cdot 10^{13}$ | $8.0 \cdot 10^{13}$ | $6.0 \cdot 10^{13}$ | $6.0 \cdot 10^{13}$ | $6.0 \cdot 10^{13}$ |
| $\|B_{Unbound}\rangle \rightarrow \|X_{Unbound}\rangle$ | $k_{BX}$ | $1.8 \cdot 10^{13}$ | $2.4 \cdot 10^{13}$ | $8.0 \cdot 10^{13}$ | $6.0 \cdot 10^{13}$ | $6.0 \cdot 10^{13}$ | $6.0 \cdot 10^{13}$ |
| $\|X_{Unbound}\rangle \rightarrow \|S0_{Unbound}\rangle$ | $k_{UF-nr}$ | $2.4 \cdot 10^{13}$ | $1.0 \cdot 10^{13}$ | $4.0 \cdot 10^{13}$ | $1.6 \cdot 10^{13}$ | $1.6 \cdot 10^{13}$ | $2.0 \cdot 10^{13}$ |
| $\|X_{Unbound}\rangle \rightarrow \|T_{Unbound}\rangle$ | $k_{ISC}$ | $4.0 \cdot 10^{13}$ | $4.0 \cdot 10^{13}$ | $2.0 \cdot 10^{13}$ | $2.0 \cdot 10^{13}$ | $2.0 \cdot 10^{13}$ | $3.6 \cdot 10^{13}$ |
| $\|Y_{Unbound}\rangle \xrightarrow{\|Z0_{Unbound}\rangle, I_{Probe}(\lambda,\tau)} \|Y_{Unbound}\rangle$ | $k_{ESA_0}$ | $1.6 \cdot 10^{14}$ | $1.5 \cdot 10^{14}$ | $1.5 \cdot 10^{14}$ | $1.8 \cdot 10^{14}$ | $2.1 \cdot 10^{14}$ | $1.9 \cdot 10^{14}$ |
| $\|B_{Unbound}\rangle \xrightarrow{\|Z0_{Unbound}\rangle, I_{Probe}(\lambda,\tau)} \|B_{Unbound}\rangle$ | | | | | | | |
| $\|X_{Unbound}\rangle \xrightarrow{\|Z0_{Unbound}\rangle, I_{Probe}(\lambda,\tau)} \|X_{Unbound}\rangle$ | | | | | | | |
| $\|T_{Unbound}\rangle \xrightarrow{\|Z0_{Unbound}\rangle, I_{Probe}(\lambda,\tau)} \|T_{Unbound}\rangle$ | | | | | | | |
| $\|Y_{Unbound}\rangle \xrightarrow{\|Z1_{Unbound}\rangle, I_{Probe}(\lambda,\tau)} \|Y_{Unbound}\rangle$ | $k_{ESA_1}$ | $4.0 \cdot 10^{13}$ | $3.2 \cdot 10^{13}$ | $3.8 \cdot 10^{13}$ | $4.8 \cdot 10^{13}$ | $4.0 \cdot 10^{13}$ | $4.9 \cdot 10^{13}$ |
| $\|B_{Unbound}\rangle \xrightarrow{\|Z1_{Unbound}\rangle, I_{Probe}(\lambda,\tau)} \|B_{Unbound}\rangle$ | | | | | | | |
| $\|X_{Unbound}\rangle \xrightarrow{\|Z1_{Unbound}\rangle, I_{Probe}(\lambda,\tau)} \|X_{Unbound}\rangle$ | | | | | | | |
| $\|T_{Unbound}\rangle \xrightarrow{\|Z1_{Unbound}\rangle, I_{Probe}(\lambda,\tau)} \|T_{Unbound}\rangle$ | | | | | | | |

| Process | Rate Coefficients | RuP (s⁻¹) | RuP 2 (s⁻¹) | RuP3 (s⁻¹) | RuCP (s⁻¹) | RuCP2 (s⁻¹) | RuCP3 (s⁻¹) |
|---|---|---|---|---|---|---|---|
| $\|Y_{Unbound}\rangle \xrightarrow{\|Z2_{Unbound}\rangle, I_{Probe}(\lambda,\tau)} \|Y_{Unbound}\rangle$ | $k_{ESA_2}$ | $4.3 \cdot 10^{13}$ | $2.8 \cdot 10^{13}$ | $4.7 \cdot 10^{13}$ | $5.2 \cdot 10^{13}$ | $5.7 \cdot 10^{13}$ | $5.3 \cdot 10^{13}$ |
| $\|B_{Unbound}\rangle \xrightarrow{\|Z2_{Unbound}\rangle, I_{Probe}(\lambda,\tau)} \|B_{Unbound}\rangle$ | | | | | | | |
| $\|X_{Unbound}\rangle \xrightarrow{\|Z2_{Unbound}\rangle, I_{Probe}(\lambda,\tau)} \|X_{Unbound}\rangle$ | | | | | | | |
| $\|T_{Unbound}\rangle \xrightarrow{\|Z2_{Unbound}\rangle, I_{Probe}(\lambda,\tau)} \|T_{Unbound}\rangle$ | | | | | | | |
| $\|Y_{Unbound}\rangle \xrightarrow{\|Z3_{Unbound}\rangle, I_{Probe}(\lambda,\tau)} \|Y_{Unbound}\rangle$ | $k_{ESA_3}$ | $4.2 \cdot 10^{13}$ | $3.0 \cdot 10^{13}$ | $5.3 \cdot 10^{13}$ | $5.6 \cdot 10^{13}$ | $9.3 \cdot 10^{13}$ | $5.7 \cdot 10^{13}$ |
| $\|B_{Unbound}\rangle \xrightarrow{\|Z3_{Unbound}\rangle, I_{Probe}(\lambda,\tau)} \|B_{Unbound}\rangle$ | | | | | | | |
| $\|X_{Unbound}\rangle \xrightarrow{\|Z3_{Unbound}\rangle, I_{Probe}(\lambda,\tau)} \|X_{Unbound}\rangle$ | | | | | | | |
| $\|T_{Unbound}\rangle \xrightarrow{\|Z3_{Unbound}\rangle, I_{Probe}(\lambda,\tau)} \|T_{Unbound}\rangle$ | | | | | | | |
| $\|Y_{Unbound}\rangle \xrightarrow{\|Z4_{Unbound}\rangle, I_{Probe}(\lambda,\tau)} \|Y_{Unbound}\rangle$ | $k_{ESA_4}$ | $3.9 \cdot 10^{13}$ | $3.0 \cdot 10^{13}$ | $4.6 \cdot 10^{13}$ | $5.7 \cdot 10^{13}$ | $11 \cdot 10^{13}$ | $5.7 \cdot 10^{13}$ |
| $\|B_{Unbound}\rangle \xrightarrow{\|Z4_{Unbound}\rangle, I_{Probe}(\lambda,\tau)} \|B_{Unbound}\rangle$ | | | | | | | |
| $\|X_{Unbound}\rangle \xrightarrow{\|Z4_{Unbound}\rangle, I_{Probe}(\lambda,\tau)} \|X_{Unbound}\rangle$ | | | | | | | |
| $\|T_{Unbound}\rangle \xrightarrow{\|Z4_{Unbound}\rangle, I_{Probe}(\lambda,\tau)} \|T_{Unbound}\rangle$ | | | | | | | |
| $\|Y_{Unbound}\rangle \xrightarrow{\|Z5_{Unbound}\rangle, I_{Probe}(\lambda,\tau)} \|Y_{Unbound}\rangle$ | $k_{ESA_5}$ | $5.4 \cdot 10^{13}$ | $3.8 \cdot 10^{13}$ | $5.0 \cdot 10^{13}$ | $7.0 \cdot 10^{13}$ | $10.5 \cdot 10^{13}$ | $5.0 \cdot 10^{13}$ |
| $\|B_{Unbound}\rangle \xrightarrow{\|Z5_{Unbound}\rangle, I_{Probe}(\lambda,\tau)} \|B_{Unbound}\rangle$ | | | | | | | |
| $\|X_{Unbound}\rangle \xrightarrow{\|Z5_{Unbound}\rangle, I_{Probe}(\lambda,\tau)} \|X_{Unbound}\rangle$ | | | | | | | |
| $\|T_{Unbound}\rangle \xrightarrow{\|Z5_{Unbound}\rangle, I_{Probe}(\lambda,\tau)} \|T_{Unbound}\rangle$ | | | | | | | |
| $\|T_{Unbound}\rangle \to \|S0_{Unbound}\rangle + rad_{Unbound}$ | $k_{rad}$ | $10.9 \cdot 10^4$ | $9.6 \cdot 10^4$ | $11 \cdot 10^4$ | $10.6 \cdot 10^4$ | $10.6 \cdot 10^4$ | $10.3 \cdot 10^4$ |
| $\|T_{Unbound}\rangle \to \|S0_{Unbound}\rangle$ | $k_{nr}$ | $3.2 \cdot 10^6$ | $2.6 \cdot 10^6$ | $2.0 \cdot 10^6$ | $2.0 \cdot 10^6$ | $2.2 \cdot 10^6$ | $2.3 \cdot 10^6$ |
| $\|Y_{Bound}\rangle + \|TiO_2\rangle \to \|Ru^3\rangle + \|TiO_2^*\rangle$ | $k_{Injection}$ | $2.4 \cdot 10^{12}$ | $1.0 \cdot 10^{12}$ | $4.0 \cdot 10^{11}$ | $1.6 \cdot 10^{11}$ | $1.6 \cdot 10^{11}$ | $1.0 \cdot 10^{11}$ |
| $\|B_{Bound}\rangle + \|TiO_2\rangle \to \|Ru^3\rangle + \|TiO_2^*\rangle$ | | | | | | | |
| $\|X_{Bound}\rangle + \|TiO_2\rangle \to \|Ru^3\rangle + \|TiO_2^*\rangle$ | | | | | | | |
| $\|T_{Bound}\rangle + \|TiO_2\rangle \to \|Ru^3\rangle + \|TiO_2^*\rangle$ | | | | | | | |
| $\|Ru^3\rangle \xrightarrow{\|Ru^{3*}\rangle, I\ (\lambda,\tau)} \|Ru^3\rangle + ESA_{\|Ru^3\rangle}$ | $k_{ESA_{\|Ru^3\rangle}}$ | $6.3 \cdot 10^{14}$ | $1.1 \cdot 10^{15}$ | $8.2 \cdot 10^{14}$ | $7.3 \cdot 10^{14}$ | $9.7 \cdot 10^{14}$ | $2.4 \cdot 10^{15}$ |
| $\|Ru^3\rangle \xrightarrow{\|Ru^{3*}\rangle, I_{Reflect}(\lambda,\tau)} \|Ru^3\rangle$ | $k_{ESA_{\|Ru^3\rangle, Reflect}}$ | $1.0 \cdot 10^{14}$ | $1.5 \cdot 10^{14}$ | $1.3 \cdot 10^{14}$ | $1.1 \cdot 10^{14}$ | $1.5 \cdot 10^{14}$ | $3.5 \cdot 10^{14}$ |
| $\|S0_{Bound}\rangle \xrightarrow{\|Y_{Bound}\rangle, I_{Probe}(\lambda,\tau)} \|S0_{Bound}\rangle$ | $k_{GSB_Y, Reflect}$ | $1.4 \cdot 10^{10}$ | $5.4 \cdot 10^9$ | $5.7 \cdot 10^9$ | $3.7 \cdot 10^9$ | $1.2 \cdot 10^6$ | $2.5 \cdot 10^{10}$ |
| $\|S0_{Bound}\rangle \xrightarrow{I_{Probe}(\lambda,\tau)} \|Y_{Bound}\rangle$ | $k_{ABS_Y, Reflect}$ | | | | | | |
| $\|Y_{Bound}\rangle \xrightarrow{I_{Probe}(\lambda,\tau)} \|S0_{Bound}\rangle$ | $k_{ESE_Y, Reflect}$ | | | | | | |
| $\|S0_{Bound}\rangle \xrightarrow{\|B_{Bound}\rangle, I_{Probe}(\lambda,\tau)} \|S0_{Bound}\rangle$ | $k_{GSB_B, Reflect}$ | $2.4 \cdot 10^{12}$ | $2.5 \cdot 10^{12}$ | $2.5 \cdot 10^{12}$ | $1.6 \cdot 10^{12}$ | $9.1 \cdot 10^{11}$ | $3.3 \cdot 10^{12}$ |
| $\|S0_{Bound}\rangle \xrightarrow{I_{Probe}(\lambda,\tau)} \|B_{Bound}\rangle$ | $k_{ABS_B, Reflect}$ | | | | | | |
| $\|B_{Bound}\rangle \xrightarrow{I_{Probe}(\lambda,\tau)} \|S0_{Bound}\rangle$ | $k_{ESE_B, Reflect}$ | | | | | | |
| $\|S0_{Bound}\rangle \xrightarrow{\|X_{Bound}\rangle, I_{Probe}(\lambda,\tau)} \|S0_{Bound}\rangle$ | $k_{GSB_X, Reflect}$ | $9.8 \cdot 10^{12}$ | $1.0 \cdot 10^{13}$ | $8.1 \cdot 10^{12}$ | $8.0 \cdot 10^{12}$ | $9.1 \cdot 10^{12}$ | $8.9 \cdot 10^{12}$ |
| $\|S0_{Bound}\rangle \xrightarrow{I_{Probe}(\lambda,\tau)} \|X_{Bound}\rangle$ | $k_{ABS_X, Reflect}$ | | | | | | |
| $\|X_{Bound}\rangle \xrightarrow{I_{Probe}(\lambda,\tau)} \|S0_{Bound}\rangle$ | $k_{ESE_X, Reflect}$ | | | | | | |
| $\|S0_{Bound}\rangle \xrightarrow{\|T_{Bound}\rangle, I_{Probe}(\lambda,\tau)} \|S0_{Bound}\rangle$ | $k_{GSB_T, Reflect}$ | $2.7 \cdot 10^{12}$ | $2.7 \cdot 10^{12}$ | $2.3 \cdot 10^{12}$ | $1.6 \cdot 10^{12}$ | $3.6 \cdot 10^{12}$ | $2.6 \cdot 10^{12}$ |
| $\|S0_{Bound}\rangle \xrightarrow{I_{Probe}(\lambda,\tau)} \|T_{Bound}\rangle$ | $k_{ABS_T, Reflect}$ | | | | | | |
| $\|T_{Bound}\rangle \xrightarrow{I_{Probe}(\lambda,\tau)} \|S0_{Bound}\rangle$ | $k_{ESE_T, Reflect}$ | | | | | | |
| $\|Y_{Bound}\rangle \xrightarrow{\|Z0_{Bound}\rangle, I_{Reflect}(\lambda,\tau)} \|Y_{Bound}\rangle$ | $k_{ESA_0, Reflect}$ | $6.6 \cdot 10^8$ | $1.8 \cdot 10^7$ | $7.2 \cdot 10^9$ | $6.8 \cdot 10^8$ | $2.1 \cdot 10^8$ | $2.7 \cdot 10^8$ |
| $\|B_{Bound}\rangle \xrightarrow{\|Z0_{Bound}\rangle, I_{Reflect}(\lambda,\tau)} \|B_{Bound}\rangle$ | | | | | | | |
| $\|X_{Bound}\rangle \xrightarrow{\|Z0_{Bound}\rangle, I_{Reflect}(\lambda,\tau)} \|X_{Bound}\rangle$ | | | | | | | |
| $\|T_{Bound}\rangle \xrightarrow{\|Z0_{Bound}\rangle, I_{Reflect}(\lambda,\tau)} \|T_{Bound}\rangle$ | | | | | | | |

| | Rate Coefficients | RuP (s⁻¹) | RuP 2 (s⁻¹) | RuP3 (s⁻¹) | RuCP (s⁻¹) | RuCP2 (s⁻¹) | RuCP3 (s⁻¹) |
|---|---|---|---|---|---|---|---|
| $\|Y_{Bound}\rangle \xrightarrow{\|Z1_{Bound}\rangle, I_{Reflect}(\lambda,\tau)} \|Y_{Bound}\rangle$ <br> $\|B_{Bound}\rangle \xrightarrow{\|Z1_{Bound}\rangle, I_{Reflect}(\lambda,\tau)} \|B_{Bound}\rangle$ <br> $\|X_{Bound}\rangle \xrightarrow{\|Z1_{Bound}\rangle, I_{Reflect}(\lambda,\tau)} \|X_{Bound}\rangle$ <br> $\|T_{Bound}\rangle \xrightarrow{\|Z1_{Bound}\rangle, I_{Reflect}(\lambda,\tau)} \|T_{Bound}\rangle$ | $k_{ESA_1,Reflect}$ | 5.3·10¹² | 4.0·10¹² | 4.9·10¹² | 6.2·10¹² | 5.2·10¹² | 5.8·10¹² |
| $\|Y_{Bound}\rangle \xrightarrow{\|Z2_{Bound}\rangle, I_{Reflect}(\lambda,\tau)} \|Y_{Bound}\rangle$ <br> $\|B_{Bound}\rangle \xrightarrow{\|Z2_{Bound}\rangle, I_{Reflect}(\lambda,\tau)} \|B_{Bound}\rangle$ <br> $\|X_{Bound}\rangle \xrightarrow{\|Z2_{Bound}\rangle, I_{Reflect}(\lambda,\tau)} \|X_{Bound}\rangle$ <br> $\|T_{Bound}\rangle \xrightarrow{\|Z2_{Bound}\rangle, I_{Reflect}(\lambda,\tau)} \|T_{Bound}\rangle$ | $k_{ESA_2,Reflect}$ | 3.6·10¹² | 2.3·10¹² | 3.8·10¹² | 4.2·10¹² | 4.4·10¹² | 4.1·10¹² |
| $\|Y_{Bound}\rangle \xrightarrow{\|Z3_{Bound}\rangle, I_{Reflect}(\lambda,\tau)} \|Y_{Bound}\rangle$ <br> $\|B_{Bound}\rangle \xrightarrow{\|Z3_{Bound}\rangle, I_{Reflect}(\lambda,\tau)} \|B_{Bound}\rangle$ <br> $\|X_{Bound}\rangle \xrightarrow{\|Z3_{Bound}\rangle, I_{Reflect}(\lambda,\tau)} \|X_{Bound}\rangle$ <br> $\|T_{Bound}\rangle \xrightarrow{\|Z3_{Bound}\rangle, I_{Reflect}(\lambda,\tau)} \|T_{Bound}\rangle$ | $k_{ESA_3,Reflect}$ | 1.8·10¹² | 1.2·10¹² | 2.0·10¹² | 2.1·10¹² | 4.1·10¹² | 2.2·10¹² |
| $\|Y_{Bound}\rangle \xrightarrow{\|Z4_{Bound}\rangle, I_{Reflect}(\lambda,\tau)} \|Y_{Bound}\rangle$ <br> $\|B_{Bound}\rangle \xrightarrow{\|Z4_{Bound}\rangle, I_{Reflect}(\lambda,\tau)} \|B_{Bound}\rangle$ <br> $\|X_{Bound}\rangle \xrightarrow{\|Z4_{Bound}\rangle, I_{Reflect}(\lambda,\tau)} \|X_{Bound}\rangle$ <br> $\|T_{Bound}\rangle \xrightarrow{\|Z4_{Bound}\rangle, I_{Reflect}(\lambda,\tau)} \|T_{Bound}\rangle$ | $k_{ESA_4,Reflect}$ | 1.1·10¹² | 7.9·10¹² | 1.3·10¹² | 1.6·10¹² | 2.9·10¹² | 1.6·10¹² |
| $\|Y_{Bound}\rangle \xrightarrow{\|Z5_{Bound}\rangle, I_{Reflect}(\lambda,\tau)} \|Y_{Bound}\rangle$ <br> $\|B_{Bound}\rangle \xrightarrow{\|Z5_{Bound}\rangle, I_{Reflect}(\lambda,\tau)} \|B_{Bound}\rangle$ <br> $\|X_{Bound}\rangle \xrightarrow{\|Z5_{Bound}\rangle, I_{Reflect}(\lambda,\tau)} \|X_{Bound}\rangle$ <br> $\|T_{Bound}\rangle \xrightarrow{\|Z5_{Bound}\rangle, I_{Reflect}(\lambda,\tau)} \|T_{Bound}\rangle$ | $k_{ESA_5,Reflect}$ | 1.4·10¹² | 1.0·10¹² | 1.3·10¹² | 1.7·10¹² | 3.1·10¹² | 1.2·10¹² |

**Table S5.** Base mechanistic scheme and rate coefficients for dyes in set 6-Ru on 4 μm film of TiO$_2$ for dataset Ru-Z with broadband probe spanning 350 nm to 700 nm. Simulated surface concentration was 2.5·10$^{-8}$ mol cm$^{-2}$.

| | Rate Coefficients | RuP (s$^{-1}$) | RuP 2 (s$^{-1}$) | RuP3 (s$^{-1}$) | RuCP (s$^{-1}$) | RuCP2 (s$^{-1}$) | RuCP3 (s$^{-1}$) |
|---|---|---|---|---|---|---|---|
| $\|S0_{Bound}\rangle \xrightarrow{h\nu_{Pump}} \|Y_{Bound}\rangle$ | $k_{Pump,Y}$ | Variable | Variable | Variable | Variable | Variable | Variable |
| $\|S0_{Bound}\rangle \xrightarrow{h\nu_{Pump}} \|B_{Bound}\rangle$ | $k_{Pump,B}$ | Variable | Variable | Variable | Variable | Variable | Variable |
| $\|S0_{Bound}\rangle \xrightarrow{h\nu_{Pump}} \|X_{Bound}\rangle$ | $k_{Pump,X}$ | Variable | Variable | Variable | Variable | Variable | Variable |
| $\|S0_{Bound}\rangle \xrightarrow{h\nu_{Pump}} \|T_{Bound}\rangle$ | $k_{Pump,T}$ | Variable | Variable | Variable | Variable | Variable | Variable |
| $\|S0_{Bound}\rangle \xrightarrow{\|Y_{Bound}\rangle,I_{Probe}(\lambda,\tau)} \|S0_{Bound}\rangle$ | $k_{GSB_Y}$ | | | | | | |
| $\|S0_{Bound}\rangle \xrightarrow{I_{Probe}(\lambda,\tau)} \|Y_{Bound}\rangle + ABS_{Y.Bound}$ | $k_{ABS_Y}$ | 1.8·10$^{13}$ | 2.0·10$^{13}$ | 2.0·10$^{13}$ | 1.9·10$^{13}$ | 1.8·10$^{13}$ | 3.4·10$^{13}$ |
| $\|Y_{Bound}\rangle \xrightarrow{I_{Probe}(\lambda,\tau)} \|S0_{Bound}\rangle + ESE_{Y.Bound}$ | $k_{ESE_Y}$ | | | | | | |
| $\|S0_{Bound}\rangle \xrightarrow{\|B_{Bound}\rangle,I_{Probe}(\lambda,\tau)} \|S0_{Bound}\rangle$ | $k_{GSB_B}$ | | | | | | |
| $\|S0_{Bound}\rangle \xrightarrow{I_{Probe}(\lambda,\tau)} \|B_{Bound}\rangle + ABS_{B.Bound}$ | $k_{ABS_B}$ | 3.8·10$^{13}$ | 4.1·10$^{13}$ | 3.3·10$^{13}$ | 3.2·10$^{13}$ | 3.2·10$^{13}$ | 3.7·10$^{13}$ |
| $\|B_{Bound}\rangle \xrightarrow{I_{Probe}(\lambda,\tau)} \|S0_{Bound}\rangle + ESE_{B.Bound}$ | $k_{ESE_B}$ | | | | | | |
| $\|S0_{Bound}\rangle \xrightarrow{\|X_{Bound}\rangle,I_{Probe}(\lambda,\tau)} \|S0_{Bound}\rangle$ | $k_{GSB_X}$ | | | | | | |
| $\|S0_{Bound}\rangle \xrightarrow{I_{Probe}(\lambda,\tau)} \|X_{Bound}\rangle + ABS_{X.Bound}$ | $k_{ABS_X}$ | 4.8·10$^{13}$ | 5.0·10$^{13}$ | 4.0·10$^{13}$ | 4.1·10$^{13}$ | 4.1·10$^{13}$ | 4.5·10$^{13}$ |
| $\|X_{Bound}\rangle \xrightarrow{I_{Probe}(\lambda,\tau)} \|S0_{Bound}\rangle + ESE_{X.Bound}$ | $k_{ESE_X}$ | | | | | | |
| $\|S0_{Bound}\rangle \xrightarrow{\|T_{Bound}\rangle,I_{Probe}(\lambda,\tau)} \|S0_{Bound}\rangle$ | $k_{GSB_T}$ | | | | | | |
| $\|S0_{Bound}\rangle \xrightarrow{I_{Probe}(\lambda,\tau)} \|T_{Bound}\rangle + ABS_{T.Bound}$ | $k_{ABS_T}$ | 1.8·10$^{13}$ | 1.8·10$^{13}$ | 1.8·10$^{13}$ | 1.8·10$^{13}$ | 1.8·10$^{13}$ | 1.8·10$^{13}$ |
| $\|T_{Bound}\rangle \xrightarrow{I_{Probe}(\lambda,\tau)} \|S0_{Bound}\rangle + ESE_{T.Bound}$ | $k_{ESE_T}$ | | | | | | |
| $\|Y_{Bound}\rangle \to \|B_{Bound}\rangle$ | $k_{YB}$ | 1.8·10$^{13}$ | 2.4·10$^{13}$ | 8.0·10$^{13}$ | 6.0·10$^{13}$ | 6.0·10$^{13}$ | 6.0·10$^{13}$ |
| $\|B_{Bound}\rangle \to \|X_{Bound}\rangle$ | $k_{BX}$ | 1.8·10$^{13}$ | 2.4·10$^{13}$ | 8.0·10$^{13}$ | 6.0·10$^{13}$ | 6.0·10$^{13}$ | 6.0·10$^{13}$ |
| $\|X_{Bound}\rangle \to \|S0_{Bound}\rangle$ | $k_{UF-nr}$ | N.A | NA | 4.0·10$^{13}$ | 1.6·10$^{13}$ | 1.6·10$^{13}$ | 2.0·10$^{13}$ |
| $\|X_{Bound}\rangle \to \|T_{Bound}\rangle$ | $k_{ISC}$ | 4.0·10$^{13}$ | 4.0·10$^{13}$ | 2.0·10$^{13}$ | 2.0·10$^{13}$ | 2.0·10$^{13}$ | 3.6·10$^{13}$ |
| $\|Y_{Bound}\rangle \xrightarrow{\|Z0_{Bound}\rangle,I_{Probe}(\lambda,\tau)} \|Y_{Bound}\rangle$ $\|B_{Bound}\rangle \xrightarrow{\|Z0_{Bound}\rangle,I_{Probe}(\lambda,\tau)} \|B_{Bound}\rangle$ $\|X_{Bound}\rangle \xrightarrow{\|Z0_{Bound}\rangle,I_{Probe}(\lambda,\tau)} \|X_{Bound}\rangle$ $\|T_{Bound}\rangle \xrightarrow{\|Z0_{Bound}\rangle,I_{Probe}(\lambda,\tau)} \|T_{Bound}\rangle$ | $k_{ESA_0}$ | 1.6·10$^{14}$ | 1.5·10$^{14}$ | 1.5·10$^{14}$ | 1.8·10$^{14}$ | 2.1·10$^{14}$ | 1.9·10$^{14}$ |
| $\|Y_{Bound}\rangle \xrightarrow{\|Z1_{Bound}\rangle,I_{Probe}(\lambda,\tau)} \|Y_{Bound}\rangle$ $\|B_{Bound}\rangle \xrightarrow{\|Z1_{Bound}\rangle,I_{Probe}(\lambda,\tau)} \|B_{Bound}\rangle$ $\|X_{Bound}\rangle \xrightarrow{\|Z1_{Bound}\rangle,I_{Probe}(\lambda,\tau)} \|X_{Bound}\rangle$ $\|T_{Bound}\rangle \xrightarrow{\|Z1_{Bound}\rangle,I_{Probe}(\lambda,\tau)} \|T_{Bound}\rangle$ | $k_{ESA_1}$ | 4.0·10$^{13}$ | 3.2·10$^{13}$ | 3.8·10$^{13}$ | 4.8·10$^{13}$ | 4.0·10$^{13}$ | 4.9·10$^{13}$ |
| $\|Y_{Bound}\rangle \xrightarrow{\|Z2_{Bound}\rangle,I_{Probe}(\lambda,\tau)} \|Y_{Bound}\rangle$ $\|B_{Bound}\rangle \xrightarrow{\|Z2_{Bound}\rangle,I_{Probe}(\lambda,\tau)} \|B_{Bound}\rangle$ $\|X_{Bound}\rangle \xrightarrow{\|Z2_{Bound}\rangle,I_{Probe}(\lambda,\tau)} \|X_{Bound}\rangle$ $\|T_{Bound}\rangle \xrightarrow{\|Z2_{Bound}\rangle,I_{Probe}(\lambda,\tau)} \|T_{Bound}\rangle$ | $k_{ESA_2}$ | 4.3·10$^{13}$ | 2.8·10$^{13}$ | 4.7·10$^{13}$ | 5.2·10$^{13}$ | 5.7·10$^{13}$ | 5.3·10$^{13}$ |
| $\|Y_{Bound}\rangle \xrightarrow{\|Z3_{Bound}\rangle,I_{Probe}(\lambda,\tau)} \|Y_{Bound}\rangle$ $\|B_{Bound}\rangle \xrightarrow{\|Z3_{Bound}\rangle,I_{Probe}(\lambda,\tau)} \|B_{Bound}\rangle$ $\|X_{Bound}\rangle \xrightarrow{\|Z3_{Bound}\rangle,I_{Probe}(\lambda,\tau)} \|X_{Bound}\rangle$ $\|T_{Bound}\rangle \xrightarrow{\|Z3_{Bound}\rangle,I_{Probe}(\lambda,\tau)} \|T_{Bound}\rangle$ | $k_{ESA_3}$ | 4.2·10$^{13}$ | 3.0·10$^{13}$ | 5.3·10$^{13}$ | 5.6·10$^{13}$ | 9.3·10$^{13}$ | 5.7·10$^{13}$ |

| | Rate Coefficients | RuP (s⁻¹) | RuP 2 (s⁻¹) | RuP3 (s⁻¹) | RuCP (s⁻¹) | RuCP2 (s⁻¹) | RuCP3 (s⁻¹) |
|---|---|---|---|---|---|---|---|
| $\|Y_{Bound}\rangle \xrightarrow{\|Z4_{Bound}\rangle, I_{Probe}(\lambda,\tau)} \|Y_{Bound}\rangle$ <br> $\|B_{Bound}\rangle \xrightarrow{\|Z4_{Bound}\rangle, I_{Probe}(\lambda,\tau)} \|B_{Bound}\rangle$ <br> $\|X_{Bound}\rangle \xrightarrow{\|Z4_{Bound}\rangle, I_{Probe}(\lambda,\tau)} \|X_{Bound}\rangle$ <br> $\|T_{Bound}\rangle \xrightarrow{\|Z4_{Bound}\rangle, I_{Probe}(\lambda,\tau)} \|T_{Bound}\rangle$ | $k_{ESA_4}$ | $3.9 \cdot 10^{13}$ | $3.0 \cdot 10^{13}$ | $4.6 \cdot 10^{13}$ | $5.7 \cdot 10^{13}$ | $11 \cdot 10^{13}$ | $5.7 \cdot 10^{13}$ |
| $\|Y_{Bound}\rangle \xrightarrow{\|Z5_{Bound}\rangle, I_{Probe}(\lambda,\tau)} \|Y_{Bound}\rangle$ <br> $\|B_{Bound}\rangle \xrightarrow{\|Z5_{Bound}\rangle, I_{Probe}(\lambda,\tau)} \|B_{Bound}\rangle$ <br> $\|X_{Bound}\rangle \xrightarrow{\|Z5_{Bound}\rangle, I_{Probe}(\lambda,\tau)} \|X_{Bound}\rangle$ <br> $\|T_{Bound}\rangle \xrightarrow{\|Z5_{Bound}\rangle, I_{Probe}(\lambda,\tau)} \|T_{Bound}\rangle$ | $k_{ESA_5}$ | $5.4 \cdot 10^{13}$ | $3.8 \cdot 10^{13}$ | $5.0 \cdot 10^{13}$ | $7.0 \cdot 10^{13}$ | $10.5 \cdot 10^{13}$ | $5.0 \cdot 10^{13}$ |
| $\|T_{Bound}\rangle \to \|S0_{Bound}\rangle + rad_{Bound}$ | $k_{rad}$ | $10.9 \cdot 10^4$ | $9.6 \cdot 10^4$ | $11 \cdot 10^4$ | $10.6 \cdot 10^4$ | $10.6 \cdot 10^4$ | $10.3 \cdot 10^4$ |
| $\|T_{Bound}\rangle \to \|S0_{Bound}\rangle$ | $k_{nr}$ | $3.2 \cdot 10^6$ | $2.6 \cdot 10^6$ | $2.0 \cdot 10^6$ | $2.0 \cdot 10^6$ | $2.2 \cdot 10^6$ | $2.3 \cdot 10^6$ |
| $\|S0_{Unbound}\rangle \xrightarrow{h\nu_{Pump}} \|Y_{Unbound}\rangle$ | $k_{Pump,Y}$ | Variable | Variable | Variable | Variable | Variable | Variable |
| $\|S0_{Unbound}\rangle \xrightarrow{h\nu_{Pump}} \|B_{Unbound}\rangle$ | $k_{Pump,B}$ | Variable | Variable | Variable | Variable | Variable | Variable |
| $\|S0_{Unbound}\rangle \xrightarrow{h\nu_{Pump}} \|X_{Unbound}\rangle$ | $k_{Pump,X}$ | Variable | Variable | Variable | Variable | Variable | Variable |
| $\|S0_{Unbound}\rangle \xrightarrow{h\nu_{Pump}} \|T_{Unbound}\rangle$ | $k_{Pump,T}$ | Variable | Variable | Variable | Variable | Variable | Variable |
| $\|S0_{Unbound}\rangle \xrightarrow{\|Y_{Unbound}\rangle, I_{Probe}(\lambda,\tau)} \|S0_{Unbound}\rangle$ | $k_{GSB_Y}$ | $1.8 \cdot 10^{13}$ | $2.0 \cdot 10^{13}$ | $2.0 \cdot 10^{13}$ | $1.9 \cdot 10^{13}$ | $1.8 \cdot 10^{13}$ | $3.4 \cdot 10^{13}$ |
| $\|S0_{Unbound}\rangle \xrightarrow{I_{Probe}(\lambda,\tau)} \|Y_{Unbound}\rangle$ | $k_{ABS_Y}$ | | | | | | |
| $\|Y_{Unbound}\rangle \xrightarrow{I_{Probe}(\lambda,\tau)} \|S0_{Unbound}\rangle$ | $k_{ESE_Y}$ | | | | | | |
| $\|S0_{Unbound}\rangle \xrightarrow{\|B_{Unbound}\rangle, I_{Probe}(\lambda,\tau)} \|S0_{Unbound}\rangle$ | $k_{GSB_B}$ | $3.8 \cdot 10^{13}$ | $4.1 \cdot 10^{13}$ | $3.3 \cdot 10^{13}$ | $3.2 \cdot 10^{13}$ | $3.2 \cdot 10^{13}$ | $3.7 \cdot 10^{13}$ |
| $\|S0_{Unbound}\rangle \xrightarrow{I_{Probe}(\lambda,\tau)} \|B_{Unbound}\rangle$ | $k_{ABS_B}$ | | | | | | |
| $\|B_{Unbound}\rangle \xrightarrow{I_{Probe}(\lambda,\tau)} \|S0_{Unbound}\rangle$ | $k_{ESE_B}$ | | | | | | |
| $\|S0_{Unbound}\rangle \xrightarrow{\|X_{Unbound}\rangle, I_{Probe}(\lambda,\tau)} \|S0_{Unbound}\rangle$ | $k_{GSB_X}$ | $4.8 \cdot 10^{13}$ | $5.0 \cdot 10^{13}$ | $4.0 \cdot 10^{13}$ | $4.1 \cdot 10^{13}$ | $4.1 \cdot 10^{13}$ | $4.5 \cdot 10^{13}$ |
| $\|S0_{Unbound}\rangle \xrightarrow{I_{Probe}(\lambda,\tau)} \|X_{Unbound}\rangle$ | $k_{ABS_X}$ | | | | | | |
| $\|X_{Unbound}\rangle \xrightarrow{I_{Probe}(\lambda,\tau)} \|S0_{Unbound}\rangle$ | $k_{ESE_X}$ | | | | | | |
| $\|S0_{Unbound}\rangle \xrightarrow{\|T_{Unbound}\rangle, I_{Probe}(\lambda,\tau)} \|S0_{Unbound}\rangle$ | $k_{GSB_T}$ | $1.8 \cdot 10^{13}$ | $1.8 \cdot 10^{13}$ | $1.8 \cdot 10^{13}$ | $1.8 \cdot 10^{13}$ | $1.8 \cdot 10^{13}$ | $1.8 \cdot 10^{13}$ |
| $\|S0_{Unbound}\rangle \xrightarrow{I_{Probe}(\lambda,\tau)} \|T_{Unbound}\rangle$ | $k_{ABS_T}$ | | | | | | |
| $\|T_{Unbound}\rangle \xrightarrow{I_{Probe}(\lambda,\tau)} \|S0_{Unbound}\rangle$ | $k_{ESE_T}$ | | | | | | |
| $\|Y_{Unbound}\rangle \to \|B_{Unbound}\rangle$ | $k_{YB}$ | $1.8 \cdot 10^{13}$ | $2.4 \cdot 10^{13}$ | $8.0 \cdot 10^{13}$ | $6.0 \cdot 10^{13}$ | $6.0 \cdot 10^{13}$ | $6.0 \cdot 10^{13}$ |
| $\|B_{Unbound}\rangle \to \|X_{Unbound}\rangle$ | $k_{BX}$ | $1.8 \cdot 10^{13}$ | $2.4 \cdot 10^{13}$ | $8.0 \cdot 10^{13}$ | $6.0 \cdot 10^{13}$ | $6.0 \cdot 10^{13}$ | $6.0 \cdot 10^{13}$ |
| $\|X_{Unbound}\rangle \to \|S0_{Unbound}\rangle$ | $k_{UF-nr}$ | $2.4 \cdot 10^{13}$ | $1.0 \cdot 10^{13}$ | $4.0 \cdot 10^{13}$ | $1.6 \cdot 10^{13}$ | $1.6 \cdot 10^{13}$ | $2.0 \cdot 10^{13}$ |
| $\|X_{Unbound}\rangle \to \|T_{Unbound}\rangle$ | $k_{ISC}$ | $4.0 \cdot 10^{13}$ | $4.0 \cdot 10^{13}$ | $2.0 \cdot 10^{13}$ | $2.0 \cdot 10^{13}$ | $2.0 \cdot 10^{13}$ | $3.6 \cdot 10^{13}$ |
| $\|Y_{Unbound}\rangle \xrightarrow{\|Z0_{Unbound}\rangle, I_{Probe}(\lambda,\tau)} \|Y_{Unbound}\rangle$ <br> $\|B_{Unbound}\rangle \xrightarrow{\|Z0_{Unbound}\rangle, I_{Probe}(\lambda,\tau)} \|B_{Unbound}\rangle$ <br> $\|X_{Unbound}\rangle \xrightarrow{\|Z0_{Unbound}\rangle, I_{Probe}(\lambda,\tau)} \|X_{Unbound}\rangle$ <br> $\|T_{Unbound}\rangle \xrightarrow{\|Z0_{Unbound}\rangle, I_{Probe}(\lambda,\tau)} \|T_{Unbound}\rangle$ | $k_{ESA_0}$ | $1.6 \cdot 10^{14}$ | $1.5 \cdot 10^{14}$ | $1.5 \cdot 10^{14}$ | $1.8 \cdot 10^{14}$ | $2.1 \cdot 10^{14}$ | $1.9 \cdot 10^{14}$ |
| $\|Y_{Unbound}\rangle \xrightarrow{\|Z1_{Unbound}\rangle, I_{Probe}(\lambda,\tau)} \|Y_{Unbound}\rangle$ <br> $\|B_{Unbound}\rangle \xrightarrow{\|Z1_{Unbound}\rangle, I_{Probe}(\lambda,\tau)} \|B_{Unbound}\rangle$ <br> $\|X_{Unbound}\rangle \xrightarrow{\|Z1_{Unbound}\rangle, I_{Probe}(\lambda,\tau)} \|X_{Unbound}\rangle$ <br> $\|T_{Unbound}\rangle \xrightarrow{\|Z1_{Unbound}\rangle, I_{Probe}(\lambda,\tau)} \|T_{Unbound}\rangle$ | $k_{ESA_1}$ | $4.0 \cdot 10^{13}$ | $3.2 \cdot 10^{13}$ | $3.8 \cdot 10^{13}$ | $4.8 \cdot 10^{13}$ | $4.0 \cdot 10^{13}$ | $4.9 \cdot 10^{13}$ |

| Transition | Rate Coefficients | RuP (s⁻¹) | RuP 2 (s⁻¹) | RuP3 (s⁻¹) | RuCP (s⁻¹) | RuCP2 (s⁻¹) | RuCP3 (s⁻¹) |
|---|---|---|---|---|---|---|---|
| $\|Y_{Unbound}\rangle \xrightarrow{\|Z2_{Unbound}\rangle, I_{Probe}(\lambda,\tau)} \|Y_{Unbound}\rangle$ <br> $\|B_{Unbound}\rangle \xrightarrow{\|Z2_{Unbound}\rangle, I_{Probe}(\lambda,\tau)} \|B_{Unbound}\rangle$ <br> $\|X_{Unbound}\rangle \xrightarrow{\|Z2_{Unbound}\rangle, I_{Probe}(\lambda,\tau)} \|X_{Unbound}\rangle$ <br> $\|T_{Unbound}\rangle \xrightarrow{\|Z2_{Unbound}\rangle, I_{Probe}(\lambda,\tau)} \|T_{Unbound}\rangle$ | $k_{ESA_2}$ | $4.3 \cdot 10^{13}$ | $2.8 \cdot 10^{13}$ | $4.7 \cdot 10^{13}$ | $5.2 \cdot 10^{13}$ | $5.7 \cdot 10^{13}$ | $5.3 \cdot 10^{13}$ |
| $\|Y_{Unbound}\rangle \xrightarrow{\|Z3_{Unbound}\rangle, I_{Probe}(\lambda,\tau)} \|Y_{Unbound}\rangle$ <br> $\|B_{Unbound}\rangle \xrightarrow{\|Z3_{Unbound}\rangle, I_{Probe}(\lambda,\tau)} \|B_{Unbound}\rangle$ <br> $\|X_{Unbound}\rangle \xrightarrow{\|Z3_{Unbound}\rangle, I_{Probe}(\lambda,\tau)} \|X_{Unbound}\rangle$ <br> $\|T_{Unbound}\rangle \xrightarrow{\|Z3_{Unbound}\rangle, I_{Probe}(\lambda,\tau)} \|T_{Unbound}\rangle$ | $k_{ESA_3}$ | $4.2 \cdot 10^{13}$ | $3.0 \cdot 10^{13}$ | $5.3 \cdot 10^{13}$ | $5.6 \cdot 10^{13}$ | $9.3 \cdot 10^{13}$ | $5.7 \cdot 10^{13}$ |
| $\|Y_{Unbound}\rangle \xrightarrow{\|Z4_{Unbound}\rangle, I_{Probe}(\lambda,\tau)} \|Y_{Unbound}\rangle$ <br> $\|B_{Unbound}\rangle \xrightarrow{\|Z4_{Unbound}\rangle, I_{Probe}(\lambda,\tau)} \|B_{Unbound}\rangle$ <br> $\|X_{Unbound}\rangle \xrightarrow{\|Z4_{Unbound}\rangle, I_{Probe}(\lambda,\tau)} \|X_{Unbound}\rangle$ <br> $\|T_{Unbound}\rangle \xrightarrow{\|Z4_{Unbound}\rangle, I_{Probe}(\lambda,\tau)} \|T_{Unbound}\rangle$ | $k_{ESA_4}$ | $3.9 \cdot 10^{13}$ | $3.0 \cdot 10^{13}$ | $4.6 \cdot 10^{13}$ | $5.7 \cdot 10^{13}$ | $11 \cdot 10^{13}$ | $5.7 \cdot 10^{13}$ |
| $\|Y_{Unbound}\rangle \xrightarrow{\|Z5_{Unbound}\rangle, I_{Probe}(\lambda,\tau)} \|Y_{Unbound}\rangle$ <br> $\|B_{Unbound}\rangle \xrightarrow{\|Z5_{Unbound}\rangle, I_{Probe}(\lambda,\tau)} \|B_{Unbound}\rangle$ <br> $\|X_{Unbound}\rangle \xrightarrow{\|Z5_{Unbound}\rangle, I_{Probe}(\lambda,\tau)} \|X_{Unbound}\rangle$ <br> $\|T_{Unbound}\rangle \xrightarrow{\|Z5_{Unbound}\rangle, I_{Probe}(\lambda,\tau)} \|T_{Unbound}\rangle$ | $k_{ESA_5}$ | $5.4 \cdot 10^{13}$ | $3.8 \cdot 10^{13}$ | $5.0 \cdot 10^{13}$ | $7.0 \cdot 10^{13}$ | $10.5 \cdot 10^{13}$ | $5.0 \cdot 10^{13}$ |
| $\|T_{Unbound}\rangle \to \|S0_{Unbound}\rangle + rad_{Unbound}$ | $k_{rad}$ | $10.9 \cdot 10^4$ | $9.6 \cdot 10^4$ | $11 \cdot 10^4$ | $10.6 \cdot 10^4$ | $10.6 \cdot 10^4$ | $10.3 \cdot 10^4$ |
| $\|T_{Unbound}\rangle \to \|S0_{Unbound}\rangle$ | $k_{nr}$ | $3.2 \cdot 10^6$ | $2.6 \cdot 10^6$ | $2.0 \cdot 10^6$ | $2.0 \cdot 10^6$ | $2.2 \cdot 10^6$ | $2.3 \cdot 10^6$ |
| $\|Y_{Bound}\rangle + \|TiO_2\rangle \to \|Ru^3\rangle + \|TiO_2^*\rangle$ <br> $\|B_{Bound}\rangle + \|TiO_2\rangle \to \|Ru^3\rangle + \|TiO_2^*\rangle$ <br> $\|X_{Bound}\rangle + \|TiO_2\rangle \to \|Ru^3\rangle + \|TiO_2^*\rangle$ <br> $\|T_{Bound}\rangle + \|TiO_2\rangle \to \|Ru^3\rangle + \|TiO_2^*\rangle$ | $k_{Injection}$ | $2.4 \cdot 10^{12}$ | $1.0 \cdot 10^{12}$ | $4.0 \cdot 10^{11}$ | $1.6 \cdot 10^{11}$ | $1.6 \cdot 10^{11}$ | $1.0 \cdot 10^{11}$ |
| $\|Ru^3\rangle \xrightarrow{\|Ru^{3*}\rangle, I\ (\lambda,\tau)} \|Ru^3\rangle + ESA_{\|Ru^3\rangle}$ | $k_{ESA_{\|Ru^3\rangle}}$ | $1.5 \cdot 10^{15}$ | $2.2 \cdot 10^{15}$ | $1.6 \cdot 10^{15}$ | $1.5 \cdot 10^{15}$ | $2.0 \cdot 10^{15}$ | $4.2 \cdot 10^{15}$ |
| $\|Ru^3\rangle \xrightarrow{\|Ru^{3*}\rangle, I_{Reflect}(\lambda,\tau)} \|Ru^3\rangle$ | $k_{ESA_{\|Ru^3\rangle},Reflect}$ | $2.6 \cdot 10^{14}$ | $3.5 \cdot 10^{14}$ | $2.6 \cdot 10^{14}$ | $2.5 \cdot 10^{14}$ | $3.4 \cdot 10^{14}$ | $6.6 \cdot 10^{14}$ |
| $\|S0_{Bound}\rangle \xrightarrow{\|Y_{Bound}\rangle, I_{Probe}(\lambda,\tau)} \|S0_{Bound}\rangle$ | $k_{GSB_Y,Reflect}$ | $4.8 \cdot 10^{12}$ | $5.8 \cdot 10^{12}$ | $5.6 \cdot 10^{12}$ | $5.3 \cdot 10^{12}$ | $5.2 \cdot 10^{12}$ | $9.3 \cdot 10^{12}$ |
| $\|S0_{Bound}\rangle \xrightarrow{I_{Probe}(\lambda,\tau)} \|Y_{Bound}\rangle$ | $k_{ABS_Y,Reflect}$ | | | | | | |
| $\|Y_{Bound}\rangle \xrightarrow{I_{Probe}(\lambda,\tau)} \|S0_{Bound}\rangle$ | $k_{ESE_Y,Reflect}$ | | | | | | |
| $\|S0_{Bound}\rangle \xrightarrow{\|B_{Bound}\rangle, I_{Probe}(\lambda,\tau)} \|S0_{Bound}\rangle$ | $k_{GSB_B,Reflect}$ | $7.0 \cdot 10^{12}$ | $7.6 \cdot 10^{12}$ | $6.0 \cdot 10^{12}$ | $6.2 \cdot 10^{12}$ | $6.0 \cdot 10^{12}$ | $6.5 \cdot 10^{12}$ |
| $\|S0_{Bound}\rangle \xrightarrow{I_{Probe}(\lambda,\tau)} \|B_{Bound}\rangle$ | $k_{ABS_B,Reflect}$ | | | | | | |
| $\|B_{Bound}\rangle \xrightarrow{I_{Probe}(\lambda,\tau)} \|S0_{Bound}\rangle$ | $k_{ESE_B,Reflect}$ | | | | | | |
| $\|S0_{Bound}\rangle \xrightarrow{\|X_{Bound}\rangle, I_{Probe}(\lambda,\tau)} \|S0_{Bound}\rangle$ | $k_{GSB_X,Reflect}$ | $6.2 \cdot 10^{12}$ | $6.5 \cdot 10^{12}$ | $5.1 \cdot 10^{12}$ | $5.5 \cdot 10^{12}$ | $5.3 \cdot 10^{12}$ | $5.4 \cdot 10^{12}$ |
| $\|S0_{Bound}\rangle \xrightarrow{I_{Probe}(\lambda,\tau)} \|X_{Bound}\rangle$ | $k_{ABS_X,Reflect}$ | | | | | | |
| $\|X_{Bound}\rangle \xrightarrow{I_{Probe}(\lambda,\tau)} \|S0_{Bound}\rangle$ | $k_{ESE_X,Reflect}$ | | | | | | |
| $\|S0_{Bound}\rangle \xrightarrow{\|T_{Bound}\rangle, I_{Probe}(\lambda,\tau)} \|S0_{Bound}\rangle$ | $k_{GSB_T,Reflect}$ | $1.5 \cdot 10^{12}$ | $1.2 \cdot 10^{12}$ | $1.7 \cdot 10^{12}$ | $1.9 \cdot 10^{12}$ | $2.2 \cdot 10^{12}$ | $1.8 \cdot 10^{12}$ |
| $\|S0_{Bound}\rangle \xrightarrow{I_{Probe}(\lambda,\tau)} \|T_{Bound}\rangle$ | $k_{ABS_T,Reflect}$ | | | | | | |
| $\|T_{Bound}\rangle \xrightarrow{I_{Probe}(\lambda,\tau)} \|S0_{Bound}\rangle$ | $k_{ESE_T,Reflect}$ | | | | | | |
| $\|Y_{Bound}\rangle \xrightarrow{\|Z0_{Bound}\rangle, I_{Reflect}(\lambda,\tau)} \|Y_{Bound}\rangle$ <br> $\|B_{Bound}\rangle \xrightarrow{\|Z0_{Bound}\rangle, I_{Reflect}(\lambda,\tau)} \|B_{Bound}\rangle$ <br> $\|X_{Bound}\rangle \xrightarrow{\|Z0_{Bound}\rangle, I_{Reflect}(\lambda,\tau)} \|X_{Bound}\rangle$ <br> $\|T_{Bound}\rangle \xrightarrow{\|Z0_{Bound}\rangle, I_{Reflect}(\lambda,\tau)} \|T_{Bound}\rangle$ | $k_{ESA_0,Reflect}$ | $4.5 \cdot 10^{13}$ | $4.2 \cdot 10^{13}$ | $4.0 \cdot 10^{13}$ | $4.9 \cdot 10^{13}$ | $5.7 \cdot 10^{13}$ | $5.0 \cdot 10^{13}$ |

| | | Rate Coefficients | RuP (s⁻¹) | RuP2 (s⁻¹) | RuP3 (s⁻¹) | RuCP (s⁻¹) | RuCP2 (s⁻¹) | RuCP3 (s⁻¹) |
|---|---|---|---|---|---|---|---|---|
| $\lvert Y_{Bound}\rangle \xrightarrow{\lvert Z1_{Bound}\rangle, I_{Reflect}(\lambda,\tau)} \lvert Y_{Bound}\rangle$ $\lvert B_{Bound}\rangle \xrightarrow{\lvert Z1_{Bound}\rangle, I_{Reflect}(\lambda,\tau)} \lvert B_{Bound}\rangle$ $\lvert X_{Bound}\rangle \xrightarrow{\lvert Z1_{Bound}\rangle, I_{Reflect}(\lambda,\tau)} \lvert X_{Bound}\rangle$ $\lvert T_{Bound}\rangle \xrightarrow{\lvert Z1_{Bound}\rangle, I_{Reflect}(\lambda,\tau)} \lvert T_{Bound}\rangle$ | | $k_{ESA_1,Reflect}$ | $3.0 \cdot 10^{12}$ | $2.3 \cdot 10^{12}$ | $2.8 \cdot 10^{12}$ | $3.6 \cdot 10^{12}$ | $2.9 \cdot 10^{12}$ | $3.3 \cdot 10^{12}$ |
| $\lvert Y_{Bound}\rangle \xrightarrow{\lvert Z2_{Bound}\rangle, I_{Reflect}(\lambda,\tau)} \lvert Y_{Bound}\rangle$ $\lvert B_{Bound}\rangle \xrightarrow{\lvert Z2_{Bound}\rangle, I_{Reflect}(\lambda,\tau)} \lvert B_{Bound}\rangle$ $\lvert X_{Bound}\rangle \xrightarrow{\lvert Z2_{Bound}\rangle, I_{Reflect}(\lambda,\tau)} \lvert X_{Bound}\rangle$ $\lvert T_{Bound}\rangle \xrightarrow{\lvert Z2_{Bound}\rangle, I_{Reflect}(\lambda,\tau)} \lvert T_{Bound}\rangle$ | | $k_{ESA_2,Reflect}$ | $2.1 \cdot 10^{12}$ | $1.3 \cdot 10^{12}$ | $2.2 \cdot 10^{12}$ | $2.4 \cdot 10^{12}$ | $2.5 \cdot 10^{12}$ | $2.3 \cdot 10^{12}$ |
| $\lvert Y_{Bound}\rangle \xrightarrow{\lvert Z3_{Bound}\rangle, I_{Reflect}(\lambda,\tau)} \lvert Y_{Bound}\rangle$ $\lvert B_{Bound}\rangle \xrightarrow{\lvert Z3_{Bound}\rangle, I_{Reflect}(\lambda,\tau)} \lvert B_{Bound}\rangle$ $\lvert X_{Bound}\rangle \xrightarrow{\lvert Z3_{Bound}\rangle, I_{Reflect}(\lambda,\tau)} \lvert X_{Bound}\rangle$ $\lvert T_{Bound}\rangle \xrightarrow{\lvert Z3_{Bound}\rangle, I_{Reflect}(\lambda,\tau)} \lvert T_{Bound}\rangle$ | | $k_{ESA_3,Reflect}$ | $6.0 \cdot 10^{12}$ | $7.1 \cdot 10^{11}$ | $1.1 \cdot 10^{12}$ | $1.2 \cdot 10^{12}$ | $1.8 \cdot 10^{12}$ | $1.2 \cdot 10^{12}$ |
| $\lvert Y_{Bound}\rangle \xrightarrow{\lvert Z4_{Bound}\rangle, I_{Reflect}(\lambda,\tau)} \lvert Y_{Bound}\rangle$ $\lvert B_{Bound}\rangle \xrightarrow{\lvert Z4_{Bound}\rangle, I_{Reflect}(\lambda,\tau)} \lvert B_{Bound}\rangle$ $\lvert X_{Bound}\rangle \xrightarrow{\lvert Z4_{Bound}\rangle, I_{Reflect}(\lambda,\tau)} \lvert X_{Bound}\rangle$ $\lvert T_{Bound}\rangle \xrightarrow{\lvert Z4_{Bound}\rangle, I_{Reflect}(\lambda,\tau)} \lvert T_{Bound}\rangle$ | | $k_{ESA_4,Reflect}$ | $6.0 \cdot 10^{11}$ | $4.5 \cdot 10^{11}$ | $7.4 \cdot 10^{11}$ | $9.0 \cdot 10^{11}$ | $1.7 \cdot 10^{12}$ | $9.0 \cdot 10^{11}$ |
| $\lvert Y_{Bound}\rangle \xrightarrow{\lvert Z5_{Bound}\rangle, I_{Reflect}(\lambda,\tau)} \lvert Y_{Bound}\rangle$ $\lvert B_{Bound}\rangle \xrightarrow{\lvert Z5_{Bound}\rangle, I_{Reflect}(\lambda,\tau)} \lvert B_{Bound}\rangle$ $\lvert X_{Bound}\rangle \xrightarrow{\lvert Z5_{Bound}\rangle, I_{Reflect}(\lambda,\tau)} \lvert X_{Bound}\rangle$ $\lvert T_{Bound}\rangle \xrightarrow{\lvert Z5_{Bound}\rangle, I_{Reflect}(\lambda,\tau)} \lvert T_{Bound}\rangle$ | | $k_{ESA_5,Reflect}$ | $8.0 \cdot 10^{11}$ | $5.6 \cdot 10^{11}$ | $7.1 \cdot 10^{11}$ | $9.8 \cdot 10^{11}$ | $1.7 \cdot 10^{12}$ | $7.1 \cdot 10^{11}$ |

## 10. Transient Absorption of Dyes on TiO$_2$

The spectra for RuP2 (Section S10.1 Figures S13A and S14A) and RuP3 (Section S10.2 Figures S15A and S16A) are similar to that of RuP, as might be expected, though the initial (~550 nm and ~575 nm) and asymptotic (~575 nm and ~590 nm) wavelengths of the isosbestic points are different. The simulated results for RuP2 in dataset **Ru-G** agree remarkably well with experimental intensities, lineshape, and the redshift of the isosbestic point, Figure S13 Panels B and C. Simulations of RuP3 in dataset Ru-G exhibit initial intensities in good agreement with experimental values, although as the isosbestic point redshifts, the GSB dominates the ESAs much more than expected, Figure S15 Panels B and C. The dominance of the GSB over the ESAs deep into the red region of the spectrum is also observed in simulations of RuP3 in dataset **Ru-Z**, Figure S16 Panels B and C. Simulations of both dyes from dataset **Ru-Z** exhibit faster decay of the ~380 nm ESA compared to experiment, Figures S14 and S16 respectively, though the initial and final intensities are well matched. While RuP2 on TiO$_2$, like RuP, does not have an ultrafast relaxation pathway from the $|X\rangle$ to $|S_0\rangle$, we found that for RuP3 simulations without that relaxation pathway do not agree with experimental results. Refining the model for RuP3 in solution using higher resolution experiments could resolve this mismatch and improve agreement with simulations of RuP3 on substrates.

Like RuP3, the methyl-phosphonated derivatives of RuBPY were only in good agreement with measured spectra by having the ultrafast relaxation pathway from the $|X\rangle$ to $|S_0\rangle$. RuCP shows an experimentally observed redshift of the isosbestic point on the picosecond time scale rather than the femtosecond timescale found with the phosphonated derivatives in set **6-Ru** (Section S10.3 Figure S17A). The simulated TA spectrum of RuCP, Figure S17 Panels B and C, are in excellent agreement with this observation. The existence of the ultrafast decay pathway allows for population to return to $|S_0\rangle$, reducing the intensity of the GSB and ESE signal components as well as the excited state population entering the triplet manifold. The lower concentration of excited state population in the triplet manifold reduces the observed rate of charge injection from $|T\rangle$.

The ultrafast relaxation pathway in RuCP introduces a challenge to using the redshift of the isosbestic point to extract the rate of charge injection. For dataset **Ru-G**, Section S10.3 Figure S18 shows in Panel A) the change in the difference between the final isosbestic point frequency ω(t) and the time-dependent isosbestic point frequency ω(-∞) in wavenumbers for the experimental (black) and simulated (blue) spectra, B) the fraction of the total rate of charge injection from each excited state, and C) the relation between the total rate of charge injection and the change in the isosbestic point frequency. The change in the frequency of the isosbestic point in Figure S18A is much slower than observed for RuP (Main Document Figure 6A), though the fraction of triplet contribution to the total rate of injection dominates much earlier in the simulation in RuCP (Figure S18B) than in the simulation of RuP (Main Document Figure 6B). The relation in Figure S18C suggest that the total rate of charge injection and the change in the isosbestic point frequency are much less correlated in RuCP than in RuP (Main Document Figure 6C).

The simulation of RuCP for dataset **Ru-Z**, Section S10.3 Figure S19, slightly overestimates the isosbestic point redshift on the 10s to 100s of picoseconds timescales, though is well within quantitative agreement with the experimental spectrum. The high energy ESA and the GSB intensities agree at initially and at long delay times. Analysis of the decay of the high energy ESA to extract the total rate of charge injection as done in Ref 3 could be a viable alternative to analysis of the lower energy isosbestic point, though resolving the relative contribution of the excited states to this signal component would be necessary to make such analysis feasible from our simulations.

The remaining methyl-phosphonated derivatives, RuCP2 and RuCP3, similarly to RuCP, show a redshift of the low energy isosbestic point on the 10s of picosecond timescale in dataset **Ru-G** (Sections S10.4 and S10.5, Panel A of Figures S20 and S22). The initial and final wavelength of the zero intensity wavelength in the simulated spectra (Panels B and C in Figures S20 and S22) are slightly to the red of those observed in experiment, also seen with RuCP, though the timescale of the redshift are in agreement. Simulations of both dyes from dataset **Ru-Z** overestimate the intensity of the GSB in the 500 nm to 700 nm region of the spectra, though within experimental error (Figures S21 and S23). Experimental and simulated intensities of the high energy ESAs agree well at all delay points.

## 10.1. RuP2

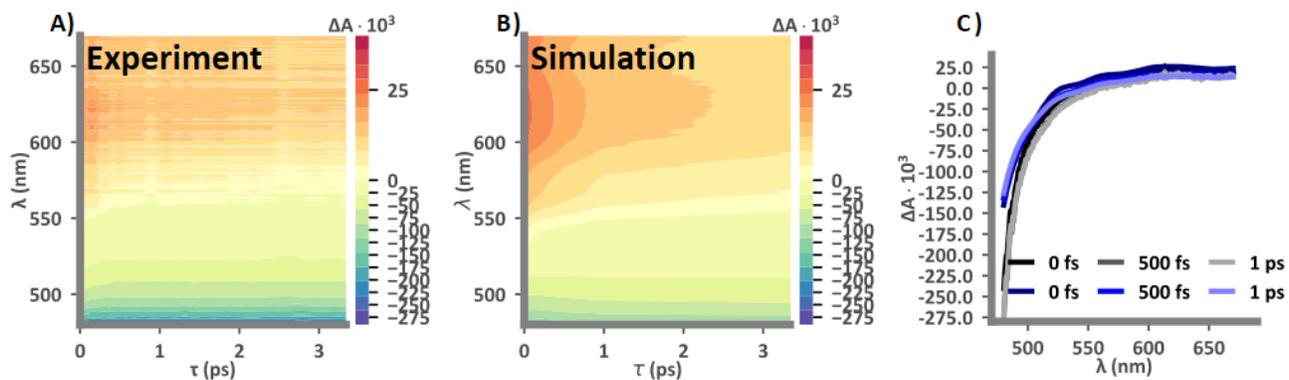

Figure S13. A) Experimental and B) simulated TA spectra of dye RuP2 on TiO$_2$ from data set **Ru-G**. C) Direct comparison of experimental (grays) and simulated (blues) TA lineshapes at delay times of 0 fs, 500 fs, and 1 ps.

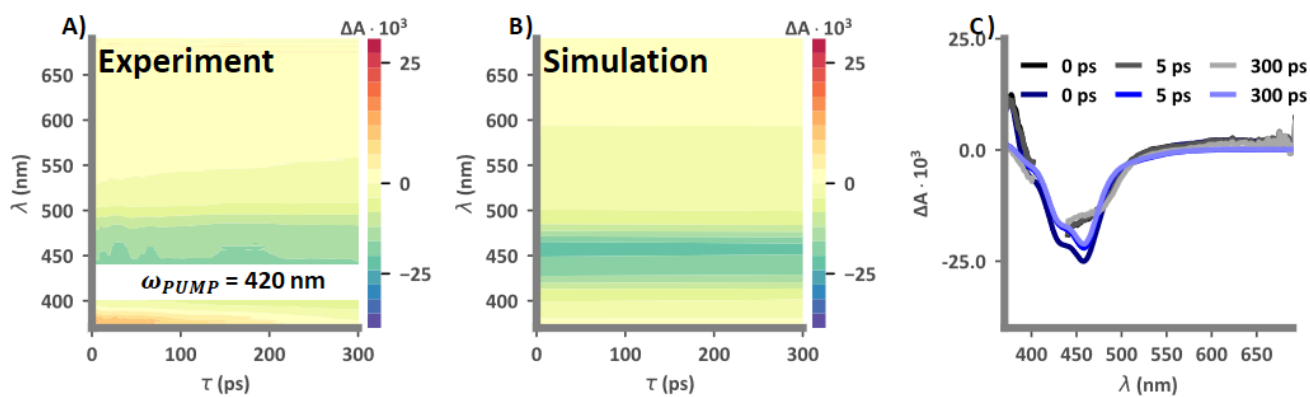

Figure S14. A) Experimental and B) simulated TA spectra of dye RuP2 on TiO$_2$ from data set **Ru-Z**. C) Direct comparison of experimental (grays) and simulated (blues) TA lineshapes at delay times of 0 fs, 5 ps, and 300 ps.

## 10.2. RuP3

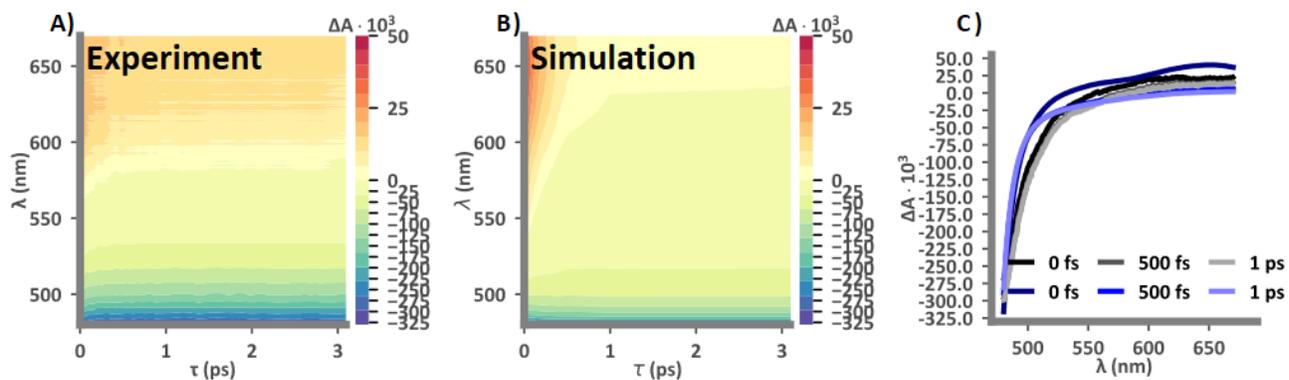

Figure S15. A) Experimental and B) simulated TA spectra of dye RuP3 on TiO$_2$ from data set **Ru-G**. C) Direct comparison of experimental (grays) and simulated (blues) TA lineshapes at delay times of 0 fs, 500 fs, and 1 ps.

v

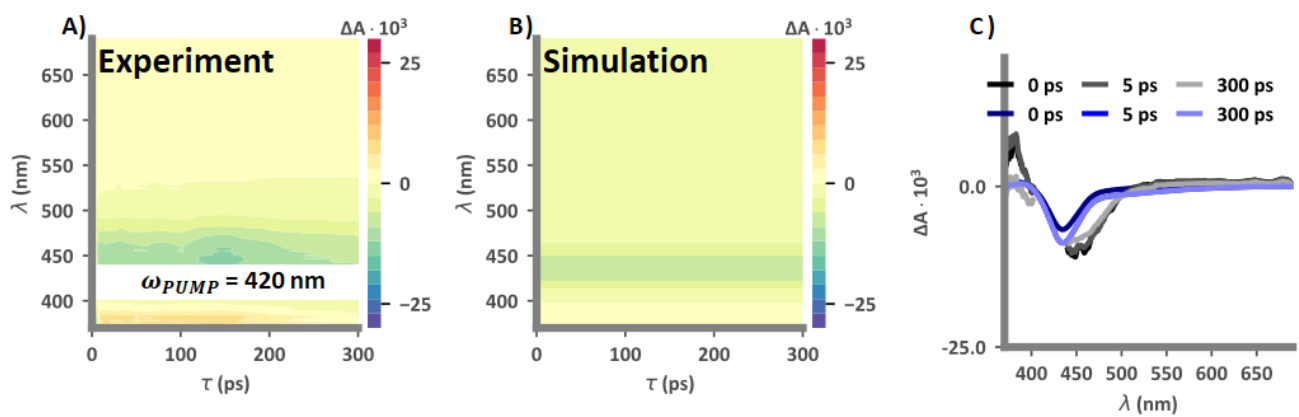

Figure S16. A) Experimental and B) simulated TA spectra of dye RuP3 on TiO$_2$ from data set **Ru-Z**. C) Direct comparison of experimental (grays) and simulated (blues) TA lineshapes at delay times of 0 fs, 5 ps, and 300 ps.

## 10.3. RuCP

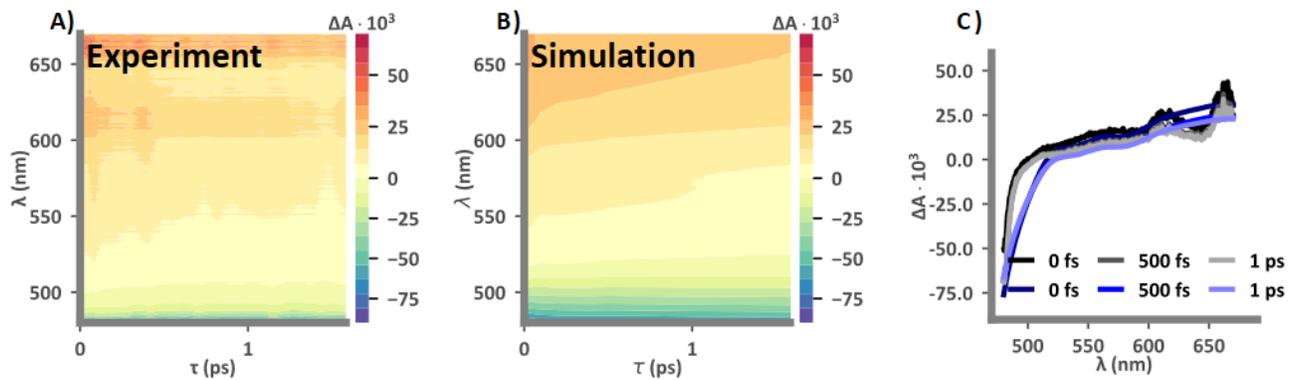

Figure S17. A) Experimental and B) simulated TA spectra of dye RuCP on $TiO_2$ from data set **Ru-G**. C) Direct comparison of experimental (grays) and simulated (blues) TA lineshapes at delay times of 0 fs, 500 fs, and 1 ps.

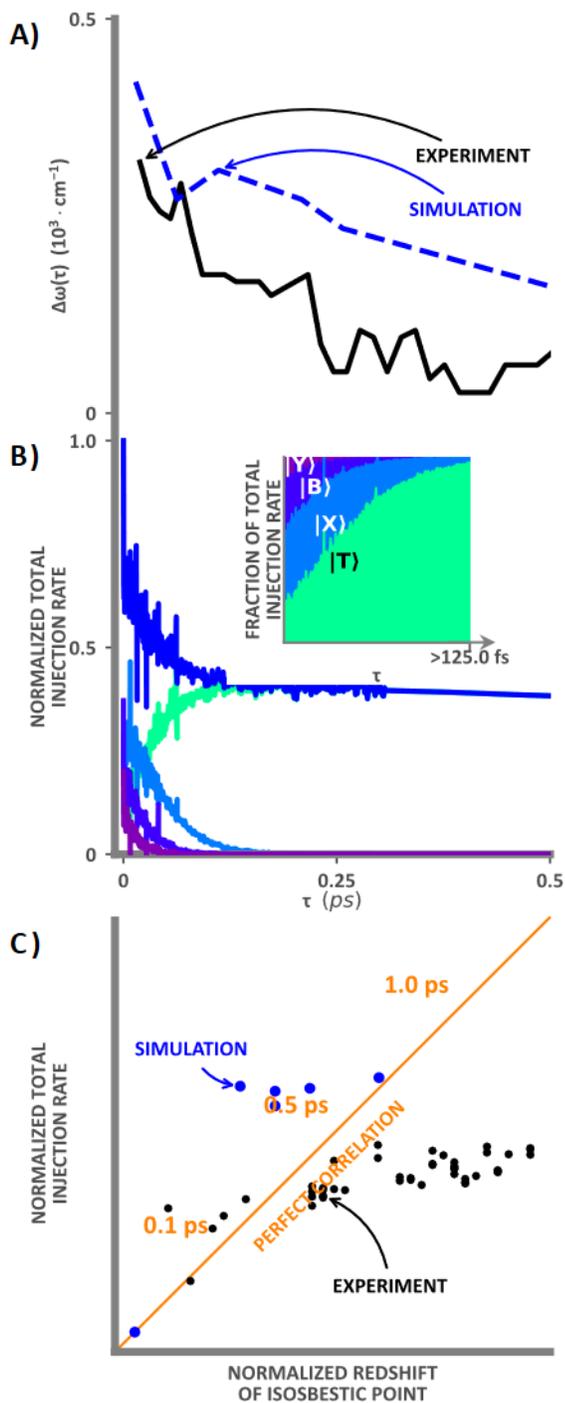

Figure S18. A) The change in experimental (black) and simulated (blue) difference between the final isosbestic point frequency ω(t) and the time-dependent isosbestic point frequency ω(-∞) in wavenumbers for RuCP. B) Normalized total simulated injection rate (blue) and contributions of the total normalized injection rate for states $|Y\rangle$ (violet), $|B\rangle$ (indigo), $|X\rangle$ (azure), and $|T\rangle$ (cyan). The inset shows the fraction of the total normalized injection rate for states $|Y\rangle$ (violet), $|B\rangle$ (indigo), $|X\rangle$ (azure), and $|T\rangle$ (cyan). C) Normalized total simulated injection rate plotted against the experimental (black) and simulated (blue) normalized redshift of the isosbestic point. The points have a slope askew of the diagonal (orange), suggesting low correlation compared to that of RuP.

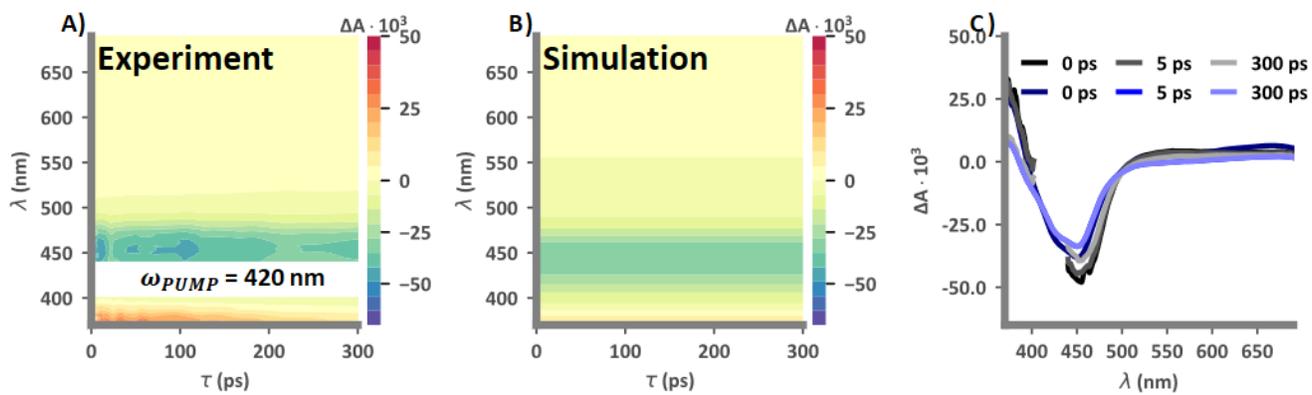

Figure S19. A) Experimental and B) simulated TA spectra of dye RuCP on TiO$_2$ from data set **Ru-Z**. C) Direct comparison of experimental (grays) and simulated (blues) TA lineshapes at delay times of 0 fs, 5 ps, and 300 ps.

## 10.4. RuCP2

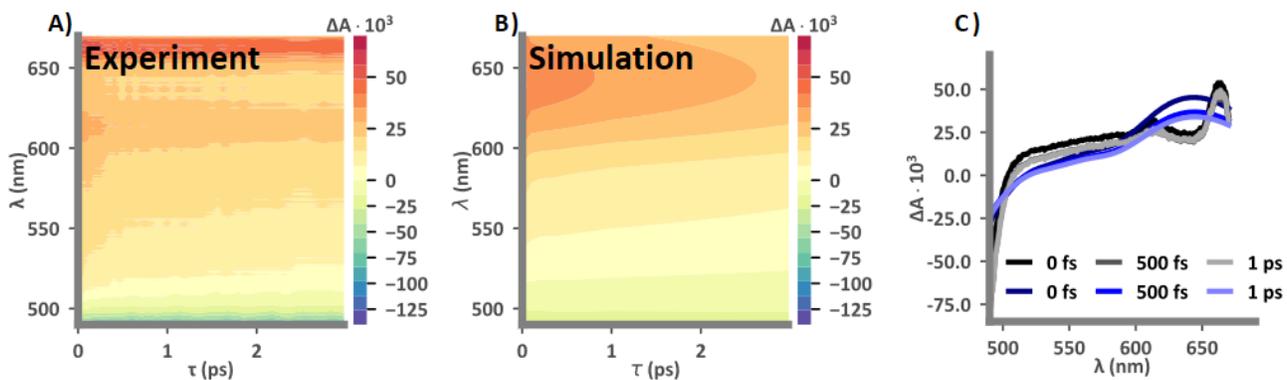

Figure S20. A) Experimental and B) simulated TA spectra of dye RuCP2 on TiO$_2$ from data set **Ru-G**. C) Direct comparison of experimental (grays) and simulated (blues) TA lineshapes at delay times of 0 fs, 500 fs, and 1 ps.

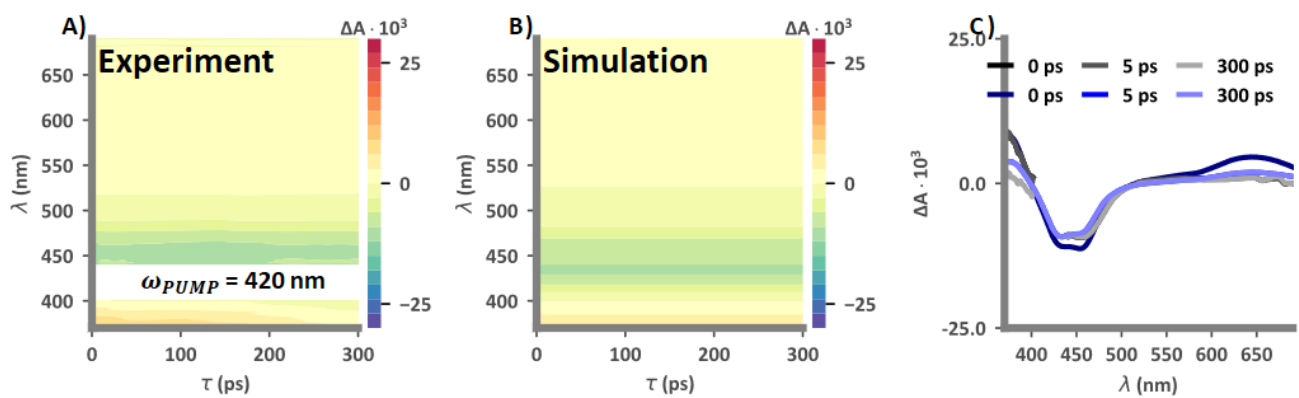

Figure S21. A) Experimental and B) simulated TA spectra of dye RuCP2 on TiO$_2$ from data set **Ru-Z**. C) Direct comparison of experimental (grays) and simulated (blues) TA lineshapes at delay times of 0 fs, 5 ps, and 300 ps.

## 10.5. RuCP3

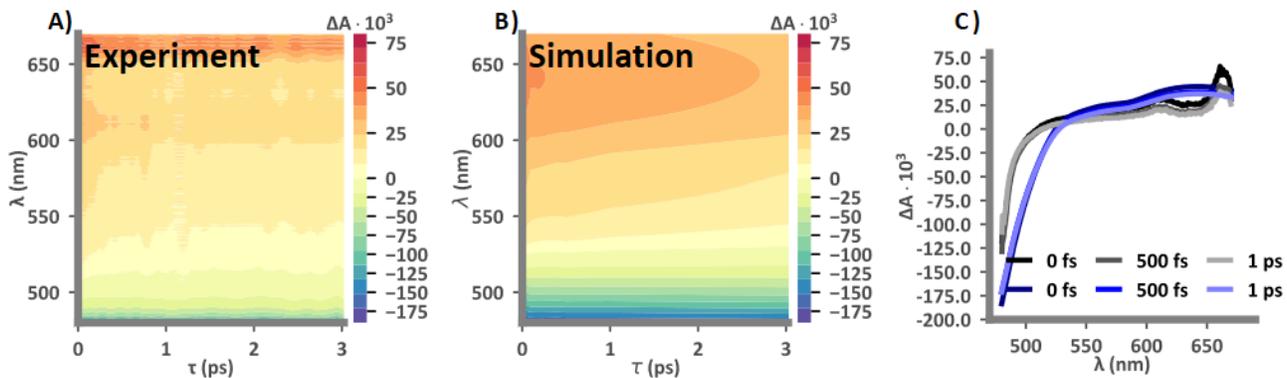

Figure S22. A) Experimental and B) simulated TA spectra of dye RuCP3 on TiO$_2$ from data set **Ru-G**. C) Direct comparison of experimental (grays) and simulated (blues) TA lineshapes at delay times of 0 fs, 500 fs, and 1 ps.

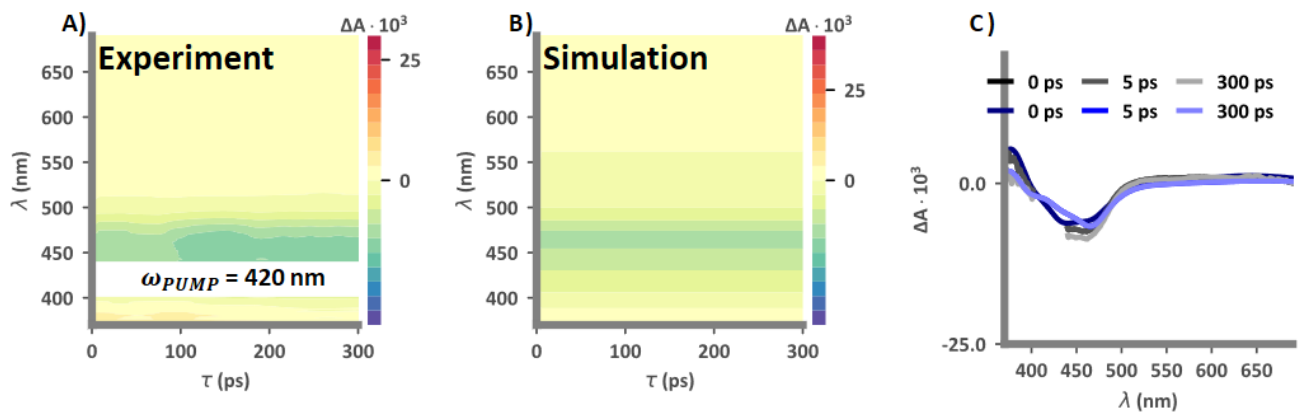

Figure S23. A) Experimental and B) simulated TA spectra of dye RuCP3 on TiO$_2$ from data set **Ru-Z**. C) Direct comparison of experimental (grays) and simulated (blues) TA lineshapes at delay times of 0 fs, 5 ps, and 300 ps.

## S11. Populations of Ru$^{II}$ Excited States Under Solar Illumination

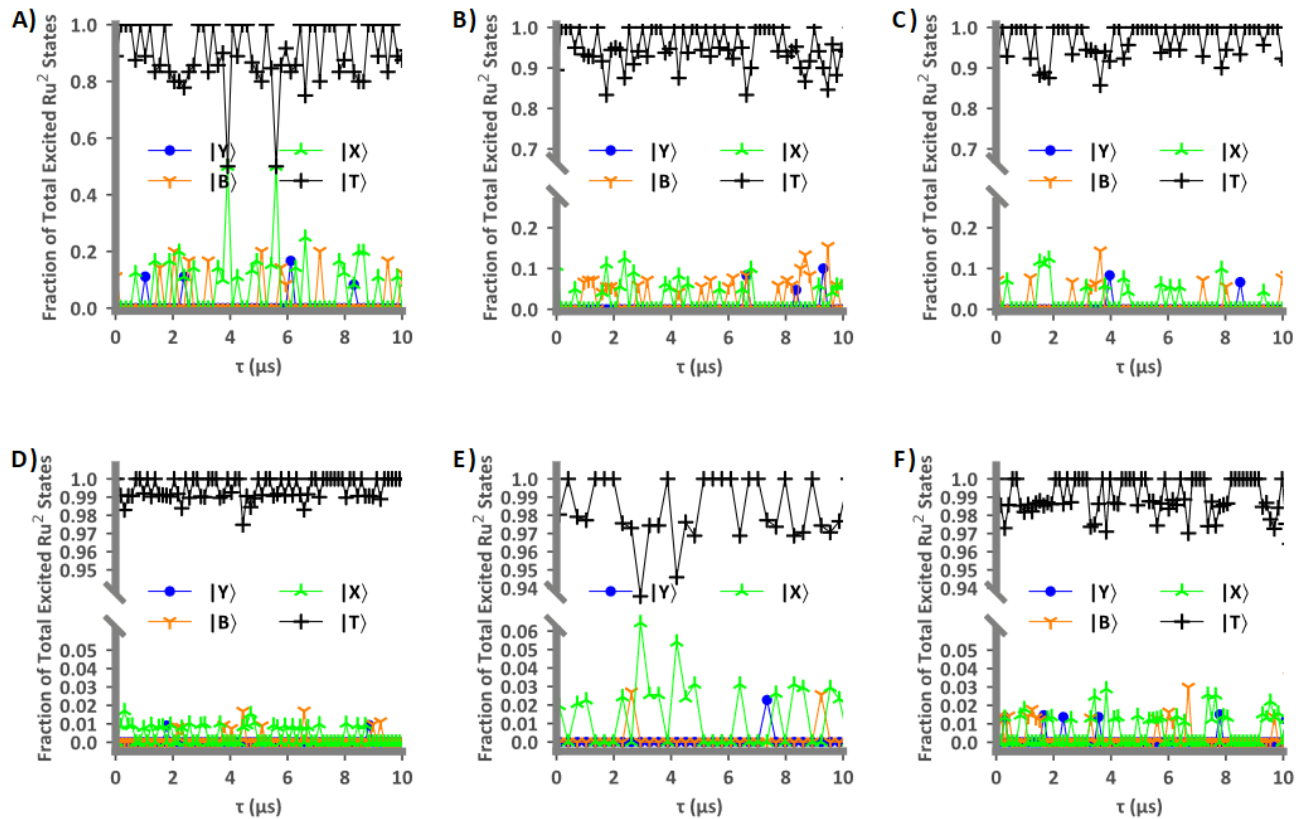

Figure S24. Fraction of Ru$^{II}$ excited state populations to total number of Ru$^{II}$ excited states during 0.1 μs of solar illumination for dye A) RuP, B) RuP2, C) RuP3, D) RuCP, E) RuCP2, and F) RuCP3.

## 12. Computed Errors of Simulation Values

The standard deviation of the injected electrons per dye per second, injected electrons per $cm^2$ per second, and the charge injection efficiencies for each state and total charge injection efficiency were calculated for each dye from four simulations run at a minimum of 1 μs. The errors are orders of magnitude smaller than the associated value and can be considered negligible.

Table S6. Computed errors of simulation values in Table 3 and Table 4 of Main Document

|  | Table 3 | | Table 4 | | | | |
|---|---|---|---|---|---|---|---|
| Dye | $e^-$ dye$^{-1}$ s$^{-1}$ | $e^-$ cm$^{-2}$ s$^{-1}$ | $|Y\rangle$ | $|B\rangle$ | $|X\rangle$ | $|T\rangle$ | Total Charge Injection Efficiency |
| RuP | ±0.003 | 6·10$^{13}$ | 2·10$^{-5}$ | 8·10$^{-6}$ | 8·10$^{-6}$ | 3·10$^{-5}$ | 3·10$^{-8}$ |
| RuP2 | ±0.01 | 2·10$^{14}$ | 2·10$^{-6}$ | 6·10$^{-6}$ | 5·10$^{-6}$ | 5·10$^{-5}$ | 7·10$^{-8}$ |
| RuP3 | ±0.007 | 1·10$^{14}$ | 3·10$^{-6}$ | 3·10$^{-6}$ | 3·10$^{-6}$ | 1·10$^{-5}$ | 1·10$^{-5}$ |
| RuCP | ±0.003 | 5·10$^{13}$ | 1·10$^{-6}$ | 3·10$^{-6}$ | 2·10$^{-6}$ | 3·10$^{-5}$ | 5·10$^{-6}$ |
| RuCP2 | ±0.004 | 7·10$^{13}$ | 5·10$^{-6}$ | 2·10$^{-6}$ | 6·10$^{-6}$ | 3·10$^{-5}$ | 7·10$^{-6}$ |
| RuCP3 | ±0.002 | 3·10$^{13}$ | 6·10$^{-6}$ | 4·10$^{-6}$ | 2·10$^{-6}$ | 7·10$^{-5}$ | 2·10$^{-6}$ |

References

[1] T. P. Cheshire *et al.*, J Phys Chem B **124** (2020) 5971.
[2] P. G. Giokas *et al.*, J Phys Chem C **117** (2013) 812.
[3] D. F. Zigler *et al.*, J Am Chem Soc **138** (2016) 4426.